\documentclass[useAMS,referee,usenatbib]{biom}
\usepackage{amsmath}
\usepackage{amsfonts}
\usepackage{graphicx}
\usepackage{algorithm}
\usepackage{algpseudocode}
\usepackage{booktabs}
\usepackage{url}

\usepackage[figuresright]{rotating}

\newcommand{\code}[1]{\texttt{#1}}

\newcommand{\Normal}{\mathcal{N}}

\newtheorem{proposition}{Proposition}

\title[TVSurvBART for Longitudinal Survival Data]{Bayesian Nonparametric Survival Regression with Time-Varying Covariates and Random Effects}

\author{
Keming Zhang$^{1}$,
Yudan Zou$^{2}$,
Fan Li$^{3}$, and
Liangyuan Hu$^{2,*}$\email{liangyuan.hu@rutgers.edu} \\
$^{1}$Department of Biostatistics, Brown University, Providence, Rhode Island, U.S.A. \\
$^{2}$Department of Biostatistics and Epidemiology, Rutgers University, Piscataway, New Jersey, U.S.A. \\
$^{3}$Department of Biostatistics, Yale University, New Haven, Connecticut, U.S.A.
}

\begin{document}


\label{firstpage}


\begin{abstract}
In longitudinal survival studies, biomarkers and treatments change across visits, their associations with event risk may be nonlinear and change over time, and subjects with identical measurements may differ in risk. Existing tree-ensemble survival models address these features separately: discrete-time formulations accommodate covariate updates without subject-specific random effects, and hazard-based ensembles with frailty take covariates as fixed at baseline. We propose Time-Varying Survival Bayesian Additive Regression Trees (TVSurvBART), a continuous-time hazard model combining a flexible baseline, a log-linear tree ensemble, and subject-specific random intercepts and slopes; covariates enter as predictable step processes, and interval-start time among the tree inputs lets effects change across visits. Integrated exposure offsets absorb the within-interval variation of the baseline and random slopes, so the tree leaves retain conjugate updates. Re-centering moves along likelihood-invariant directions and alternating centered and non-centered covariance updates address the resulting posterior dependence. In two simulation families, TVSurvBART's landmark predictions from held-out subjects' histories up to the landmark time have higher time-dependent AUC than the Cox, frailty Cox, and forest benchmarks in every setting and lower Brier scores in all but one comparison, although population-summary gains are not uniform. In the CARRA juvenile idiopathic arthritis registry, discrimination for recorded flares is similar to Cox regression, while a secondary interval-probability score favors TVSurvBART.
\end{abstract}

%
%

\maketitle
\section{Introduction}
\label{sec:introduction}

Longitudinal survival studies record biomarkers, clinical states, and treatment indicators at repeated visits, and a central question is how a subject's event risk changes as these measurements evolve. In the motivating Childhood Arthritis and Rheumatology Research Alliance (CARRA) registry study, patients whose juvenile idiopathic arthritis has become inactive on adalimumab are followed for flare at clinic visits that update disease activity, pain, joint findings, and dosing. The associations between these evolving measurements and event risk may be nonlinear, involve complex interactions, and change over time. The Cox model accommodates predictable time-varying covariates \citep{Cox1972}, but nonlinear terms, interactions, and time-dependent coefficients generally require advance specification, which becomes difficult with many longitudinal predictors. The survival forest method of \citet{Yao2022} offers a flexible alternative that accommodates time-varying covariates and learns nonlinear associations and interactions without prespecifying their functional forms.

Beyond the effects of measured covariates, subjects with similar observed histories may still differ in their underlying event risk. This unobserved heterogeneity motivates frailty models \citep{vaupel1979impact}. A random intercept captures a persistent shift in a subject's log-hazard, while a random slope allows this shift to evolve over time. Combining flexible, data-adaptive covariate effects with structured subject-specific heterogeneity can account for both sources of variation within a unified survival model. However, with at most one event per subject, random effects are weakly informed by the data and require regularization. Their dependence on the baseline hazard and flexible covariate effects also creates substantial challenges for posterior computation.

Bayesian additive regression trees (BART), a sum of trees with priors that keep each tree simple \citep{chipman2010bart}, learn nonlinear associations and interactions without prespecified forms. Later work added sparsity-inducing priors for variable selection \citep{linero2018bayesian} and soft trees \citep{LineroYang2018}. \citet{Hill2020Review} review the methodology. For survival data, \citet{sparapani2016nonparametric} discretize time and model the conditional event indicators with probit BART, using time as a predictor. Related constructions handle competing risks \citep{sparapani2020nonparametric} and recurrent events with time-varying covariates \citep{Sparapani2020Recurrent}. This formulation allows covariate updates but ties the hazard to a time grid and has no subject-specific random effects. Continuous-time ensembles include accelerated failure time models \citep{Henderson2020AFT, Sparapani2023NFT, Hu2022Clustered} and hazard-based models \citep{Linero2022Survival, Basak2022SBART}, among them an excess-hazard model with conjugate leaf updates based on event counts and exposures \citep{Basak02012026}. These models take covariates as fixed at baseline, and the random effects of \citet{Hu2022Clustered} and \citet{Basak2022SBART} are shared within a cluster and constant over time.

To our knowledge, existing continuous-time Bayesian survival tree ensembles do not jointly combine covariates updated during follow-up, time-varying effects learned from the data, and subject-specific random intercepts and slopes with tractable posterior computation. Much of the difficulty is computational: between visits the tree contribution is constant, but a smooth baseline and a random-slope term keep varying, so each interval's cumulative hazard involves every component.

We propose time-varying survival BART (TVSurvBART), which combines a piecewise-constant or exposure-centered spline baseline, a log-linear BART risk surface \citep{Murray03042021}, and correlated subject-specific random intercepts and slopes. Covariates enter as predictable step processes, and interval-start time can enter the trees so that effects may change over follow-up. Given the baseline and random effects, the likelihood is a product of Poisson-type kernels whose exposure offsets, computed by one-dimensional quadrature, absorb all within-interval variation, so the leaves retain conjugate updates. The overall hazard level, and with a spline baseline a common time trend, can shift among components without changing any hazard (Proposition~\ref{prop:invariance}). To improve mixing, we add exact re-centering moves along these directions, paired with centered and non-centered covariance updates. The contribution is this longitudinal formulation and its posterior computation; the tree prior and frailty distribution are standard.

The model yields a posterior survival distribution along each subject's covariate path. In two simulation families, TVSurvBART's landmark predictions attain, in every setting, a higher time-dependent area under the receiver operating characteristic curve (AUC) than time-varying Cox regression, a frailty Cox model, and the survival forest method of \citet{Yao2022}, and lower Brier scores in all but one comparison. TVSurvBART's errors for population-level restricted mean and median survival, measured against latent event times, are not uniformly smaller. In the registry, discrimination for recorded flares is similar to that of Cox regression, whereas a secondary interval-probability score favors TVSurvBART.

Section~\ref{sec1:model} defines the model. Section~\ref{sec:priors} specifies the priors, and Section~\ref{sec:posterior} describes posterior computation. Section~\ref{sec:simulation} compares TVSurvBART with the benchmarks in two simulation families. Section~\ref{sec:application} makes the same comparison for recorded flares in the registry. Section~\ref{sec:discussion} discusses limitations and extensions.

\section{The TVSurvBART Model}\label{sec1:model}

\subsection{Notation}\label{sec:notation}

Consider $N$ independent subjects. For subject $i$, let $T_i^*$ be the event time, $C_i$ the censoring time, $T_i=\min(T_i^*,C_i)$, and $\delta_i=\mathbb{I}(T_i^*\le C_i)$. Let $t_{i0}=0<t_{i1}<\cdots<t_{i,n_i}<T_i$ be the measurement times strictly before the end of follow-up, and append $t_{i,n_i+1}=T_i$. Write $\bm{x}_{i,j-1}$ for the measurement obtained at $t_{i,j-1}$ and use $\bm{x}_i(t)=\bm{x}_{i,j-1}$ on $(t_{i,j-1},t_{ij}]$. Thus, a measurement at an event time is not used to explain that event. Let $\bm{U}_i(t)=\bm{U}_{ij}$ contain $\bm{x}_{i,j-1}$ and, optionally, the scaled interval start $t_{i,j-1}/\tau$, where $\tau>0$ is a fixed time scale.

\subsection{Model Specification}\label{sec:model}

TVSurvBART specifies the conditional hazard through an additive log-hazard with a tree ensemble and Gaussian random effects:
$
\lambda_i(t\mid\bm{b}_i)=\lambda_i(t)=\exp\{\eta_i(t)\},
$
where
\begin{equation}
\eta_i(t)
=
\alpha_0(t)
+\sum_{h=1}^{m}g_h\bigl(\bm{U}_i(t);R_h,M_h\bigr)
+\bm{b}_i^{\top}\bm{z}_i(t),
\qquad
\bm{b}_i\sim\mathcal{N}(\bm{\mu}_b,\Sigma_b).
\label{eq:eta}
\end{equation}
Here $\alpha_0(t)=\log\lambda_0(t)$ is the log-baseline hazard, and $R_h$ and $M_h$ are the structure and terminal-node values of tree $h$. We fix $\bm\mu_b=\bm0$ and, unless stated otherwise, use the random-intercept-and-slope design $\bm z_i(t)=(1,t/\tau)^{\top}$, so that $q=2$. A random-intercept-only model, $\bm z_i(t)=1$ with $q=1$, is the special case obtained by removing the slope. The computations below apply to it with the slope coordinate dropped. Fixing the population mean of the random effects does not constrain their realized sample mean. The conditional event likelihood cannot separate certain common shifts among the baseline, trees, and random effects. The priors regularize these directions, and Section~\ref{sec:level-moves} uses them to construct joint updates.

We consider two alternative representations for the log-baseline hazard: an exposure-weighted, centered natural cubic spline and a piecewise-constant function. For the natural cubic spline specification, let $\kappa_0<\kappa_1<\cdots<\kappa_{K-1}$ be the knots, the first and last of which are the boundary knots, and let $B_0(t)=1,B_1(t)=t,B_2(t),\ldots,B_{K-1}(t)$ be the corresponding $K$-dimensional natural cubic spline basis, given explicitly in Web Appendix~B.1. The nonintercept basis functions are centered at their means over the observed at-risk time, $\widetilde B_\ell(t)=B_\ell(t)-\bar B_\ell$ with $\bar B_\ell=\sum_{i=1}^{N}\int_0^{T_i}B_\ell(u)\,du/\sum_{i=1}^{N}T_i$ for $\ell=1,\ldots,K-1$.
The corresponding log-baseline hazard is
$
\alpha_0(t)
=
\beta_0+\sum_{\ell=1}^{K-1}
\beta_\ell\widetilde B_\ell(t).
$
By construction,
\(\sum_{i=1}^{N}\int_0^{T_i}\widetilde B_\ell(u)\,du=0\), so \(\beta_0\) represents the
exposure-weighted baseline log-hazard level, while
\(\{\beta_\ell\}_{\ell=1}^{K-1}\) determine its temporal shape. Alternatively, the log-baseline hazard may be modeled as a step function.
Let $0=\tau_0<\tau_1<\cdots<\tau_J$ define a partition of the follow-up period. The piecewise-constant specification is
$
\alpha_0(t)
=
\sum_{j=1}^{J}
\gamma_j
\mathbb{I}\!\left(\tau_{j-1}<t\leq\tau_j\right),
$
where \(\gamma_j\) is the log-baseline hazard on the \(j\)th interval.
Equivalently,
$
\lambda_0(t)
=
\sum_{j=1}^{J}
\exp(\gamma_j)
\mathbb{I}\!\left(\tau_{j-1}<t\leq\tau_j\right),
$
so the baseline hazard is constant within each interval but may vary
between intervals.

\subsection{Log-Linear BART}\label{sec-bart}

The tree component of (\ref{eq:eta}) is $F(\bm{u})=\sum_{h=1}^m g_h(\bm{u};R_h,M_h)$, with $f(\bm{u})=\exp\{F(\bm{u})\}$.
The \(h\)th tree is characterized by its structure \(R_h\) and terminal-node
parameters \(M_h=(\mu_{h1},\ldots,\mu_{hb_h})\). Its tree structure partitions
the covariate space into \(b_h\) disjoint terminal regions
\(\{\mathcal{A}_{h1},\allowbreak\ldots,\allowbreak\mathcal{A}_{hb_h}\}\), such that
$
g_h(\bm{u};R_h,M_h)
=
\mu_{hk}
$
if
$
\bm{u}\in\mathcal{A}_{hk}.
$

Following the log-linear BART construction of
\citet{Murray03042021}, a positive-valued regression function is obtained by
placing the additive tree ensemble on the log scale:
$
f(\bm{u})
=
\exp\!\left\{
\sum_{h=1}^{m}g_h(\bm{u};R_h,M_h)
\right\}
=
\prod_{h=1}^{m}
G_h(\bm{u};R_h,\bm{\theta}_h),
$
where
$
\theta_{hk}=\exp(\mu_{hk}),
$
$
\bm{\theta}_h
=
(\theta_{h1},\ldots,\theta_{hb_h}),
$
and
$
G_h(\bm{u};R_h,\bm{\theta}_h)
=
\exp\!\left\{g_h(\bm{u};R_h,M_h)\right\}
=
\theta_{hk}
$
if
$
\bm{u}\in\mathcal{A}_{hk}, k=1,\ldots,b_h.
$
Applying this construction to the interval-constant tree input
\(\bm{U}_i(t)\), the hazard $\lambda_i(t)$ can equivalently be written as
$
\lambda_i(t)
=
\exp\!\left\{
\alpha_0(t)+\bm{b}_i^{\top}\bm{z}_i(t)
\right\}
\prod_{h=1}^{m}
G_h\!\left(
\bm{U}_i(t);R_h,\bm{\theta}_h
\right).
$
The ensemble is thus additive on the log-hazard scale and multiplicative on the hazard scale, which keeps the hazard positive while retaining the flexibility of BART for nonlinear effects and interactions.

\subsection{Event-process likelihood}\label{sec:likelihood}

We assume that, given the observed history before $t$, the censoring, visit, and covariate processes at $t$ depend neither on $\bm b_i$ nor on the baseline, tree, and random-effect covariance parameters, and that the parameters of these processes are distinct from and a priori independent of those parameters. The factors of these processes are omitted, and with the partition and inputs of Section~\ref{sec:notation}, the event-process likelihood given the random effects is
\begin{equation}
\begin{aligned}
L
&=
\prod_{i=1}^{N}
\lambda_i(T_i)^{\delta_i}
\exp\!\left\{
-\int_0^{T_i}\lambda_i(u)\,du
\right\}
=
\prod_{i=1}^{N}
w_i\,
f\!\left(\bm{U}_{i,n_i+1}\right)^{\delta_i}
\prod_{j=1}^{n_i+1}
\exp\!\left\{
-v_{ij}f(\bm{U}_{ij})
\right\},
\end{aligned}
\label{eq:likelihood}
\end{equation}
where
$
w_i
=
\exp\!\left\{
\delta_i
\left[
\alpha_0(T_i)
+
\bm{b}_i^{\top}\bm{z}_i(T_i)
\right]
\right\}
$
and
$
v_{ij}
=
\int_{t_{i,j-1}}^{t_{ij}}
\exp\!\left\{
\alpha_0(u)
+
\bm{b}_i^{\top}\bm{z}_i(u)
\right\}
\,du.
$

Conditional on the current baseline-hazard and random-effects parameters,
\(w_i\) does not depend on the tree ensemble, and the \(v_{ij}\) are fixed
nonnegative exposure offsets. Defining
$
y_{ij}
=
\delta_i\,
\mathbb{I}\!\left(j=n_i+1\right),
$
the likelihood kernel for \(f\) can therefore be written as
$
L_f
\propto
\prod_{i=1}^{N}
\prod_{j=1}^{n_i+1}
f(\bm{U}_{ij})^{y_{ij}}
\exp\!\left\{
-v_{ij}f(\bm{U}_{ij})
\right\}.
$
This is a Poisson likelihood kernel in $f$ only when the baseline and random effects are fixed. It gives conjugate tree updates, but its omitted factor depends on the other blocks and must be retained in their updates (Web Appendix~A.3). The intervals do not represent independent recurrent events.

\section{Prior Specification}\label{sec:priors}

The tree ensemble, the random-effects covariance, and the baseline parameters receive independent priors. Web Appendix~B.2 gives their full forms, the calibration of the leaf prior, and the choice of knots and intervals.

\subsection{Tree structures and terminal-node values}

The trees are independent a priori: each structure $R_h$ has the BART structure prior $p(R_h)$ of \citet{chipman2010bart}, and, given the structures, the terminal-node values $\theta_{hk}=\exp(\mu_{hk})$ are independent with the balanced mixture prior
\begin{equation}
p_\theta(\theta_{hk}\mid c,d)=\frac{1}{2}\,p_{\mathrm{GIG}}(\theta_{hk}\mid -c,2d,0)+\frac{1}{2}\,p_{\mathrm{GIG}}(\theta_{hk}\mid c,0,2d),
\label{eq:theta-prior}
\end{equation}
where $p_{\mathrm{GIG}}(x\mid\eta,\chi,\psi)\propto x^{\eta-1}\exp\{-(\chi/x+\psi x)/2\}$, $x>0$, is the generalized inverse Gaussian (GIG) density; equivalently, with probability $1/2$ each, $\theta_{hk}$ is a $\mathrm{Gamma}(c,d)$ variable, in the shape--rate parameterization used throughout, or its reciprocal. This prior is symmetric in $\mu_{hk}$ about zero and conditionally conjugate under the likelihood kernel $L_f$ of Section~\ref{sec:likelihood}. Following \citet{Murray03042021}, we set $c=m/s_F^2+1/2$ and $d=m/s_F^2$, where $s_F>0$ is the ensemble scale, so that $F(\bm u)$ is approximately $\mathcal{N}(0,s_F^2)$ at any fixed $\bm u$ when $m$ is large. We use $m=100$ trees and $s_F\approx2.492$, fixed before the analyses reported here as a deliberately broad calibration under which two inputs with uncorrelated tree values have a central $95\%$ prior hazard-ratio interval of $[0.001,1000]$.

\subsection{Random-effects covariance}

The random-effects covariance has the hierarchical prior of \citet{huang2013simple}: given auxiliary scales $a_1,\ldots,a_q$ with inverse-gamma priors, $\Sigma_b$ is inverse-Wishart, and each standard deviation $\sqrt{\Sigma_{b,kk}}$, on the log-hazard scale, then has a half-$t_\nu(A_k)$ prior with $\nu$ degrees of freedom and scale $A_k$. We use $\nu=2$, which makes the prior on the intercept--slope correlation uniform, and $A_1=A_2=5$, which gives each standard deviation a prior median of about~$4$, far above plausible values. As the slope multiplies $t/\tau$, its prior scale depends on $\tau$: the administrative cutoff in the simulations ($3$ and $5$) and $5$ years in the case study.

\subsection{Baseline parameters}

The spline coefficients $\bm\beta=(\beta_0,\ldots,\beta_{K-1})^{\top}$ have independent $\mathcal{N}(\mu_\beta,\sigma_\beta^2)$ priors in the centered basis, and the piecewise hazard heights $\lambda_{0j}=\exp(\gamma_j)$ have independent $\mathrm{Gamma}(a_\lambda,b_\lambda)$ priors, with $\mu_\beta=0$, $\sigma_\beta^2=1$, and $a_\lambda=b_\lambda=0.1$. How informative these priors are depends on the time unit, the scaling of the basis, and the interval partition. The values given in this section are the defaults used in all analyses below unless stated otherwise.

\section{Posterior Computation}\label{sec:posterior}

Posterior computation uses a Markov chain Monte Carlo (MCMC) sampler as in Algorithm~\ref{alg:full}. Two features of the model make posterior sampling difficult. First, the random effects have no conjugate update, and with at most one event per subject they are weakly informed, so centered updates of their covariance can mix slowly. Second, the overall hazard level, and with a spline baseline a common time trend, can shift among the components without changing the likelihood (Proposition~\ref{prop:invariance}), which induces strong posterior dependence along these directions. Each iteration is also costly. Because the baseline and the random-slope term vary within each visit interval, the exposure offsets $v_{ij}$ in (\ref{eq:likelihood}) are integrals of $\exp\{\alpha_0(u)+\bm b_i^{\top}\bm z_i(u)\}$ over the intervals, and the baseline update, the $N$ random-effect updates, and each of the three non-centered covariance updates need such integrals at new parameter values. Integrals over all $\sum_{i=1}^{N}(n_i+1)$ intervals are therefore recomputed by one-dimensional quadrature at least five times per iteration. Section~\ref{sec:tree-updates} shows that the tree updates remain conjugate given these offsets, and Sections~\ref{sec:random-updates} and~\ref{sec:level-moves} address the two difficulties. Web Appendices~A and B give the full conditionals, acceptance ratios, and tuning rules.

\subsection{Tree updates}\label{sec:tree-updates}

For the $h$th tree, hold the baseline and the random effects, and hence $w_i$ and $v_{ij}$ in (\ref{eq:likelihood}), at their current values, and let $f_{(h)}(\bm u)=\prod_{\ell\ne h}G_\ell(\bm u;R_\ell,\bm\theta_\ell)$ be the product of the other trees. Let
$
r_{hk}=\sum_{i=1}^{N}\delta_i\,\mathbb{I}(\bm U_{i,n_i+1}\in\mathcal A_{hk}),
$
and
$
s_{hk}=\sum_{i=1}^{N}\sum_{j=1}^{n_i+1}\mathbb{I}(\bm U_{ij}\in\mathcal A_{hk})\,v_{ij}f_{(h)}(\bm U_{ij})
$
be the number of observed events assigned to terminal node $k$ and its partial-ensemble-weighted exposure. As a function of $\bm\theta_h$, the likelihood is then proportional to $\prod_{k=1}^{b_h}\theta_{hk}^{r_{hk}}\exp(-s_{hk}\theta_{hk})$, and integrating the terminal-node values under the prior (\ref{eq:theta-prior}) gives the marginal likelihood of the structure in closed form,
\begin{equation}
\mathcal{M}_h(R_h)=c_h\prod_{k=1}^{b_h}\frac{Z(-c+r_{hk},2d,2s_{hk})+Z(c+r_{hk},0,2[d+s_{hk}])}{2Z(c,0,2d)},
\label{eq:tree-marginal}
\end{equation}
where $c_h$ does not depend on $R_h$ and $Z(\eta,\chi,\psi)=\int_0^\infty x^{\eta-1}\exp\{-(\chi/x+\psi x)/2\}\,dx$ is the normalizing constant of the GIG density in (\ref{eq:theta-prior}), expressible through gamma functions and the modified Bessel function of the second kind (Web Appendix~B.3). A structure proposal $R_h^\ast$ is accepted by Metropolis--Hastings with target $p(R_h)\mathcal M_h(R_h)$. Every terminal-node value of the tree is then redrawn independently from its full conditional, a mixture of the GIG densities with parameters $(-c+r_{hk},2d,2s_{hk})$ and $(c+r_{hk},0,2[d+s_{hk}])$ whose weights are proportional to the two $Z$ terms in (\ref{eq:tree-marginal}). Because the leaves are redrawn also after a rejection, the structure move and the leaf draw together preserve the joint posterior (Web Appendix~A.2).

\subsection{Random effects, covariance, and baseline}\label{sec:random-updates}

Each $\bm b_i$ is updated by a Metropolis-adjusted Langevin algorithm (MALA) step with proposal $\bm b_i^\ast=\bm b_i+\tfrac12\varsigma_i^2\widetilde{\bm g}_i(\bm b_i)+\varsigma_i\bm\varepsilon_i$, $\bm\varepsilon_i\sim\mathcal N(\bm0,I_q)$, where $\widetilde{\bm g}_i(\bm b_i)=\bm g_i(\bm b_i)-\Sigma_b^{-1}\bm b_i$ is the gradient of its log full conditional and the score $\bm g_i$ of the subject's log-likelihood has an event-minus-exposure form (Web Appendix~A.2). Under the hierarchical prior of \citet{huang2013simple}, the auxiliary scales $a_1,\ldots,a_q$ and the covariance matrix $\Sigma_b$ are drawn from their inverse-gamma and inverse-Wishart full conditionals. This centered update can mix slowly when the random effects are weakly informed \citep{PapaspiliopoulosRobertsSkold2007,BetancourtGirolami2015}, as they are here with at most one event per subject, so each centered draw is followed by three non-centered Metropolis updates \citep{YuMeng2011}. Writing $\Sigma_b=L_bL_b^{\top}$ with $L_b$ lower triangular, a Gaussian random walk on the log-diagonal and below-diagonal entries of $L_b$ moves $\Sigma_b$ and every random effect jointly, $\bm b_i^\ast=L_b^\ast L_b^{-1}\bm b_i$, so the whitened effects $L_b^{-1}\bm b_i$ stay fixed and the acceptance ratio involves only the likelihood, the covariance prior, and the log-Cholesky Jacobian.

The spline coefficients $\bm\beta$ are updated jointly by a preconditioned MALA step whose likelihood score has the same event-minus-exposure form (Web Appendix~B.4). For the piecewise-constant baseline, with heights $\lambda_{0j}=\exp(\gamma_j)$ on $I_j=(\tau_{j-1},\tau_j]$, the gamma prior is conjugate: $\lambda_{0j}\mid\cdot\sim\mathrm{Gamma}(a_\lambda+d_j,b_\lambda+E_j)$, where $d_j$ counts the events in $I_j$ and $E_j$ is the corresponding exposure (Web Appendix~A.2). The MALA scales are adapted during burn-in, those of the random effects toward the acceptance rate $0.574$ of \citet{RobertsRosenthal1998}, and all proposal parameters are held fixed afterwards (Web Appendix~B.5).

\subsection{Likelihood-invariant re-centering moves}\label{sec:level-moves}

The overall hazard level can be redistributed among the baseline, the ensemble, and the random effects without changing any hazard (Section~\ref{sec:model}), which induces strong posterior dependence. Two re-centering moves therefore shift the parameters along these directions.

The ensemble--baseline level shift draws a mean-zero Gaussian increment $c_f$ and sets $\theta_{hk}^\ast=\theta_{hk}\exp(-c_f/m)$ for every terminal node, which multiplies the ensemble by $\exp(-c_f)$. The baseline compensates through $\beta_0^\ast=\beta_0+c_f$ under the spline baseline or $\lambda_{0j}^\ast=\lambda_{0j}\exp(c_f)$, $j=1,\ldots,J$, under the piecewise-constant baseline. Because the likelihood is unchanged and the proposal is symmetric, the Metropolis--Hastings ratio contains only prior ratios and the Jacobian, $\exp(-c_fH/m)$ for the spline and $\exp(c_fJ-c_fH/m)$ for the piecewise baseline, where $H=\sum_{h=1}^{m}b_h$ is the total number of terminal nodes.

The group shift moves all random effects against the baseline. With the spline baseline, write $\alpha_0(t)=\widetilde{\bm B}(t)^{\top}\bm\beta$ with $\widetilde{\bm B}(t)=(1,\widetilde B_1(t),\ldots,\widetilde B_{K-1}(t))^{\top}$, and let $A$ be the $K\times q$ matrix with $\widetilde{\bm B}(t)^{\top}A=\bm z_i(t)^{\top}$ for all $t$, which exists because the basis contains an intercept and linear time (Web Appendix~A.1). The shift $\bm b_i^\ast=\bm b_i-\bm c$, $i=1,\ldots,N$, with $\bm\beta^\ast=\bm\beta+A\bm c$, preserves $\alpha_0(t)+\bm b_i^{\top}\bm z_i(t)$, so a common shift of the random intercepts and slopes is absorbed by the baseline intercept and the linear spline coefficient. Drawing $\bm c$ from its Gaussian conditional distribution along this orbit makes the move an exact parameter-expansion Gibbs update \citep{LiuWu1999,MengvanDyk1999}. A piecewise-constant baseline cannot absorb a common slope shift, so with that baseline a Metropolis step shifts only the random intercepts against $\lambda_{01},\ldots,\lambda_{0J}$. Both moves change two components and rely on the following identity.

\begin{proposition}[Likelihood invariance of the re-centering moves]\label{prop:invariance}
Fix $c_f,c_g\in\mathbb{R}$ and $\bm c\in\mathbb{R}^q$. (i)~The level shift $\theta_{hk}^\ast=\theta_{hk}\exp(-c_f/m)$, $h=1,\ldots,m$, $k=1,\ldots,b_h$, together with $\beta_0^\ast=\beta_0+c_f$ under the spline baseline or $\lambda_{0j}^\ast=\lambda_{0j}\exp(c_f)$, $j=1,\ldots,J$, under the piecewise-constant baseline, leaves $\eta_i(t)$ in (\ref{eq:eta}) unchanged for every subject $i$ and every $t$ in the follow-up period. (ii)~Under the spline baseline, the group shift $\bm b_i^\ast=\bm b_i-\bm c$, $i=1,\ldots,N$, together with $\bm\beta^\ast=\bm\beta+A\bm c$ for the matrix $A$ defined above, likewise leaves every $\eta_i(t)$ unchanged. (iii)~Under the piecewise-constant baseline, so does the common-intercept shift $\bm b_i^\ast=\bm b_i-c_g\bm e_0$, $i=1,\ldots,N$, together with $\lambda_{0j}^\ast=\lambda_{0j}\exp(c_g)$, $j=1,\ldots,J$, where ${\bm e_0=(1,0,\ldots,0)^{\top}\in\mathbb{R}^q}$. Consequently, under each move every subject-specific hazard $\lambda_i(t)=\exp\{\eta_i(t)\}$, and therefore the likelihood (\ref{eq:likelihood}), is unchanged.
\end{proposition}

The likelihood thus leaves the split of the overall hazard level between the ensemble and the baseline to the broad leaf prior and the baseline prior. The data determine the level itself, so the posterior forms a long, narrow ridge. All other updates in Algorithm~\ref{alg:full} change at most one of the three components, so any step they take along the ridge changes $\eta_i(t)$ and must be short. Such updates alone can therefore mix slowly along the ridge \citep{LiuWu1999}, whereas the re-centering moves follow the invariant directions as far as the priors allow. Web Appendix~A.1 proves Proposition~\ref{prop:invariance} and derives the acceptance ratios and the Gaussian draw of $\bm c$.

\subsection{The sampler}\label{sec:full-sampler}

Algorithm~\ref{alg:full} lists one iteration. With proposals fixed after burn-in and exact integrals and normalizing constants, every complete block preserves the posterior (Web Appendix~A.2).

\begin{algorithm}[tb]
\caption{One iteration of the TVSurvBART sampler}
\label{alg:full}
\raggedright
\begin{algorithmic}[1]
\State \textbf{Baseline.} Spline: update $\bm\beta$ by a preconditioned MALA step targeting
\Statex \hspace*{\algorithmicindent}$\pi(\bm\beta\mid\cdot)\propto L\,\mathcal N(\bm\beta;\mu_\beta\bm1_K,\sigma_\beta^2I_K)$, with $L$ in (\ref{eq:likelihood}).\hfill(Web Appendix~B.4)
\Statex Piecewise: draw $\lambda_{01},\ldots,\lambda_{0J}$ from their gamma full conditionals
\Statex \hspace*{\algorithmicindent}$\lambda_{0j}\mid\cdot\sim\mathrm{Gamma}(a_\lambda+d_j,b_\lambda+E_j)$.\hfill(Web Appendix~A.2)
\State \textbf{Trees.} For $h=1,\ldots,m$, update $R_h$ by Metropolis--Hastings with target $p(R_h)\mathcal M_h(R_h)$ in (\ref{eq:tree-marginal}), then redraw each $\theta_{hk}$ from its GIG-mixture full conditional
\Statex \hspace*{\algorithmicindent}$\pi(\theta_{hk}\mid R_h,\cdot)\propto p_\theta(\theta_{hk}\mid c,d)\,\theta_{hk}^{r_{hk}}e^{-s_{hk}\theta_{hk}}$.\hfill(Web Appendix~B.3)
\State \textbf{Level shift.} Draw a mean-zero Gaussian increment $c_f$, multiply every $\theta_{hk}$ by $e^{-c_f/m}$, raise $\alpha_0(t)$ by $c_f$, and accept by a Metropolis step.\hfill(Web Appendix~A.1)
\State \textbf{Random effects.} For $i=1,\ldots,N$, update $\bm b_i$ by a MALA step targeting
\Statex \hspace*{\algorithmicindent}$\pi(\bm b_i\mid\cdot)\propto L\,\mathcal N(\bm b_i;\bm0,\Sigma_b)$.\hfill(Web Appendix~A.2)
\State \textbf{Covariance.} Draw $a_1,\ldots,a_q$ and $\Sigma_b$ from their full conditionals
\Statex \hspace*{\algorithmicindent}$a_k\mid\cdot\sim\mathrm{Inv\text{-}Gamma}\{(\nu+q)/2,\nu(\Sigma_b^{-1})_{kk}+A_k^{-2}\}$,
\Statex \hspace*{\algorithmicindent}$\Sigma_b\mid\cdot\sim\mathrm{IW}\{2\nu\,\mathrm{diag}(a_1^{-1},\ldots,a_q^{-1})+\sum_{i}\bm b_i\bm b_i^{\top},\nu+q-1+N\}$,
\Statex then apply three non-centered Metropolis updates that move $\Sigma_b$ and all $\bm b_i$ jointly,
\Statex \hspace*{\algorithmicindent}$\bm b_i^\ast=L_b^\ast L_b^{-1}\bm b_i$.\hfill(Web Appendix~A.2)
\State \textbf{Group shift.} Spline: draw a common shift $\bm c$ of all $\bm b_i$ against $\bm\beta$ by Gibbs from
\Statex \hspace*{\algorithmicindent}$\pi(\bm c\mid\cdot)\propto\mathcal N(\bm\beta+A\bm c;\mu_\beta\bm1_K,\sigma_\beta^2I_K)\prod_{i}\mathcal N(\bm b_i-\bm c;\bm0,\Sigma_b)$, its Gaussian full conditional,
\Statex and set $\bm b_i\leftarrow\bm b_i-\bm c$ and $\bm\beta\leftarrow\bm\beta+A\bm c$.
\Statex Piecewise: shift the random intercepts against $\lambda_{01},\ldots,\lambda_{0J}$ by a Metropolis step.
\Statex \hfill(Web Appendix~A.1)
\end{algorithmic}
\end{algorithm}

Each fit uses one chain of $120{,}000$ iterations; the first $20{,}000$ are discarded and every $50$th subsequent state is retained, giving $2{,}000$ draws. On one core of our computing cluster (Intel Xeon Gold 6538Y+; $4$ GB of memory), a Simulation~I fit took a median of $7.0$ hours and a piecewise case-study fit $2.6$ hours (Web Appendix~B.6).

\section{Simulation Studies}\label{sec:simulation}

We consider two scenarios to evaluate our proposed model. Simulation~I has nonlinear Cox-type hazards with a random intercept and slope: proportional hazard (PH) and non-proportional hazard (non-PH models) compare the methods under time-constant and time-varying covariate effects and assess recovery of $\Sigma_b$, and a non-PH model with irregular visits, which mirrors the registry visits of Section~\ref{sec:application}. Simulation~II, adapted from \citet{wen2021}, generates no frailty. Each setting has $200$ replicates of $1{,}200$ training and $300$ test subjects, at censoring fractions $r=0.20$ and $0.40$ ($0.20$ only with irregular visits). The landmark AUC and Brier score are the primary criteria, and population summaries, interval coverage, and the test log score secondary. Fitting with smaller prior scales or fewer random effects assess prior sensitivity and what the random effects contribute.

\subsection{Comparison methods and evaluation}\label{sec-comparison}

We compare TVSurvBART with three benchmarks: time-varying Cox regression (CoxPH), fitted with \code{survival} \citep{survival-package}; an adapted frailty Cox model with a log-normal random intercept, fitted with \code{coxme} \citep{coxme-package}; and a left-truncated, right-censored random survival forest (RSF) without random effects \citep{Yao2022}, fitted with \code{LTRCforests} \citep{LTRCforests-package}. Web Appendix~C.2 gives their settings and survival-curve construction.

For evaluation, each method predicted survival to $t=s+1$ at landmark time $s$ ($1,2$ in Simulation~I and $1,2,3$ in Simulation~II), conditional on survival to $s$ and the history available by then. The last visit was carried forward, with the trees' time input held at that visit. TVSurvBART used $\widehat S_i(t)/\widehat S_i(s)$, where $\widehat S_i$ averages survival along the path over posterior draws and a shared random effect for the entire path (Web Appendix~A.4). Among test subjects at risk at $s$, we evaluated cumulative/dynamic AUC \citep[\code{timeROC};][]{blanche2013timeROC} and the Brier score at $t$, using inverse probability of censoring weights (IPCW) estimated in that risk set. Population summaries instead used complete generated paths independent of event occurrence: restricted mean survival time (RMST) at $\tau/2$ and median survival from the average predicted curve, compared with the latent event times, and survival-curve RMSE ($\mathrm{RMSE}_S$) against the true curves (Web Appendix~C.1). Web Appendix~E.3 compares the methods along carried-forward paths (Web Tables~7--11).

\subsection{Simulation designs}\label{sec:sim-designs}

In all scenarios, event times are continuous. The censoring time is $C_i=\min(C_i^{\mathrm e},\tau)$, where $C_i^{\mathrm e}$ is independent exponential with its rate calibrated in each data set, conditional on the latent event times $T_i^*$, so that the expected overall censoring proportion, including random and administrative censoring, equals $r$.

\noindent\textit{Simulation~I (PH).}
Visits occur at $t_j=0.5j$, $j=0,\ldots,5$, with follow-up ending at $t_6=\tau=3$. Three continuous covariates $x_1,x_2,x_3$ follow quadratic mixed-model trajectories with measurement error; a binary covariate $x_4$ follows a logistic mixed model; and $x_5$ is time-invariant and uniform on $\{1,2,3,4\}$ (Web Appendix~D.1, Web Table~1). On $(t_{j-1},t_j]$, the hazard is
$
\lambda_i(t\mid\bm b_i)=\xi\,\lambda_0(t)\exp\{\eta_i^0(t_{j-1})+b_{i0}+b_{i1}t/\tau\},
$
where
$
\eta_i^0(t_{j-1})=-\tfrac13x_1-\tfrac13x_2+x_4+\beta_{\mathrm{EDU}}(x_5)-\tfrac12x_3^2+x_1x_2.
$
The baseline is Weibull, $\lambda_0(t)=0.05\times3t^2$, with independent $b_{i0}\sim\Normal(0,1)$ and $b_{i1}\sim\Normal(0,4)$, and $(\beta_{\mathrm{EDU}}(1),\ldots,\beta_{\mathrm{EDU}}(4))=(0,1,2,1)$. We set $\xi=3.45$ to obtain a marginal event probability of $85\%$ by $\tau$, as in the other Simulation~I scenarios.

\noindent\textit{Simulation~I (non-PH).}
We use the PH design with $\xi=2.96$ and replace $\eta_i^0(t_{j-1})$ by $\eta_i^0(t_{j-1})+\beta_3(t_{j-1})x_3$, where $\beta_3(s)=1$ on $[0,1)$, $0$ on $[1,2)$, and $-1$ on $[2,3)$.

\noindent\textit{Simulation~I (non-PH, irregular visits).}
We maintain the non-PH design but replace scheduled visits with $t_{ij}=t_{i,j-1}+\Delta_{ij}<\tau$, where the gaps $\Delta_{ij}$ are independently uniform on $[0.25,0.75]$ and unaffected by the covariate processes, $\bm b_i$, or the event time. We set $\xi=2.84$ (Web Appendix~D.2).

\noindent\textit{Simulation~II.}
Follow-up consists of five unit intervals $(j-1,j]$, with $\tau=5$. At visits $t=j\in\{0,\ldots,4\}$, we observe covariates $L_j=(L_{1j},\ldots,L_{6j})$, binary for $L_1$--$L_3$ and continuous for $L_4$--$L_6$, and a binary predictor $X_j$ (transitions in Web Appendix~D.3). Conditional on survival to interval $j$, the event probability is $p_j=\operatorname{expit}\{\varphi_j-2X_{j-1}+\Psi(L_{j-1})\}$, where
$
\Psi(L)=L_4/3-L_5/3+L_1+0.5(L_4-L_6)^2+0.5(L_5-3)^2.
$
The hazard is $-\log(1-p_j)$ on $(j-1,j]$, with intercepts $(\varphi_1,\ldots,\varphi_5)=(-2.32,\allowbreak-3.52,\allowbreak-3.92,\allowbreak-3.52,\allowbreak-2.32)$ chosen to yield a marginal event probability of $85\%$ by $\tau$.

For TVSurvBART, we use the spline baseline, the default priors of Section~\ref{sec:priors}, and the MCMC settings of Web Appendix~B.6. In the non-PH scenarios, the trees additionally receive the scaled interval start $t_{i,j-1}/\tau$.

\subsection{Results}\label{sec:sim-results}

\begin{figure}
\centering
\includegraphics[width=\textwidth]{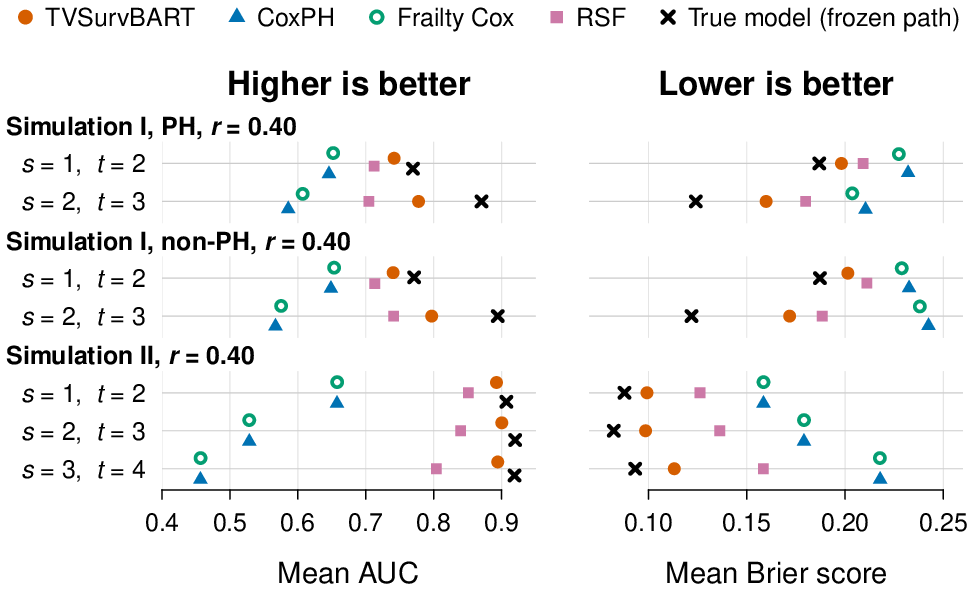}
\caption{Landmark prediction in the simulations with censoring fraction $r=0.40$, means over 200 replicates. For test subjects at risk at the landmark time $s$, each method predicts survival to $t=s+1$ given survival to $s$ along the frozen path, the history up to $s$ with the last visit at or before $s$ carried forward (Section~\ref{sec-comparison}). True model (frozen path) is the generating model's conditional survival along this path, and RSF is the survival forest of \citet{Yao2022}. The cumulative/dynamic AUC and Brier score at $t$ use IPCW estimated within the risk set. Monte Carlo standard errors are at most $0.011$ (AUC) and $0.004$ (Brier score). Web Figure~1 (Web Appendix~E.1) shows all settings, and Web Tables~2 and~3 give the means to three decimals.}
\par\medskip\noindent\parbox{\linewidth}{\linespread{1}\footnotesize Alt text: Dot plot of mean AUC (left, higher is better) and mean Brier score (right, lower is better) in seven rows: landmark times s = 1 and 2 for Simulation I with proportional and non-proportional hazards, and s = 1, 2, and 3 for Simulation II. In every row TVSurvBART has the highest AUC and lowest Brier score among fitted methods, followed by RSF, with CoxPH and frailty Cox last, and the true model (frozen path) beats all four on both measures.}
\label{fig:sim-landmark}
\end{figure}

\begin{figure}
\centering
\includegraphics[width=\textwidth]{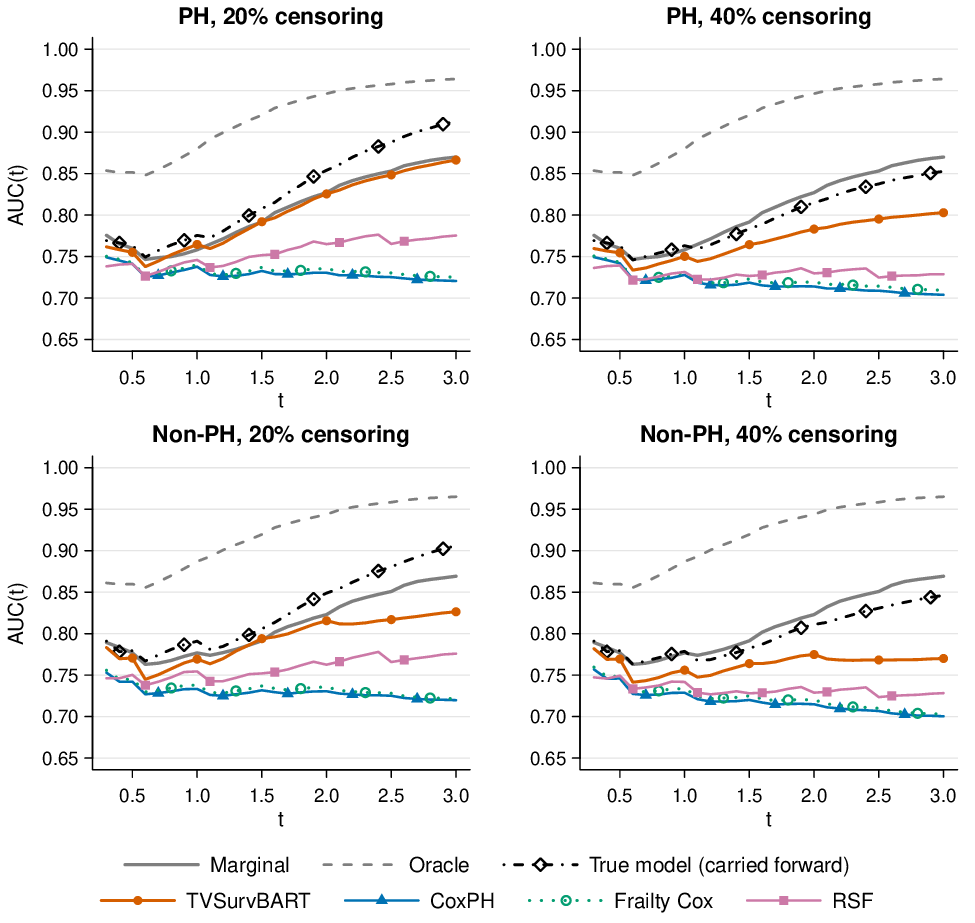}
\caption{Simulation I, mean out-of-sample AUC$(t)$ along carried-forward paths over 200 replicates under PH (top) and non-PH (bottom) observed-covariate effects with $20\%$ (left) and $40\%$ (right) nominal censoring. The black dot-dashed line with open diamonds is the true model, with frailty integrated out, along the same carried-forward paths of the same test subjects (labeled ``True model (carried forward)''). The solid and dashed gray lines are the population references along complete paths with frailty integrated out (labeled ``Marginal'') and with realized frailty (labeled ``Oracle''). CoxPH (solid line, triangles) and frailty Cox (dotted line, open circles) nearly coincide. Web Table~8 (Web Appendix~E.3) gives the values at $t=1,2,3$.}
\par\medskip\noindent\parbox{\linewidth}{\linespread{1}\footnotesize Alt text: Four line plots of mean AUC(t) against t. In every plot the Oracle curve lies above all others, and TVSurvBART's curve is the highest of the four fitted methods at every t.}
\label{fig:sim1-auc-curves}
\end{figure}

\begin{figure}
\centering
\includegraphics[width=\textwidth]{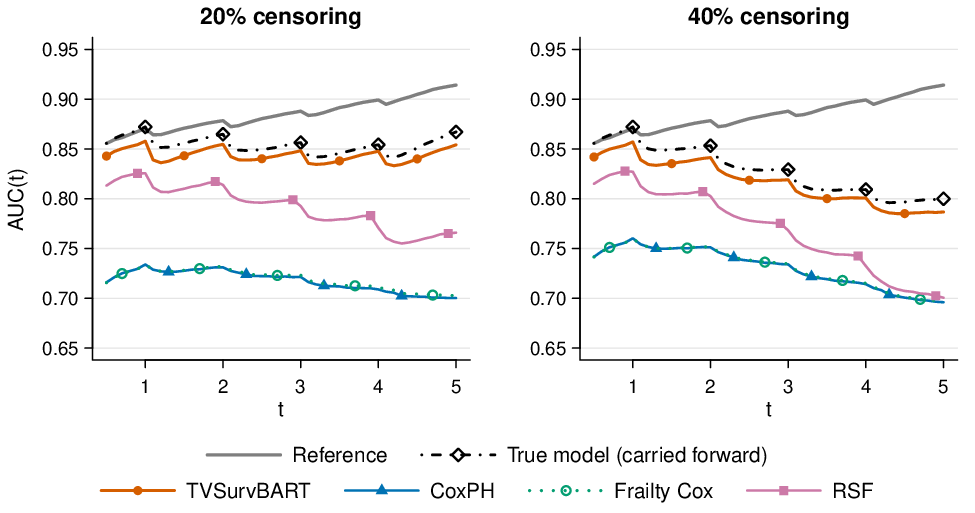}
\caption{Simulation II, mean out-of-sample AUC$(t)$ along carried-forward paths over 200 replicates with $20\%$ (left) and $40\%$ (right) nominal censoring. The black dot-dashed line with open diamonds is the true model along the same carried-forward paths of the same test subjects (labeled ``True model (carried forward)''); like the fitted curves, it depends on the censoring fraction. The solid gray line is the population reference computed from complete generated paths (labeled ``Reference''). CoxPH (solid line, triangles) and frailty Cox (dotted line, open circles) nearly coincide. Web Table~11 (Web Appendix~E.3) gives the values at $t=1,\ldots,5$.}
\par\medskip\noindent\parbox{\linewidth}{\linespread{1}\footnotesize Alt text: Two line plots of mean AUC(t) against t. TVSurvBART is the highest fitted method at every t. RSF is second, and CoxPH and frailty Cox are lowest.}
\label{fig:sim2-auc-curves}
\end{figure}

With regular visits, TVSurvBART has the highest AUC and lowest Brier score among fitted methods at every landmark, with all paired improvements exceeding two Monte Carlo standard errors (Figure~\ref{fig:sim-landmark} for $r=0.40$; Web Appendix~E.1, Web Figure~1 and Web Tables~2 and~3). In these settings, its AUC exceeds RSF's by $0.027$ to $0.090$. Irregular visits reduce AUC for all methods, but TVSurvBART remains the best in AUC. Its Brier score matches RSF's at $s=2$. Along carried-forward paths, TVSurvBART's mean AUC$(t)$ is the highest of the fitted methods at every tabulated horizon (Figures~\ref{fig:sim1-auc-curves} and~\ref{fig:sim2-auc-curves}; Web Appendix~E.3, Web Tables~8 and~11).

Along complete paths, TVSurvBART has the smallest $\mathrm{RMSE}_S$ in every setting and the largest AUC of the fitted methods at every horizon (Web Appendix~E.2, Web Tables~4--6). Its RMST and median RMSEs are below those of the Cox models and at most $0.001$ above RSF's (Web Tables~4 and~5). Last-visit carry-forward causes even the true model to overestimate RMST in Simulation~II by $0.014$ and $0.031$ at $r=0.20$ and $0.40$ (Web Appendix~E.3). In Simulation~I, TVSurvBART's $95\%$ survival-probability intervals along complete paths cover the truth for $0.940$ to $0.971$ of test subjects at $t=1$, $0.676$ to $0.763$ at $t=3$, and $0.491$ to $0.608$ for survival from $s=2$ to $3$. Misses concentrate at extreme true probabilities, where predictions shrink toward the average (Web Appendix~E.4, Web Tables~12 and~13). TVSurvBART has the highest test log score in every setting (Web Appendix~E.5, Web Tables~14 and~15).

Random-effect ablations in the PH setting at $r=0.20$ show that the intercept improves prediction while the slope adds little. Removing random effects increases the landmark Brier score at $s=1$ by $0.0017$ (SE $0.0003$). An intercept-only fitting matches the full model at $s=1$ and improves it at $s=2$ by $0.0018$ (SE $0.0002$). Neither refit matters in Simulation~II, and both still outperform the benchmarks at every landmark (Web Appendix~E.6, Web Table~16). In Simulation~I, mean posterior medians of the intercept and slope standard deviations $\sigma_0$ and $\sigma_1$ fall below their true values $1$ and $2$ (Table~\ref{tab:sim-re}(a)). The default prior mildly penalizes these values, but at most one event per subject limits information about $\Sigma_b$ (Web Appendix~E.7, Web Tables~17 and~18). With no frailty in Simulation~II, the medians are at most $0.139$ and $0.319$, well below the prior median of about $4$ (Web Appendix~E.7).

\begin{table}
\caption{Simulation~I, posterior of the TVSurvBART random-effect covariance $\Sigma_b$: intercept standard deviation $\sigma_0$ and slope standard deviation $\sigma_1$, the square roots of the diagonal entries of $\Sigma_b$ (the slope multiplies $t/\tau$), and correlation $\rho$. Entries are the mean over replicates of the posterior median followed by the proportion of replicates whose $95\%$ credible interval contains the true value, in brackets. (a) Over 200 replicates per setting. Irregular is the non-PH model with irregular visits. (b) Prior sensitivity in the PH setting at $r=0.20$, over replicates 1--100 with the data and chain seeds of the main analysis. The default prior has half-$t_2(5)$ priors on both random-effect standard deviations ($A_1=A_2=5$) and ensemble scale $s_F\approx2.492$. The next two columns change the common half-$t$ scale, and the last column changes $s_F$ only, which narrows the central $95\%$ prior hazard-ratio interval for two inputs with uncorrelated tree values from $[0.001,1000]$ to $[0.01,100]$. Prediction entries, including the integrated Brier score (IBS), follow carried-forward paths, with the AUC computed from the latent event times (Web Appendix~C.1), and are means over replicates, rounded to three decimals. The log score is the mean paired difference from the default (Monte Carlo standard error), positive values favoring the column prior. Monte Carlo standard errors, the bias and RMSE of the posterior medians, and interval widths for (a) are in Web Table~17, and the remaining criteria for (b) in Web Table~18, both in Web Appendix~E.7.}
\label{tab:sim-re}
\centering
\begin{tabular}{@{}lccccc@{}}
\toprule
\multicolumn{6}{@{}l}{(a) By setting}\\
 & \multicolumn{2}{c}{PH} & \multicolumn{2}{c}{Non-PH} & Irregular\\
\cmidrule(lr){2-3}\cmidrule(lr){4-5}\cmidrule(lr){6-6}
 & $r=.20$ & $r=.40$ & $r=.20$ & $r=.40$ & $r=.20$\\
\midrule
$\sigma_0$ (true value 1) & .741 [.755] & .640 [.605] & .720 [.635] & .605 [.485] & .710 [.645]\\
$\sigma_1$ (true value 2) & .800 [.270] & 1.322 [.810] & .595 [.120] & 1.174 [.725] & .708 [.205]\\
$\rho$ (true value 0) & .322 [.980] & .384 [.990] & .190 [1.000] & .287 [1.000] & .244 [1.000]\\
\midrule
\multicolumn{6}{@{}l}{(b) Prior sensitivity, PH, $r=.20$}\\
 & Default & $A_1=A_2=2$ & $A_1=A_2=1$ & $s_F\approx1.661$ & \\
\midrule
$\sigma_0$ & .738 [.740] & .737 [.710] & .734 [.680] & .639 [.350] & \\
$\sigma_1$ & .800 [.260] & .760 [.210] & .681 [.150] & .731 [.220] & \\
$\rho$ & .326 [1.000] & .353 [.980] & .412 [.970] & .270 [1.000] & \\
IBS & .123 & .123 & .123 & .124 & \\
$\mathrm{RMSE}_S$ & .098 & .098 & .098 & .100 & \\
AUC, $t=3$ & .867 & .867 & .868 & .856 & \\
Log score difference & -- & $+0.02$ (0.08) & $+0.14$ (0.06) & $-2.8$ (0.1) & \\
\bottomrule
\end{tabular}
\end{table}

Prior sensitivity in the PH setting at $r=0.20$ (replicates 1--100; Table~\ref{tab:sim-re}(b)) shows little predictive change when the random-effect half-$t$ scale falls from $5$ to $2$ or $1$. The integrated Brier score (IBS) and $\mathrm{RMSE}_S$ remain unchanged to three decimals, and AUC along carried-forward paths changes little. The mean posterior median of $\sigma_1$ decreases from $0.800$ to $0.760$ and $0.681$. Reducing the ensemble scale to $s_F\approx1.661$ has more impact: AUC at $t=3$ falls from $0.867$ to $0.856$, and the log score decreases by $2.8$. All other fits use the default prior, whose $s_F$ was fixed before these analyses, not selected by held-out performance.

\section{Case Study}
\label{sec:application}

\subsection{Cohort and data}
\label{sec:app-data}

We analyze CARRA Registry data for subjects of all ages with non-systemic juvenile idiopathic arthritis whose disease had become inactive on adalimumab, followed from an index date set by inactive disease at two consecutive visits (Web Appendix~F.1). Index dates run from April 2016 to August 2022. The endpoint is the first recorded flare under the registry's composite definition (a therapy change recorded for flare or active disease, or a clinic visit with active disease). 

From the $862$ subjects in the extract, we removed 5 with inconsistent outcome records and $126$ with zero at-risk follow-up, mostly entering in 2021 or 2022 (Web Appendix~F.1). The remaining $731$ contribute $361$ flares ($49.4\%$), $2{,}765$ records, and $2{,}034$ at-risk intervals, with medians of $3$ visit records per subject and $183$ days (interquartile range $140$ to $231$) between consecutive irregular visits. Predictors include lagged weekly adalimumab dose; visit-level disease activity, pain, body mass index, and joint findings; concurrent disease-modifying antirheumatic drugs; and baseline characteristics. Within-subject carry-forward fills missing visit values, and leading missing values (not yet recorded for the subject) are set to training-fold medians or a separate categorical level (Web Appendix~F.2).

\subsection{Models and evaluation}
\label{sec:app-eval}

TVSurvBART with a piecewise (primary) or spline baseline is compared with the CoxPH, adapted frailty Cox, and RSF benchmarks of Section~\ref{sec-comparison} on the same at-risk intervals and covariates. It uses $m=100$ trees with the default priors of Section~\ref{sec:priors}. The random slope multiplies $t/\tau$ with $\tau=5$ years, and the trees receive the covariates only (Web Appendix~F.3). Evaluation uses $100$ repetitions of $10$-fold cross-validation, stratified by flare status and split by subject. The primary criteria are those of the landmark evaluation of Section~\ref{sec-comparison}. At landmarks $s=1$ and $2$ years, each method predicts survival to $t=s+1$, and from $s=1$ also to $t=3$, for the held-out subjects at risk at $s$, given survival to $s$ and the covariate history up to $s$ with the record in force at $s$ carried forward; the cumulative/dynamic AUC and Brier score at $t$ use IPCW estimated within the risk set. They are computed in each repetition on the pooled held-out subjects and averaged over repetitions, with standard errors from a bootstrap over the subjects at risk that holds the fitted models fixed (Web Appendix~F.3).

\subsection{Landmark prediction}
\label{sec:app-landmark}

Discrimination after the landmark is modest for every method (Table~\ref{tab:app-landmark}). From $s=1$ with a $1$-year window, piecewise TVSurvBART has AUC $0.572$ and CoxPH $0.572$ (difference $+0.0003$, SE $0.012$), and RSF has the highest AUC, $0.617$. For $t=3$ from $s=1$ and $s=2$, CoxPH has AUC $0.585$ and $0.583$ against $0.577$ and $0.570$ for piecewise TVSurvBART (differences $-0.008$, SE $0.014$, and $-0.012$, SE $0.019$), and RSF falls to $0.544$ and $0.455$. With the spline baseline, TVSurvBART has AUC $0.566$ to $0.568$. Piecewise TVSurvBART has a lower Brier score than CoxPH in all three windows, by $0.004$, $0.004$, and $0.002$ (SE $0.003$, $0.004$, and $0.004$), and frailty Cox matches CoxPH in AUC and Brier score to three decimals. RSF has the lowest Brier score in the first window and the highest in the other two. These AUC and Brier score differences between TVSurvBART and Cox models are small relative to their standard errors (Web Appendix~F.3, Web Table~19, which also gives the index of prediction accuracy and the calibration).

\begin{table}[tb]
\centering
\caption{Landmark prediction in the case study, averaged over $100$ repetitions of $10$-fold cross-validation. For the held-out subjects at risk at the landmark time $s$, each method predicts survival to $t$ given survival to $s$ and the covariate history up to $s$. Cumulative/dynamic AUC and Brier score at $t$ use IPCW estimated within the risk set, with time from $s$. Standard errors in parentheses come from $1{,}000$ bootstrap resamples of the subjects at risk with the fitted models held fixed. At risk counts the subjects at risk at $s$, Flares those with a recorded flare in $(s,t]$, and Flare-free those without a flare at $t$; the others are censored before $t$. PW: piecewise baseline; NCS: natural-cubic-spline baseline; RSF: random survival forest.}
\label{tab:app-landmark}
\begin{tabular}{@{}llccc@{}}
\toprule
 & & $s=1$, $t=2$ & $s=1$, $t=3$ & $s=2$, $t=3$\\
\midrule
\multicolumn{2}{@{}l}{At risk / Flares / Flare-free} & 438 / 127 / 194 & 438 / 178 / 75 & 194 / 51 / 75\\
\midrule
AUC & TVSurvBART (PW) & .572 (.033) & .577 (.037) & .570 (.049)\\
 & TVSurvBART (NCS) & .567 (.032) & .568 (.037) & .566 (.048)\\
 & CoxPH & .572 (.033) & .585 (.037) & .583 (.048)\\
 & Frailty Cox & .572 (.033) & .585 (.037) & .583 (.048)\\
 & RSF & .617 (.030) & .544 (.037) & .455 (.050)\\
\midrule
Brier score & TVSurvBART (PW) & .219 (.010) & .246 (.007) & .214 (.016)\\
 & TVSurvBART (NCS) & .220 (.008) & .246 (.006) & .215 (.017)\\
 & CoxPH & .224 (.011) & .250 (.010) & .216 (.017)\\
 & Frailty Cox & .224 (.011) & .250 (.010) & .216 (.017)\\
 & RSF & .211 (.010) & .252 (.007) & .228 (.019)\\
\bottomrule
\end{tabular}
\end{table}

\subsection{Retrospective comparison}
\label{sec:app-results}

\begin{figure}
\centering
\includegraphics[width=0.8\textwidth]{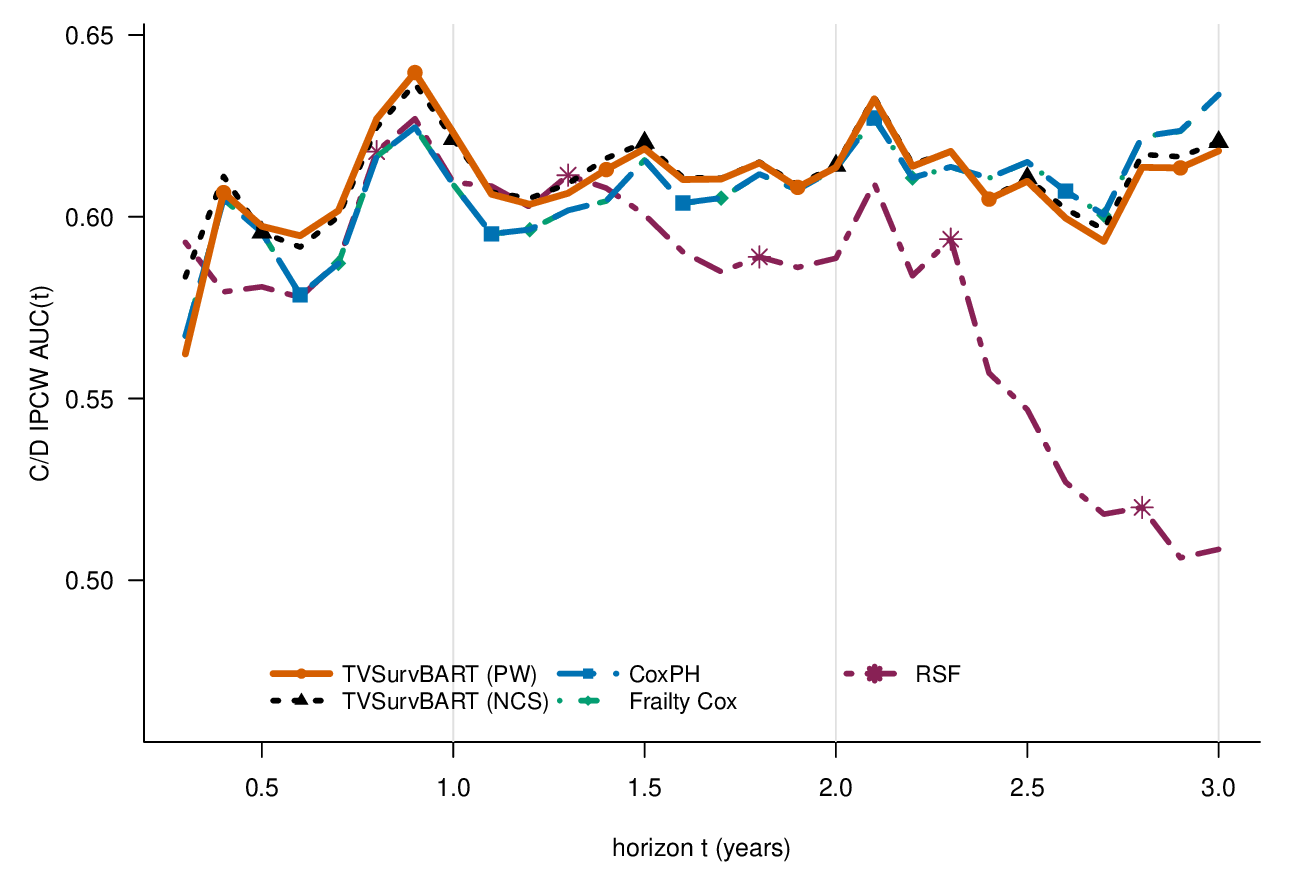}
\caption{Case study, out-of-fold cumulative/dynamic IPCW AUC$(t)$ along observed, carried-forward histories, averaged over $100$ repetitions of $10$-fold cross-validation. PW: piecewise baseline; NCS: natural-cubic-spline baseline; RSF: random survival forest.}
\par\medskip\noindent\parbox{\linewidth}{\linespread{1}\footnotesize Alt text: Line plot of AUC(t) against horizon t for TVSurvBART (PW), TVSurvBART (NCS), CoxPH, frailty Cox, and RSF. The first four lie between about 0.56 and 0.64. CoxPH and frailty Cox coincide, and the TVSurvBART curves are slightly above them from 0.4 to 2.3 years and slightly below them after. RSF crosses the other curves up to about 1.4 years, lies below all of them from 1.5 years, and falls steeply after 2.3 years, to about 0.51 at 3 years.}
\label{fig:app-auc}
\end{figure}

A secondary comparison scores the held-out subjects' observed histories, which stop at the flare or censoring time and are carried forward beyond the last visit. Its cumulative/dynamic AUC, the path AUC, uses IPCW from the observed outcomes with the marker $1-\widehat S_i(t)$ along these histories (Web Appendix~C.1), and an interval-probability score (Web Appendix~A.3) gives each flare the log predicted probability of falling between the last flare-free visit and the recorded date, and each censored subject its log survival probability. Because both depend on the observed visits and events, the comparison is retrospective, and we interpret it descriptively (Web Appendix~F.3). At one year, the path AUC is $0.623$ for piecewise TVSurvBART ($0.621$ with the spline baseline) and $0.609$ for CoxPH (difference $+0.014$, SE $0.012$). At three years, CoxPH is higher ($0.634$ versus $0.618$; difference $-0.015$, SE $0.017$), and RSF falls to $0.509$ (Figure~\ref{fig:app-auc}). Piecewise TVSurvBART has the highest interval-probability score ($-947.2$), exceeding spline TVSurvBART by $8.8$ (SE $6.0$), RSF by $34.1$ (SE $9.8$), CoxPH by $35.8$ (SE $7.6$), and frailty Cox by $35.9$ (SE $7.6$; Web Appendix~F.3, Web Table~20). All standard errors in this subsection are approximate Nadeau--Bengio standard errors \citep{NadeauBengio2003}. These score contrasts favor TVSurvBART under this shared retrospective scoring rule without establishing better prospective prediction or recovery of flare onset.

\subsection{Random-effect covariance}
\label{sec:app-re}
In four pooled full-cohort chains with the piecewise baseline (Web Appendix~F.4), the posterior medians of the intercept and slope standard deviations $\sigma_0$ and $\sigma_1$ are $0.372$ ($95\%$ credible interval $0.024$ to $0.887$) and $0.650$ ($0.038$ to $2.236$), and that of their correlation $\rho$ is $0.059$ ($-0.779$ to $0.831$; Web Appendix~F.5, Web Table~21, Default column). Relative to the half-$t_2(5)$ prior, with median $4.08$ and central $95\%$ interval $0.18$ to $31.0$ for each standard deviation, the data rule out very large standard deviations but do not separate small values from zero, and carry little information about $\rho$. Likewise, the frailty Cox variance estimate is coxme's smallest default starting value, $0.0004$.

\subsection{Sensitivity analyses}
\label{sec:app-sens}

Refitting the piecewise model on the ten first-repetition folds and the full cohort with half-$t$ scales $A_1=A_2=2$ or $1$ in place of $5$, or $s_F\approx1.661$ in place of $2.492$ (Web Appendix~F.5), changes the path AUC at $t=1,2,3$ by at most $0.004$, the landmark AUC by at most $0.004$, the landmark Brier score by at most $0.002$, and the interval-probability score by $-0.2$ to $+2.2$ (largest SE $1.4$; Web Table~21). All other case-study fits use the default prior. Across the half-$t$ scales, with prior medians from $0.82$ to $4.08$, the posterior of $\sigma_0$ is nearly unchanged (medians $0.345$ to $0.372$), whereas that of $\sigma_1$ moves with the prior. On the same folds, multiple imputation of leading missing values by chained equations \citep{vanBuurenGroothuisOudshoorn2011}, with the flare indicator and Nelson--Aalen cumulative hazard in the imputation model \citep{WhiteRoyston2009}, lowers the path AUC at $t=2$ and $3$ by at most $0.008$ for TVSurvBART and $0.018$ for CoxPH, changes the landmark AUC by $-0.011$ to $+0.001$ for TVSurvBART and lowers it by $0.007$ to $0.009$ for CoxPH, and TVSurvBART's score advantage over CoxPH falls from $33.8$ (SE $7.8$) to $30.5$ (SE $6.9$; Web Appendix~F.6, Web Table~22).

\section{Discussion}\label{sec:discussion}

We have developed TVSurvBART, a flexible Bayesian framework for continuous-time survival regression with longitudinally updated covariates, potentially time-varying effects, and subject-specific random intercepts and slopes. A central contribution is the integration of these components within a computationally tractable full-likelihood model. By incorporating within-interval variation in the baseline hazard and random effects into exposure offsets, we retain conjugate tree-leaf updates while requiring only one-dimensional numerical integration. Likelihood-invariant re-centering moves and a non-centered covariance update further address the posterior dependence among the tree ensemble, baseline hazard, and random effects without imposing empirical centering constraints. These developments extend log-linear BART to longitudinal survival settings while preserving efficient, posterior-invariant computation with irregular visits and evolving covariate histories.

Our simulations demonstrate consistent gains in landmark prediction across nonlinear, non-proportional hazards, and irregular-visit settings. TVSurvBART achieved the highest AUC among the fitted methods at every landmark and matched or improved the Brier scores of CoxPH, frailty Cox, and RSF. The gains were particularly apparent in individual-level prediction, whereas population RMST and median survival, which aggregate predicted curves across subjects, showed smaller differences among methods. Ablation results confirmed that the flexible tree component remained competitive without random effects, while the random intercept provided additional predictive benefit in Simulation~I. Although random-effect covariance components were difficult to estimate precisely, predictions were largely insensitive to moderate changes in their prior scales (Sections~\ref{sec:sim-results} and~\ref{sec:app-sens}).

The CARRA Registry application further illustrates the framework's practical use with irregular longitudinal clinical data. TVSurvBART and the Cox models had similar landmark discrimination for recorded flares (AUC $0.570$--$0.585$), but TVSurvBART performed better under the secondary retrospective interval-probability score. These findings highlight the importance of assessing survival methods using complementary measures of predictive performance.

TVSurvBART models survival conditional on observed covariate histories and does not model how the covariates evolve, so landmark prediction requires an assumption about future covariates. Carrying the last visit forward, as our landmark predictions do, treats the covariates after the landmark as external. For internal covariates such as disease activity, a joint model of the covariate and event processes \citep{wulfsohn1997joint} is an alternative (Web Appendix~A.4). The likelihood also treats the covariate, visit, and censoring processes as ignorable given the observed history (Section~\ref{sec:likelihood}), which may fail when visits are prompted by symptoms. In the application, the model describes the recorded flare date, which follows the unobserved onset and may depend on the visit schedule (Section~\ref{sec:app-data}). Modeling the onset as interval-censored between the last flare-free visit and the recorded date would target the onset itself if visit times do not depend on it. Extending TVSurvBART to a joint model of the covariate, visit, and event processes is a natural next step.

\section*{Acknowledgements}
AI tools were used to assist with code optimization to improve computational efficiency. The authors independently tested and validated the revised code using simulated data.

\section*{Supplementary Materials}

Web Appendices, Tables, and Figures referenced in Sections~2--7 are available with this paper at the Biometrics website on Oxford Academic. The supplementary materials contain proofs and technical clarifications, computational details and timings, benchmark implementations, simulation designs and results, case-study details, and supplementary tables and figures.

\section*{Funding}
This research was supported by the Patient-Centered Outcomes Research Institute (PCORI; Award No. ME-2021C2-23685) and the National Heart, Lung, and Blood Institute (NHLBI; Grant No. 1R01HL159077-01). The content of this article is solely the responsibility of the authors and does not necessarily represent the official views of PCORI or NHLBI.

\section*{Conflicts of interest.}
None declared.

\section*{Data Availability Statement}

Individual-level data from the CARRA Registry are subject to controlled access. Code implementing the proposed Bayesian nonparametric survival regression methodology and simulation is publicly available at \url{https://github.com/gosicksky/TVSurvBART}.

\bibliographystyle{biom}
\bibliography{coxbart_references}

\end{document}


\title{Supplementary Materials for ``Bayesian Nonparametric Survival Regression with Time-Varying Covariates and Random Effects''}

\author{Keming Zhang$^{1}$, Yudan Zou$^{2}$, Fan Li$^{3}$, Liangyuan Hu$^{*2}$\\
        \small $^{1}$Department of Biostatistics, Brown University \\
        \small $^{2}$Department of Biostatistics and Epidemiology, Rutgers University\\
        \small $^{3}$Department of Biostatistics, Yale University\\
        \small $^{*}$Corresponding author: liangyuan.hu@rutgers.edu}
\date{}

\maketitle
\appendix

\section{Proofs and technical clarifications}\label{app:proofs}

\subsection{Likelihood-preserving re-centering}
\label{supp:recentering}

Throughout this appendix, the random-effects design is $\bm z_i(t)=\bm z(t)=(1,t/\tau)^{\trans}$, where $\tau>0$ is the fixed time scale, and $\bm b_i\sim\mathcal N(\bm0,\Sigma_b)$, with $q=2$. The random-intercept-only model is the case $q=1$ with $\bm z(t)=1$. The arguments below apply to it with the slope coordinate removed, and we indicate the one place where the construction changes. For fixed tree structures, let $k_h(\bm u)$ denote the terminal node containing $\bm u$ in tree $h$, and write
\[
F(\bm u)=\sum_{h=1}^m\log\theta_{h,k_h(\bm u)},
\qquad
f(\bm u)=e^{F(\bm u)},
\qquad
\lambda_i(t)=\exp\{\alpha_0(t)+F(\bm U_i(t))+
\bm b_i^{\trans}\bm z(t)\}.
\]
Here $\theta_{hk}>0$, and $\bm U_i(t)$ is the predictable, interval-constant tree input. As in main-text Section~2.4, omitting the factors of the covariate, visit, and censoring processes assumes that, given the preceding observed history, these processes depend neither on the random effects nor on the baseline, tree, and random-effect covariance parameters, and that their parameters are distinct from and a priori independent of those parameters. The event-process likelihood given the random effects is
\[
L=\prod_{i=1}^N\lambda_i(T_i)^{\delta_i}
\exp\left\{-\int_0^{T_i}\lambda_i(u)\,du\right\}.
\]

We restate main-text Proposition~1 in the notation of this appendix.

\begin{mainpropone}[Likelihood invariance of the re-centering moves]
(i) For $c_f\in\mathbb R$, the ensemble--baseline level shift $\theta_{hk}^\ast=\theta_{hk}e^{-c_f/m}$ for every terminal node, together with $\beta_0^\ast=\beta_0+c_f$ under the spline baseline or $\lambda_{0\ell}^\ast=e^{c_f}\lambda_{0\ell}$, $\ell=1,\ldots,J$, under the piecewise-constant baseline, leaves $\lambda_i(t)$ unchanged for every subject $i$ and every $t$ in the follow-up period. (ii) Under the spline baseline $\alpha_0(t)=\widetilde{\bm B}(t)^{\trans}\bm\beta$, with $\widetilde{\bm B}(t)=(1,\widetilde B_1(t),\ldots,\widetilde B_{K-1}(t))^{\trans}$, so does the group shift $\bm b_i^\ast=\bm b_i-\bm c$, $i=1,\ldots,N$, with $\bm\beta^\ast=\bm\beta+A\bm c$ for $\bm c\in\mathbb R^q$ and a fixed matrix $A$ satisfying $\widetilde{\bm B}(t)^{\trans}A=\bm z(t)^{\trans}$ for all $t$. (iii) Under the piecewise-constant baseline, so does the common-intercept shift $\bm b_i^\ast=\bm b_i-c_g\,\bm e_0$, $i=1,\ldots,N$, with $\lambda_{0\ell}^\ast=e^{c_g}\lambda_{0\ell}$, $\ell=1,\ldots,J$, for $c_g\in\mathbb R$ and $\bm e_0=(1,0,\ldots,0)^{\trans}\in\mathbb R^q$. These moves therefore leave the likelihood $L$ unchanged.
\end{mainpropone}

\begin{proof}[Proof of main-text Proposition~1]
(i) For the ensemble--baseline level shift, set $\theta_{hk}^*=\theta_{hk}e^{-c_f/m}$ for every terminal node. Exactly one terminal node contributes from each tree, so
\[
F^*(\bm u)=\sum_{h=1}^m
\log\{\theta_{h,k_h(\bm u)}e^{-c_f/m}\}=F(\bm u)-c_f.
\]
Under the spline baseline, increasing its intercept by $c_f$ gives $\alpha_0^*(t)=\alpha_0(t)+c_f$. Under the piecewise baseline, multiplying every hazard height by $e^{c_f}$ gives the same identity on the modeled follow-up range. Thus, the change in the baseline exactly cancels the change in $F$, with $\bm b_i$ unchanged.

(ii) For the spline group shift, if the fixed matrix $A$ satisfies $\widetilde{\bm B}(t)^{\trans}A=\bm z(t)^{\trans}$, then
\[
\widetilde{\bm B}(t)^{\trans}(\bm\beta+A\bm c)
+(\bm b_i-\bm c)^{\trans}\bm z(t)
=\widetilde{\bm B}(t)^{\trans}\bm\beta
+\bm b_i^{\trans}\bm z(t).
\]
Consequently, $\bm\beta^*=\bm\beta+A\bm c$ and $\bm b_i^*=\bm b_i-\bm c$ preserve the log-hazard.

(iii) For the piecewise common-intercept shift, multiplying every hazard height by $e^{c_g}$ raises $\alpha_0(t)$ by $c_g$ on the modeled follow-up range, while $\bm b_i^{\trans}\bm z(t)$ falls by $c_g\,\bm e_0^{\trans}\bm z(t)=c_g$.

In all three cases, these transformations preserve $\lambda_i(t)$ pointwise, including its value at an event time, and hence preserve both terms of $L$.
\end{proof}

For the centered natural cubic spline basis, $\widetilde B_1(t)=t-\bar t$, where $\bar t$ is the fixed exposure-weighted mean time. With the scaled random-slope design $\bm z(t)=(1,t/\tau)^{\trans}$ used here, the required matrix is explicitly
\[
A=\begin{pmatrix}
1&\bar t/\tau\\
0&1/\tau\\
0&0\\
\vdots&\vdots
\end{pmatrix},
\qquad
A\bm c=(c_0+\bar t c_1/\tau,c_1/\tau,0,\ldots,0)^{\trans}.
\]
Thus, $\widetilde{\bm B}(t)^{\trans}A=(1,t/\tau)=\bm z(t)^{\trans}$, so a common random-slope shift is absorbed jointly by the baseline intercept and its linear spline coefficient. For the random-intercept-only model, $A$ is the first column of this matrix and the shift is absorbed by the baseline intercept alone. The centering constants $\bar B_\ell$, time scale, and spline basis are held fixed during these moves.

\paragraph{Level-shift acceptance ratios.}
Likelihood invariance does not imply prior invariance. The priors, expressed with respect to the positive leaf values and hazard heights, are
\[
\begin{aligned}
p_\theta(x\mid c,d)
&=\frac{d^c}{2\Gamma(c)}
\{x^{c-1}e^{-dx}+x^{-c-1}e^{-d/x}\},\qquad x>0,\\
p_\lambda(x)
&=\frac{b_\lambda^{a_\lambda}}{\Gamma(a_\lambda)}
x^{a_\lambda-1}e^{-b_\lambda x},\qquad x>0.
\end{aligned}
\]
The spline prior is $p_\beta(\bm\beta)= \phi_K(\bm\beta;\bm\mu_\beta,V_\beta)$, where $\bm\mu_\beta=(\mu_\beta,\ldots,\mu_\beta)^{\trans}$, $V_\beta=\sigma_\beta^2I_K$, and $\phi_q$ denotes a Gaussian density. Let $H=\sum_{h=1}^m b_h$ be the total number of terminal nodes. For a symmetric increment proposal for $c_f$, the spline move has acceptance probability $\min\{1,R_f^{\rm sp}\}$, with
\[
R_f^{\rm sp}=
\frac{p_\beta(\bm\beta+c_f\bm e_\beta)}{p_\beta(\bm\beta)}
\prod_{h=1}^m\prod_{k=1}^{b_h}
\frac{p_\theta(\theta_{hk}e^{-c_f/m}\mid c,d)}
{p_\theta(\theta_{hk}\mid c,d)}
e^{-c_fH/m},
\]
where $\bm e_\beta=(1,0,\ldots,0)^{\trans}\in\mathbb R^K$. For the piecewise move, the acceptance probability is $\min\{1,R_f^{\rm pc}\}$, with
\[
R_f^{\rm pc}=
\prod_{\ell=1}^J
\frac{p_\lambda(e^{c_f}\lambda_{0\ell})}{p_\lambda(\lambda_{0\ell})}
\prod_{h=1}^m\prod_{k=1}^{b_h}
\frac{p_\theta(\theta_{hk}e^{-c_f/m}\mid c,d)}
{p_\theta(\theta_{hk}\mid c,d)}
e^{c_fJ-c_fH/m}.
\]
These factors can be derived in log coordinates: $\mu_{hk}=\log\theta_{hk}$ and $\gamma_\ell=\log\lambda_{0\ell}$. The moves are symmetric translations in those coordinates, whose induced prior densities are $p_\theta(e^\mu)e^\mu$ and $p_\lambda(e^\gamma)e^\gamma$. Their density ratios give precisely $e^{-c_fH/m}$ and $e^{c_fJ-c_fH/m}$. Equivalently, these are the Jacobians in the positive-valued parameterization.

\paragraph{Gaussian update along the spline group orbit.}
With $\Sigma_b$ fixed, the log prior after the group shift is, up to a constant independent of $\bm c$,
\[
-\frac12(\bm\beta+A\bm c-\bm\mu_\beta)^{\trans}V_\beta^{-1}
(\bm\beta+A\bm c-\bm\mu_\beta)
-\frac12\sum_{i=1}^N(\bm b_i-\bm c)^{\trans}
\Sigma_b^{-1}(\bm b_i-\bm c).
\]
Expanding gives $-\bm c^{\trans}H_c\bm c/2+ \bm c^{\trans}\bm h_c$, where
\[
\begin{aligned}
H_c&=A^{\trans}V_\beta^{-1}A+N\Sigma_b^{-1},\\
\bm h_c&=\Sigma_b^{-1}\sum_{i=1}^N\bm b_i
-A^{\trans}V_\beta^{-1}(\bm\beta-\bm\mu_\beta).
\end{aligned}
\]
The increment is therefore drawn as
\[
\bm c\mid\cdot\sim\mathcal N(H_c^{-1}\bm h_c,H_c^{-1}),
\qquad
\bm\beta^*=\bm\beta+A\bm c,
\qquad
\bm b_i^*=\bm b_i-\bm c.
\]

To see explicitly why this is a Gibbs update, define $\bar{\bm b}=N^{-1}\sum_i\bm b_i$, $\bm d_i=\bm b_i-\bar{\bm b}$, and $\bm v=\bm\beta+A\bar{\bm b}$. The quantities $\bm v$ and $\{\bm d_i\}$ are invariant under the group shift. Using $N-1$ independent deviations, these and the orbit coordinate $\bm r=\bar{\bm b}$ form an invertible linear parameterization with constant Jacobian. Since $\bm\beta=\bm v-A\bm r$ and $\bm b_i=\bm d_i+\bm r$, the log-hazard equals $\widetilde{\bm B}(t)^{\trans}\bm v+F(\bm U_i(t))+ \bm d_i^{\trans}\bm z(t)$ and is independent of $\bm r$. Its full conditional is consequently
\[
\bm r\mid\bm v,\{\bm d_i\},\Sigma_b,\cdot
\sim\mathcal N\!\left(
H_c^{-1}A^{\trans}V_\beta^{-1}(\bm v-\bm\mu_\beta),
H_c^{-1}\right).
\]
Drawing $\bm r^*$ and setting $\bm c=\bar{\bm b}-\bm r^*$ produces the increment distribution above. This stochastic update does not impose an empirical zero-mean constraint on the random effects and does not make the separate hazard components identifiable.

\paragraph{Common-intercept shift for a piecewise baseline.}
Let $\bm e_0=(1,0,\ldots,0)^{\trans}\in\mathbb R^q$. For a symmetric scalar increment $c_g$, set $\bm b_i^*=\bm b_i-c_g\bm e_0$ and $\lambda_{0\ell}^*=e^{c_g}\lambda_{0\ell}$. Because $\bm e_0^{\trans}\bm z(t)=1$, this move preserves the hazard. Its acceptance probability is $\min\{1,R(c_g)\}$, where
\[
R(c_g)=e^{Jc_g}
\prod_{i=1}^N
\frac{\phi_q(\bm b_i-c_g\bm e_0;\bm0,\Sigma_b)}
{\phi_q(\bm b_i;\bm0,\Sigma_b)}
\prod_{\ell=1}^J
\frac{p_\lambda(e^{c_g}\lambda_{0\ell})}{p_\lambda(\lambda_{0\ell})}.
\]
The height Jacobian $e^{Jc_g}$ is required. It is equivalently included in the induced prior on log-heights. A continuously varying common slope cannot generally be absorbed by a piecewise-constant baseline, so this move shifts only the random intercepts.

\subsection{Posterior invariance of the sampler}\label{supp:invariance}

Let $\pi$ denote the joint posterior of the tree structures and leaves, baseline parameters, random effects, covariance matrix, and auxiliary scales $\bm a=(a_1,\ldots,a_q)^{\trans}$. We consider the idealized sampler with fixed proposal parameters and exact evaluation of exposure integrals and normalizing constants. Each complete block described below preserves $\pi$.

\paragraph{Baseline, random effects, and covariance hierarchy.}
For a nonconjugate block $x$, a proposal $x^*$ from the Metropolis-adjusted Langevin algorithm (MALA) density $q_x(x^*\mid x)$ is accepted with probability
\[
\min\left\{1,\
\frac{\pi(x^*\mid\text{remaining blocks})q_x(x\mid x^*)}
{\pi(x\mid\text{remaining blocks})q_x(x^*\mid x)}\right\}.
\]
The full conditional includes the prior, and the reverse proposal is evaluated at the proposed state. This establishes invariance for the spline coefficients and each subject's random effects. For $\bm b_i$, the proposal is $\bm b_i^\ast=\bm b_i+\tfrac12\varsigma_i^2\widetilde{\bm g}_i(\bm b_i)+\varsigma_i\bm\varepsilon_i$, $\bm\varepsilon_i\sim\mathcal N(\bm0,I_q)$, where $\widetilde{\bm g}_i(\bm b_i)=\bm g_i(\bm b_i)-\Sigma_b^{-1}\bm b_i$ is the gradient of its log full conditional and $\bm g_i$ is the score given below. The scales $\varsigma_i$ are tuned during burn-in (Web Appendix~\ref{supp:adapt}). For the piecewise baseline, write $I_\ell=(\tau_{\ell-1},\tau_\ell]$ and define
\[
d_\ell=\sum_i\delta_i\mathbb I\{T_i\in I_\ell\},
\qquad
E_\ell=\sum_i\sum_j f(\bm U_{ij})
\int_{(t_{i,j-1},t_{ij}]\cap I_\ell}
e^{\bm b_i^{\trans}\bm z(u)}\,du.
\]
The independent draws $\lambda_{0\ell}\mid\cdot\sim \mathrm{Gamma}(a_\lambda+d_\ell,b_\lambda+E_\ell)$ are Gibbs updates, using the shape--rate convention. Under the hierarchical inverse-Wishart prior, the centered covariance updates are the exact full-conditional draws
\[
\begin{aligned}
a_k\mid\Sigma_b
&\sim\mathrm{Inv\text{-}Gamma}\!\left(
\frac{\nu+q}{2},\nu(\Sigma_b^{-1})_{kk}+A_k^{-2}\right),\\
\Sigma_b\mid\{\bm b_i\},\bm a
&\sim\mathrm{IW}\!\left(
2\nu\,\mathrm{diag}(a_1^{-1},\ldots,a_q^{-1})
+\sum_i\bm b_i\bm b_i^{\trans},\nu+q-1+N\right).
\end{aligned}
\]
The inverse-Wishart arguments are the scale matrix and degrees of freedom, respectively.

\paragraph{Collapsed tree updates as complete blocks.}
For tree $h$, condition on all other blocks, denoted by $W$, and let $L_h(R_h,\bm\theta_h;W)$ be the likelihood as a function of tree $h$ (Web Appendix~\ref{supp:gig}). The marginal conditional target for its structure is
\begin{equation*}
\begin{aligned}
\pi_{h,R}(R_h\mid W)&\propto p(R_h)\mathcal M_h(R_h;W),\\
\mathcal M_h(R_h;W)&=
\int L_h(R_h,\bm\theta_h;W)
\prod_{k=1}^{b_h}p_\theta(\theta_{hk}\mid c,d)\,d\bm\theta_h.
\end{aligned}
\end{equation*}
Let $K_R$ be the structure Metropolis--Hastings kernel targeting this marginal. After its accept/reject decision, every leaf of tree $h$ is drawn from $\pi_h(d\bm\theta_h\mid R_h,W)$, including after a rejection. The combined kernel is
\[
K_h\bigl((R,\bm\theta),dR'\,d\bm\theta'\mid W\bigr)
=K_R(R,dR'\mid W)\,
\pi_h(d\bm\theta'\mid R',W).
\]
Integrating against the incoming conditional posterior first gives $\pi_{h,R}(dR'\mid W)$ by invariance of $K_R$, and then restores $\pi_h(dR'\,d\bm\theta'\mid W)$ by the conditional leaf draw. Thus the structure move and immediate leaf draw together preserve the joint posterior. The structure move alone need not do so if old leaf values are retained. All exposure offsets use the current baseline and random effects, and the partial ensemble is updated as preceding trees change within the sweep.

\paragraph{Non-centered covariance update.}
Write $\Sigma_b=L_bL_b^{\trans}$, with $L_b$ lower triangular and positive on its diagonal. Let $\bm\vartheta$ contain its log-diagonal and below-diagonal entries, and define $\widetilde{\bm b}_i=L_b^{-1}\bm b_i$. The Jacobian from $\bm\vartheta$ to the distinct entries of $\Sigma_b$ is
\[
J_\vartheta=2^q\prod_{k=1}^q L_{b,kk}^{q-k+2},
\qquad
p_\vartheta(\bm\vartheta\mid\bm a)
=p_{\mathrm{IW}}(\Sigma_b(\bm\vartheta)\mid\bm a)J_\vartheta.
\]
Also, $d\bm b_i=|L_b|\,d\widetilde{\bm b}_i$ and
\[
\phi_q(L_b\widetilde{\bm b}_i;\bm0,\Sigma_b)|L_b|
=\phi_q(\widetilde{\bm b}_i;\bm0,I_q).
\]
Up to terms independent of $\bm b_i$, define its log-likelihood by
\[
\ell_i(\bm b_i)=\delta_i\bm b_i^{\trans}\bm z(T_i)
-\sum_{j=1}^{n_i+1}f(\bm U_{ij})
\int_{t_{i,j-1}}^{t_{ij}}
e^{\alpha_0(u)+\bm b_i^{\trans}\bm z(u)}\,du.
\]
Its gradient, the score used by the MALA update of $\bm b_i$ in main-text Algorithm~1, is
\[
\bm g_i(\bm b_i)=\delta_i\bm z(T_i)-\sum_{j=1}^{n_i+1}f(\bm U_{ij})\int_{t_{i,j-1}}^{t_{ij}}\bm z(u)\,e^{\alpha_0(u)+\bm b_i^{\trans}\bm z(u)}\,du.
\]
The joint conditional target in the transformed coordinates is therefore proportional to
\[
p_\vartheta(\bm\vartheta\mid\bm a)
\prod_{i=1}^N\phi_q(\widetilde{\bm b}_i;\bm0,I_q)
\exp\{\ell_i(L_b(\bm\vartheta)\widetilde{\bm b}_i)\}.
\]
Holding the whitened effects fixed, propose $\bm\vartheta^*=\bm\vartheta+\bm\zeta$ with $\bm\zeta\sim\mathcal N(\bm0,\Omega)$ and fixed $\Omega$. The acceptance ratio is
\[
R_\vartheta=
\frac{p_\vartheta(\bm\vartheta^*\mid\bm a)}
{p_\vartheta(\bm\vartheta\mid\bm a)}
\prod_{i=1}^N
\exp\{\ell_i(\bm b_i^*)-\ell_i(\bm b_i)\},
\qquad
\bm b_i^*=L_b^*L_b^{-1}\bm b_i.
\]
The standard-normal factors cancel. This is an ordinary Metropolis update in an invertible parameterization and therefore preserves the original joint posterior as well. Three successive such proposals also preserve it.

A covariance update targeting a marginal posterior with the random effects integrated out cannot generally retain their current values and preserve the joint posterior. One valid repair is an exact draw of the entire random-effects block from its conditional given the new covariance. A finite number of MALA transitions generally does not restore that conditional law. The non-centered update above instead moves the covariance and random effects jointly.

Finally, the level shift and the piecewise common-intercept shift are Metropolis updates with the prior and Jacobian factors derived above, and the spline group shift is a Gibbs update of the orbit coordinate. Composing these complete blocks preserves $\pi$. This shows that $\pi$ is a stationary distribution of the idealized sampler. It does not establish convergence from an arbitrary initial state or bound the sampler's mixing rate. Adaptation is restricted to burn-in and the proposal parameters are then fixed. The implementation uses adaptive quadrature for the exposure offsets $v_{ij}$ and numerical approximations to generalized inverse Gaussian (GIG) normalizing constants when needed.

\subsection{Survival likelihood and held-out log scores}
\label{supp:scores}

For subject $i$, let $j_i=n_i+1$ and $0=t_{i0}<\cdots<t_{i,j_i}=T_i$. Measurements before $T_i$ determine the predictable tree input $\bm U_{ij}$ on $(t_{i,j-1},t_{ij}]$. A measurement at $T_i$ is not used to explain an event at $T_i$. Define
\[
\begin{aligned}
v_{ij}&=\int_{t_{i,j-1}}^{t_{ij}}
e^{\alpha_0(u)+\bm b_i^{\trans}\bm z(u)}\,du,\\
\mu_{ij}&=v_{ij}f(\bm U_{ij}),
\qquad
y_{ij}=\delta_i\mathbb I(j=j_i).
\end{aligned}
\]
Conditional on the parameters and the subject's random effect, the survival contribution and product of working Poisson masses satisfy
\[
\begin{aligned}
L_i^{\rm surv}
&=\lambda_i(T_i)^{\delta_i}
\exp\left\{-\sum_{j=1}^{j_i}\mu_{ij}\right\},\\
L_i^{\rm Pois}
&=\prod_{j=1}^{j_i}\frac{\mu_{ij}^{y_{ij}}e^{-\mu_{ij}}}{y_{ij}!}
=L_i^{\rm surv}
\left\{\frac{\mu_{i,j_i}}{\lambda_i(T_i)}\right\}^{\delta_i}\\
&=L_i^{\rm surv}
\left\{
\frac{v_{i,j_i}}
{\exp\{\alpha_0(T_i)+\bm b_i^{\trans}\bm z(T_i)\}}
\right\}^{\delta_i}.
\end{aligned}
\]
The tree contribution cancels from the extra factor. This explains the Poisson likelihood kernel used for conditional tree updates when the baseline and random effects are fixed. The factor generally depends on those other parameters and cannot be dropped from their updates or moved outside predictive integration. If the entire hazard is constant on the event interval, the factor is $(T_i-t_{i,j_i-1})^{\delta_i}$, which is parameter-free on a fixed partition. A spline baseline or a nonzero random time slope generally violates this condition; a piecewise baseline alone is insufficient. The Poisson representation does not turn one event per subject into independent recurrent events.

\paragraph{Predictive averaging and the scoring target.}
Let $\Theta$ collect the global parameters and let $\mathcal D_{\rm train}$ be the training data. For a new subject on a supplied path, write
\[
\mathcal A_i[g]=
\mathbb E_{\Theta\mid\mathcal D_{\rm train}}
\left[\int g(\Theta,\bm b)
\phi_q(\bm b;\bm0,\Sigma_b)\,d\bm b\right].
\]
The same $\bm b$ is used along the subject's entire path. The full censored-data predictive log score, summed over held-out subjects, is
\[
Q_{\rm surv}=
\sum_{i\in\mathcal I_{\rm test}}
\log\mathcal A_i\!\left[
\lambda_i(T_i;\Theta,\bm b)^{\delta_i}
e^{-H_i(T_i;\Theta,\bm b)}\right],
\qquad
H_i(t;\Theta,\bm b)=\int_0^t\lambda_i(u;\Theta,\bm b)\,du.
\]
The censoring contribution is omitted only under the assumption on the censoring, visit, and covariate processes in main-text Section~2.4. Each logarithm follows random-effect integration and posterior averaging. Averaging log-likelihoods, exponentiating an average cumulative hazard, or multiplying separately marginalized interval contributions gives a different quantity. This sum of subjectwise predictive log scores is also distinct from the joint test-set log predictive density, in which the product over subjects is taken before averaging over their shared global parameters.

The simulation tables report $Q_{\rm surv}$, with the event-time density evaluated on the fixed grid described at the end of this section, so censored test subjects enter through their predicted survival probabilities.

In the case study, fitting uses the first recorded composite-flare date, not latent biological onset. A secondary descriptive score assigns each event the log probability of the interval from its preceding no-flare visit, $t_{i,\mathrm{last}}$, to its recorded date $T_i$:
$\log\mathcal A_i[\exp\{-H_i(t_{i,\mathrm{last}};\Theta,\bm b)\}-\exp\{-H_i(T_i;\Theta,\bm b)\}]$.
Censored subjects contribute their log survival probabilities. Follow-up is truncated at three years. All methods use the same intervals, but these depend on observed visits and events. This score differs from the fitting likelihood and is not an interval-censored analysis of onset. Each subject is scored once, after whole-path random-effect integration and posterior averaging.

Interval--Poisson scores are a common alternative for counting-process data. For a fixed partition, two possible predictive aggregations are
\[
\begin{aligned}
Q_{\rm Pois,int}
&=\sum_{i\in\mathcal I_{\rm test}}\sum_{j=1}^{j_i}
\log\mathcal A_i\!\left[
\frac{\mu_{ij}^{y_{ij}}e^{-\mu_{ij}}}{y_{ij}!}\right],\\
Q_{\rm Pois,subj}
&=\sum_{i\in\mathcal I_{\rm test}}
\log\mathcal A_i[L_i^{\rm Pois}].
\end{aligned}
\]
Both take logarithms after predictive averaging, but they differ because the subject's latent effect and global parameter uncertainty couple its intervals. The aggregation convention must therefore be specified when reporting an interval--Poisson score. Neither convention is generally the full continuous-time censored-data score. The constant-hazard special case applies to the joint subject contribution on a common fixed partition. Plug-in comparator scores replace posterior averaging with their fitted predictive probabilities.

$Q_{\rm Pois,int}$ scores each interval as if it belonged to a new subject. For a plug-in model without random effects, the interval terms of subject $i$ sum to $\delta_i\log\widehat\mu_{i,j_i}-\widehat H_i(T_i)$, which differs from the subject's contribution to $Q_{\rm surv}$ only in the event term. With random effects, integrating them separately in each interval ignores that a single $\bm b$ governs all of a subject's intervals. For a scalar frailty, and typically otherwise, it then understates the probability of long event-free follow-up and overstates the conditional probability of an event after it, and the score changes when an interval is split at an additional visit with unchanged covariates. Neither reported analysis uses an interval--Poisson score.

\paragraph{Event-time densities on the evaluation grid.}
A step-function survival curve does not by itself specify a continuous-time density with respect to Lebesgue measure. The simulation scores therefore use one grid convention for every method. Let $u_\ell=0.1\ell$, $\ell=0,\ldots,L$, span the follow-up window, and let $(u_{\ell-1},u_\ell]$ be the grid cell containing $T_i$. An event contributes $\log[\{\widehat S_i(u_{\ell-1})-\widehat S_i(u_\ell)\}/(u_\ell-u_{\ell-1})]$, the average predicted density over that cell. If a step curve assigns no probability to that cell, the cell is widened by one grid step on each side at a time until it does, and the probability of the widened cell is divided by its width. In the reported simulations, this applies to fewer than $0.5\%$ of the event terms of the Cox models, fewer than $2.5\%$ of those of the forest and none of TVSurvBART. A censoring time contributes $\log\widetilde S_i(T_i)$, where $\widetilde S_i$ interpolates $\widehat S_i$ linearly between $u_{\ell-1}$ and $u_\ell$. For TVSurvBART, $\widehat S_i$ is the predictive survival curve, with the random effects integrated out over the subject's whole path and the posterior draws averaged before the logarithm is taken. The probabilities inside the logarithms are linear in $\widehat S_i$, so averaging before taking logs gives the score of the posterior-mean curve; the log score itself is not linear. The Cox models and the forest are scored on their fitted curves.

\subsection{Supplied paths and prospective prediction}
\label{supp:prediction}

Fix an externally supplied tree-input path $\bm U_i(u)$ on $[0,t]$. The model-based survival functional for a new subject is
\[
\bar S_i(t)=
\mathbb E_{\Theta\mid\mathcal D_{\rm train}}
\left[\int e^{-H_i(t;\Theta,\bm b)}
\phi_q(\bm b;\bm0,\Sigma_b)\,d\bm b\right].
\]
This averages survival probabilities over both the new subject's random effect and global parameter uncertainty. A fitted subject's own posterior random-effect draws give a different, subject-specific summary and are not used for held-out subjects.

Suppose the new subject is known to survive to $s<t$, no other informative observation is used, and the same externally fixed path is used in the numerator and denominator. If $\bar S_i(s)>0$, then
\[
\Pr(T_i^*>t\mid T_i^*>s,\bm U_i(\cdot),\mathcal D_{\rm train})
=\frac{\bar S_i(t)}{\bar S_i(s)}.
\]
In particular, survival to $s$ changes the weighting distribution to
\[
q_s(\Theta,\bm b)=
\frac{e^{-H_i(s;\Theta,\bm b)}
p(\Theta\mid\mathcal D_{\rm train})
\phi_q(\bm b;\bm0,\Sigma_b)}{\bar S_i(s)}.
\]
The ratio is the expectation of $\exp\{-H_i(t;\Theta,\bm b)+H_i(s;\Theta,\bm b)\}$ under $q_s$. Its unweighted expectation under the training posterior and the new-subject Gaussian distribution is a different quantity. Thus one cannot average draw-specific survival ratios without this survival weighting. Integrating $q_s$ over $\bm b$ shows that the ratio is also the average of the parameter-specific ratios $\bar S_i(t;\Theta)/\bar S_i(s;\Theta)$ with weights proportional to $\bar S_i(s;\Theta)\,p(\Theta\mid\mathcal D_{\rm train})$, where $\bar S_i(t;\Theta)=\int e^{-H_i(t;\Theta,\bm b)}\phi_q(\bm b;\bm0,\Sigma_b)\,d\bm b$ is the random-effect-integrated survival probability given $\Theta$. The $95\%$ intervals reported for this ratio in the simulations are the corresponding weighted quantiles of the draw-specific ratios over the retained draws, whose weighted mean is the ratio of the posterior means. Under the assumption of main-text Section~2.4, a new subject's covariate history up to $s$ carries no information about its random effects apart from survival to $s$. The weighting $q_s$ is then the full update of the random-effect distribution at $s$, so the ratio is the model's conditional survival probability given that history and the supplied path after $s$.

When trees include interval-start time, the supplied path must specify visit times as well as future covariates. Subdividing an interval solely for numerical integration does not reset its tree time input. If the future path is unknown at the prediction time, prospective prediction requires a future-path model or an explicitly stated scenario. For internal covariates or informative visit times, the supplied-path functional alone is not a joint-model forecast, and the history up to $s$ can then carry information about the random effects beyond survival to $s$, which the ratio does not use.

The landmark evaluations of main-text Sections~5 and~6.3 use this ratio along the frozen path, the observed history up to $s$ with the last visit at or before $s$ carried forward as one interval beyond $s$ (Web Appendix~\ref{supp:sim-eval}). This holds the covariates at their last observed values beyond $s$, as in landmark prediction \citep{vanHouwelingen2007,vanHouwelingenPutter2011}, and is an explicitly stated scenario in the sense above: it treats the covariates after $s$ as external, and a joint model for the covariate and event processes is the alternative. The landmark predictions use only information available at $s$ and are scored on the outcomes after $s$. The carried-forward evaluations of Web Appendix~\ref{supp:sim-pred} (Web Tables~\mbox{\ref{tab:sim1-calibration}--\ref{tab:sim2-auc}}) instead extend observed histories that stop at the event or censoring time by last observation carried forward. Such paths can contain information about the outcome or follow-up process, which neither holding out whole subjects nor inverse probability of censoring weighting (IPCW) removes, so those results are retrospective comparisons along realized paths.

\section{Additional model and computational details}\label{supp:details}

\subsection{Natural cubic spline basis}\label{supp:spline}
For the natural cubic spline baseline of main-text Section~2.2, let \(a=\kappa_0\) and
\(b=\kappa_{K-1}\) denote the boundary knots, and let
\(\kappa_1,\ldots,\kappa_{K-2}\) denote the internal knots. Define
\((u)_+^3=\{\max(u,0)\}^3\) and
$
\omega_j=\frac{b-\kappa_j}{b-a}, j=1,\ldots,K-2.
$
A \(K\)-dimensional natural cubic spline basis, including the intercept, is
given by
\[
\begin{aligned}
B_0(t)&=1,
\qquad
B_1(t)=t,\\
B_{j+1}(t)
&=
(t-\kappa_j)_+^3
-\omega_j(t-a)_+^3
-(1-\omega_j)(t-b)_+^3,
\qquad j=1,\ldots,K-2.
\end{aligned}
\]
The nonintercept basis functions are centered at their means over the observed at-risk time:
\[
\widetilde B_\ell(t)
=
B_\ell(t)-\bar B_\ell,
\qquad
\bar B_\ell
=
\frac{\sum_{i=1}^{N}\int_0^{T_i}B_\ell(u)\,du}{\sum_{i=1}^{N}T_i},
\qquad
\ell=1,\ldots,K-1.
\]
By construction, $\sum_{i=1}^{N}\int_0^{T_i}\widetilde B_\ell(u)\,du=0$.

\subsection{Prior details and calibration}\label{supp:leafprior}
\paragraph{Tree ensemble.} The tree-ensemble prior of main-text Section~3 factorizes as
\[
p\!\left(\{R_h,\bm{\theta}_h\}_{h=1}^{m}\right)
=
\prod_{h=1}^{m}
p(R_h)
\prod_{k=1}^{b_h}
p_\theta(\theta_{hk}\mid c,d).
\]
Under the structure prior $p(R_h)$ of \citet{chipman2010bart}, a node at depth $r$, with the root at depth $0$, is split with probability $0.95(1+r)^{-2}$, the default of that paper, and, conditional on splitting, the splitting variable is drawn uniformly from the available covariates and the cut point uniformly from its admissible values. The implementation uses the form of this prior in the \code{BART} package, from which its tree sampler is adapted: the splitting variable is drawn uniformly from all tree inputs, and any birth that would create a terminal node with fewer than five data records, counting one record per visit interval and one end-of-follow-up record per subject, is rejected. The sampler therefore targets the posterior under this form of the prior, restricted to trees whose terminal nodes each contain at least five such records. Under the balanced mixture prior of main-text Section~3, the leaf values can be written as
\[
\theta_{hk}
=
U_{hk}W_{hk}
+
(1-U_{hk})W_{hk}^{-1},
\qquad
W_{hk}\sim\mathrm{Gamma}(c,d),
\qquad
U_{hk}\sim\mathrm{Bernoulli}\!\left(\frac12\right),
\]
where the gamma distribution is parameterized by shape and rate.

The ensemble-scale parameter $s_F$ determines $(c,d)$ as follows. The induced prior on $\mu_{hk}=\log\theta_{hk}$ satisfies
\[
\mathbb{E}(\mu_{hk})=0,
\qquad
\operatorname{Var}(\mu_{hk})=\psi_1(c)+\{\psi_0(c)-\log d\}^2,
\]
where $\psi_0$ and $\psi_1$ denote the digamma and trigamma functions, respectively. The implementation uses the closed-form calibration $c=m/s_F^2+1/2$ and $d=m/s_F^2$. For small $s_F^2/m$, these values approximate the exact variance calibration $\psi_1(c)=s_F^2/m$ and $\log d=\psi_0(c)$. Consequently, $\mathbb{E}\{F(\bm{u})\}=0$ and $\operatorname{Var}\{F(\bm{u})\}\approx s_F^2$ at any fixed $\bm{u}$, and $F(\bm{u})$ is approximately $\mathcal{N}(0,s_F^2)$ when $m$ is large. Thus, $s_F$ controls the overall prior scale of the tree component, and each tree contributes log-scale standard deviation approximately $s_F/\sqrt{m}$.

For two inputs $\bm{u},\bm{u}'$ at the same time and random-effect value, the tree contribution to the log-hazard contrast has prior variance approximately $2s_F^2\{1-\rho(\bm{u},\bm{u}')\}$, where $\rho$ is the prior correlation of $F$ at those inputs. As a broad calibration, the zero-correlation normal approximation gives a central $95\%$ hazard-ratio interval of $[0.001,1000]$ when $s_F=\log(1000)/(z_{0.975}\sqrt{2})\approx2.492$; with $m=100$ trees, $c\approx16.601$ and $d\approx16.101$. This calibration is deliberately broad, and it refers to two inputs with uncorrelated tree values. Inputs that share terminal nodes have positively correlated values of $F$, so contrasts between similar inputs are regularized more tightly than the marginal interval suggests. The prior-sensitivity value $s_F=\log(100)/(z_{0.975}\sqrt{2})\approx1.661$ narrows the interval to $[0.01,100]$ and gives $c\approx36.727$ and $d\approx36.227$.

\paragraph{Random-effects covariance.} The hierarchical prior of \citet{huang2013simple} is
\[
\begin{aligned}
\Sigma_b\mid\bm{a}
&\sim
\mathrm{IW}\!\left(
2\nu\,\mathrm{diag}(a_1^{-1},\ldots,a_q^{-1}),
\,
\nu+q-1
\right),\\
a_k
&\overset{\mathrm{ind}}{\sim}
\mathrm{Inv\text{-}Gamma}\!\left(
\frac12,\frac{1}{A_k^2}
\right),
\qquad k=1,\ldots,q,
\end{aligned}
\]
where the two arguments of \(\mathrm{IW}(\cdot,\cdot)\) denote the scale matrix and degrees of freedom, respectively. Under this hierarchy, each marginal standard deviation satisfies $\sqrt{\Sigma_{b,kk}}\sim\mathrm{half}\text{-}t_\nu(A_k)$. Setting \(\nu=2\) induces a uniform prior on the intercept--slope correlation. With $q=1$, the hierarchy reduces to a half-$t_\nu(A_1)$ prior on the random-intercept standard deviation. The scales $A_k$ are interpreted on the log-hazard scale. The analyses use $\nu=2$ and $A_1=A_2=5$ unless stated otherwise.

\paragraph{Baseline.} Under the natural cubic spline specification (Web Appendix~\ref{supp:spline}), the coefficients have independent priors $\beta_\ell\sim\mathcal{N}(\mu_\beta,\sigma_\beta^2)$, $\ell=0,\ldots,K-1$, with $\mu_\beta=0$ and $\sigma_\beta^2=1$. These priors are defined in the centered basis, so how informative they are depends on the time unit and the scaling of the basis as well as on the coefficient variance. The knots are equally spaced: at $0,1,\ldots,\tau$ in the simulations (main-text Section~5) and at four points from $0$ to the maximum follow-up time in the case study (Web Appendix~\ref{supp:app-eval}). Under the piecewise-constant specification, the interval-specific hazard heights $\lambda_{0j}=\exp(\gamma_j)$ have independent priors $\lambda_{0j}\sim\mathrm{Gamma}(a_\lambda,b_\lambda)$, $j=1,\ldots,J$, in the shape--rate parameterization. Equivalently, the log-heights $\gamma_j$ receive the induced log-gamma priors. We use $a_\lambda=b_\lambda=0.1$. Because $\lambda_{0j}$ is a hazard per unit time, how informative this prior is depends on the time unit and the interval partition. When time is recorded on a discrete grid, consecutive grid points can serve as the interval boundaries. Otherwise the cut points are fixed before fitting, and in the case study, they are chosen from each training set (Web Appendix~\ref{supp:app-eval}).

\subsection{Tree updates}\label{supp:gig}
For \(\eta\in\mathbb{R}\) and \(\chi,\psi\geq 0\) such that the integral is finite, the GIG normalizing constant used in main-text Section~4 is
\[
\begin{aligned}
Z(\eta,\chi,\psi)
&=
\int_0^\infty
x^{\eta-1}
\exp\!\left\{
-\frac{1}{2}
\left(
\frac{\chi}{x}+\psi x
\right)
\right\}
\,dx                                                     \\
&=
\begin{cases}
\Gamma(\eta)(2/\psi)^\eta,
& \eta>0,\ \chi=0,\ \psi>0,\\[3pt]
\Gamma(-\eta)(2/\chi)^{-\eta},
& \eta<0,\ \chi>0,\ \psi=0,\\[3pt]
2(\chi/\psi)^{\eta/2}
K_\eta(\sqrt{\chi\psi}),
& \chi>0,\ \psi>0,
\end{cases}
\end{aligned}
\]
where \(K_\eta\) is the modified Bessel function of the second kind.

For the $h$th tree, let $f_{(h)}(\bm u)=\prod_{\ell\ne h}G_\ell(\bm u;R_\ell,\bm\theta_\ell)$ denote the partial ensemble. Grouping the likelihood contributions of main-text Section~2.4 according to the terminal regions of $R_h$ gives
\[
L_h(R_h,\bm\theta_h\mid\cdot)=c_h\prod_{k=1}^{b_h}\theta_{hk}^{r_{hk}}\exp(-s_{hk}\theta_{hk}),
\qquad
c_h=\prod_{i=1}^{N}w_i\bigl[f_{(h)}(\bm U_{i,n_i+1})\bigr]^{\delta_i},
\]
where
\[
r_{hk}=\sum_{i=1}^{N}\delta_i\mathbb I\{\bm U_{i,n_i+1}\in\mathcal A_{hk}\},\qquad s_{hk}=\sum_{i=1}^{N}\sum_{j=1}^{n_i+1}\mathbb I\{\bm U_{ij}\in\mathcal A_{hk}\}v_{ij}f_{(h)}(\bm U_{ij})
\]
are the number of observed events assigned to terminal node $k$ and its partial-ensemble-weighted exposure; $c_h$ does not depend on $R_h$ or $\bm\theta_h$. Integrating the terminal-node values under the leaf prior of main-text Section~3 gives the marginal likelihood $\mathcal M_h(R_h)$ of main-text Section~4. For a structure proposal $R_h^\ast\sim q_R(R_h^\ast\mid R_h)$, the Metropolis--Hastings acceptance probability is
\[
\alpha_h=\min\left\{1,\ \frac{p(R_h^\ast)\mathcal M_h(R_h^\ast)q_R(R_h\mid R_h^\ast)}{p(R_h)\mathcal M_h(R_h)q_R(R_h^\ast\mid R_h)}\right\}.
\]
After accepting or rejecting the structure proposal, all terminal-node values are redrawn independently from
\[
p(\theta_{hk}\mid\cdot)=\pi_{hk}\,p_{\mathrm{GIG}}(\theta_{hk}\mid -c+r_{hk},\,2d,\,2s_{hk})+(1-\pi_{hk})\,p_{\mathrm{GIG}}(\theta_{hk}\mid c+r_{hk},\,0,\,2[d+s_{hk}]),
\]
where
\[
\pi_{hk}=\frac{Z(-c+r_{hk},2d,2s_{hk})}{Z(-c+r_{hk},2d,2s_{hk})+Z(c+r_{hk},0,2[d+s_{hk}])}
\]
and $p_{\mathrm{GIG}}(\cdot\mid\eta,\chi,\psi)$ is the GIG density with normalizing constant $Z(\eta,\chi,\psi)$.

\subsection{Spline baseline update}\label{supp:spline-update}
For the natural cubic spline baseline, let
$
\widetilde B_0(t)\equiv 1
$
and
$
\alpha_0(t)
=
\sum_{j=0}^{K-1}
\beta_j\widetilde B_j(t).
$
The likelihood score for \(\beta_j\) is
\[
\frac{\partial\ell(\bm{\beta})}{\partial\beta_j}
=
\sum_{i=1}^{N}
\delta_i\widetilde B_j(T_i)
-
\sum_{i=1}^{N}
\sum_{r=1}^{n_i+1}
f(\bm{U}_{ir})
\int_{t_{i,r-1}}^{t_{ir}}
\widetilde B_j(u)
\exp\!\left\{
\alpha_0(u)
+
\bm{b}_i^{\trans}\bm{z}_i(u)
\right\}
\,du.
\]
Let $\bm g_\beta(\bm\beta)$ be the log full-conditional gradient, with components $\partial\ell(\bm\beta)/\partial\beta_j-(\beta_j-\mu_\beta)/\sigma_\beta^2$. The coefficients are updated jointly by the preconditioned MALA proposal
\[
\bm\beta^\ast
=
\bm\beta
+
\frac{\varsigma_\beta^2}{2}P_\beta\bm g_\beta(\bm\beta)
+
\varsigma_\beta L_\beta\bm\varepsilon,
\qquad
\bm\varepsilon\sim\mathcal{N}(\bm0,I_K),
\qquad
P_\beta=L_\beta L_\beta^{\trans},
\]
where $P_\beta$ is a positive-definite preconditioner. The proposal is accepted using the full conditional density and the asymmetric-proposal correction. The step size $\varsigma_\beta$ and the preconditioner $P_\beta$, estimated from regularized secant curvature information, are adapted only during burn-in and then held fixed.

\subsection{Adaptation of proposal scales}\label{supp:adapt}
The MALA scales $\varsigma_i$ of the random effects (Web Appendix~\ref{supp:invariance}) are adapted during burn-in using $0.574$ as a tuning target \citep{RobertsRosenthal1998} and then held fixed. This target need not be optimal for these low-dimensional blocks. The non-centered covariance update of main-text Algorithm~1 uses three proposals per iteration. During burn-in, its proposal covariance $\Omega=s_\vartheta^2P_\vartheta$ (Web Appendix~\ref{supp:invariance}) is adapted: $s_\vartheta$ every $100$ iterations toward acceptance $0.30$ for $q=2$ ($0.44$ for $q=1$), and $P_\vartheta$ every $500$ iterations using a ridge-regularized empirical covariance after the first quarter. Both are then fixed. The standard deviations of the Gaussian increments of the ensemble--baseline level shift (main-text Algorithm~1) and of the piecewise common-intercept shift (Web Appendix~\ref{supp:recentering}) are adapted every $500$ burn-in iterations toward acceptance $0.44$ and then fixed.

\subsection{Run lengths, within-chain checks, and computing times}\label{supp:mcmc}
Each production fit of TVSurvBART uses one chain of $120{,}000$ iterations, with $20{,}000$ burn-in iterations and every $50$th subsequent state retained, giving $2{,}000$ draws. Parameters, selected survival probabilities, and evaluation criteria are monitored using rank-normalized split $\widehat R$ and bulk and tail effective sample sizes, with thresholds $1.01$ and $400$. The simulation fits and the case-study cross-validation fits ran as single-threaded jobs on a computing cluster (Intel Xeon Gold 6538Y+), with $4$ GB of memory. A simulation fit took a median of $7.0$ hours in Simulation~I ($8.5$ hours with irregular visits) and $8.8$ hours in Simulation~II, and a case-study fit $2.6$ hours with the piecewise baseline and $3.9$ hours with the spline baseline.

\section{Additional simulation details}\label{supp:simulation_studies}

\subsection{Evaluation metrics}\label{supp:sim-eval}

Every method predicts a held-out subject's survival along a supplied covariate path, which is a frozen landmark path, a complete generated path, or a carried-forward path, as defined below. Let $s_{ij}$ denote the interval endpoints of the path up to the evaluation horizon, and each interval carries the covariates measured at its start. The Cox-type benchmarks accumulate the cumulative hazard record by record along this path, and the forest chains conditional survival across records (Web Appendix~\ref{supp:benchmarks}). TVSurvBART averages survival probabilities over the retained parameter states and over one random effect for the whole path, $\widehat S_i(t)=D^{-1}\sum_{d=1}^D\bar S_i^{(d)}(t)$ with $\bar S_i^{(d)}(t)=\int\exp\{-\widehat H_i^{(d)}(t;\bm b)\}\phi_q(\bm b;\bm0,\Sigma_b^{(d)})\,d\bm b$, where $\widehat H_i^{(d)}(t;\bm b)=\int_0^t\lambda_i^{(d)}(u;\bm b)\,du$ is the cumulative hazard under draw $d$, $\phi_q$ is the $\mathcal N(\bm0,\Sigma_b^{(d)})$ density, and the integral is approximated by Gauss--Hermite quadrature with $16$ nodes per dimension. This is the model's predictive survival probability for an externally fixed path (Web Appendix~\ref{supp:prediction}).

\paragraph{Landmark predictions.} The landmark times are $s=1,2$ in Simulation~I, with regular or irregular visits, and $s=1,2,3$ in Simulation~II, each with horizon $t=s+1$. The at-risk set at $s$ consists of the test subjects of a replicate whose observed follow-up time $T_i$ exceeds $s$. The frozen path of an at-risk subject is its observed history up to $s$, with the last visit at or before $s$ carried forward as one interval beyond $s$, so it uses only information available at $s$. Along it, the tree time input of TVSurvBART stays at the time of that visit, which on the regular grid is $s$ itself. This matters only in the non-PH models, whose trees receive time. 

Every method predicts the conditional survival probability $\widehat S_i(t\mid s)=\widehat S_i(t)/\widehat S_i(s)$ along the same path: $\exp[-\{\widehat H_i(t)-\widehat H_i(s)\}]$ for CoxPH, and the ratio at $t$ and $s$ of the frailty-integrated survival probability for frailty Cox and of the chained survival curve for RSF, which equals the forest's chained conditional survival from $s$ when $s$ is a record boundary, as on the regular grids. For TVSurvBART, it is the ratio of the posterior means, $\sum_d\bar S_i^{(d)}(t)/\sum_d\bar S_i^{(d)}(s)$, and its $95\%$ interval is formed by the $2.5\%$ and $97.5\%$ quantiles of the draw ratios $\bar S_i^{(d)}(t)/\bar S_i^{(d)}(s)$ weighted by $\bar S_i^{(d)}(s)$ (Web Appendix~\ref{supp:prediction}).

The landmark criteria use the observed outcomes of the at-risk subjects and inverse probability of censoring weights from the product-limit estimate $\widehat G$ of their censoring distribution after $s$. The cumulative/dynamic AUC is computed with \code{timeROC}, with time measured from $s$, from the marker $1-\widehat S_i(t\mid s)$. The IPCW Brier score averages over the at-risk set $\widehat S_i(t\mid s)^2/\widehat G(T_i-)$ for subjects with an event before $t$, $\{1-\widehat S_i(t\mid s)\}^2/\widehat G(t)$ for subjects event-free at $t$, and $0$ for subjects censored before $t$. The index of prediction accuracy is $\mathrm{IPA}=1-\mathrm{BS}/\mathrm{BS}_0$, where $\mathrm{BS}_0$ is the Brier score of the product-limit estimate of survival in the at-risk set, and calibration-in-the-large is the mean of $\widehat S_i(t\mid s)$ minus that estimate at $t$, positive when survival is over-predicted. All landmark criteria are evaluated at $t-10^{-6}$, so that at $s=2$ in Simulation~I, where $t=3$ is the administrative censoring time $\tau$, subjects censored at $\tau$ count as event-free at the horizon. Each criterion is averaged over the $200$ replicates, with Monte Carlo standard errors from the variation between replicates and with differences between methods paired within replicates.

Two true-model references are computed for the same at-risk subjects. ``True model (frozen path)'' is the generating model's conditional survival probability along the frozen path itself: on its last interval, which starts at the last visit at or before $s$, the covariates and, in the non-PH models, the coefficient $\beta_3$ of $x_3$ keep their values at that visit, and in Simulation~I, the random effects are integrated out over their true distribution. On the regular grid, that visit is at $s$ and $\beta_3$ is constant on $(s,t]$. For irregular visits, the last visit precedes $s$, and $\beta_3$ keeps its value at that visit as if no further visit came. ``True model (complete path)'' is the same conditional survival probability along the complete generated path defined below. In Simulation~II, the two references coincide for every at-risk subject, because the visits are at the integers and the hazard on $(s,s+1]$ depends on the covariates measured at $s$. The bias and RMSE of every method, and the coverage and mean width of TVSurvBART's $95\%$ intervals, are computed against the frozen-path truth over the at-risk subjects of each replicate and then averaged over replicates.

\paragraph{Complete generated paths.} The complete generated path of a test subject contains every visit generated before $\tau$, including those after its event or censoring time, with the covariates generated there. The covariate processes and the irregular visit times are generated without reference to the event and censoring times, so these paths are well defined for every test subject. They are not observed in practice, and predictions along them separate the hazard model from the extension of observed paths. Along these paths, Web Appendix~\ref{supp:sim-coverage} (Web Tables~\ref{tab:sim1-coverage} and~\ref{tab:sim-landmark-coverage}) assesses the coverage of TVSurvBART's intervals for $S_i(t)$ and for the landmark conditional survival of every test subject, against targets in which the random effects of Simulation~I are integrated out over their true distribution rather than set to their realized values, and Web Appendix~\ref{supp:sim-complete} (Web Tables~\mbox{\ref{tab:sim1-complete}--\ref{tab:sim-complete-auc}}) reports RMST at $\tau/2$, median survival, IBS, $\mathrm{RMSE}_S$, and AUC, defined as for carried-forward paths below, with every method's curves, and those of the true model, computed along the complete paths.

\paragraph{Carried-forward paths.} A carried-forward path follows a held-out subject's observed covariate history, which ends at its event or censoring time, and extends it beyond the last observed visit by last observation carried forward (LOCF). Summaries along these paths are retrospective (Web Appendix~\ref{supp:prediction}). In the two non-PH models of Simulation~I, the carried-forward predictions of TVSurvBART also hold the tree time input $t_{i,j-1}/\tau$ at its value at the last observed visit, whereas the coefficient of $x_3$ in the generating model continues to change after that visit, a limitation of these predictions.

Each simulated cohort is split by subject into training and test sets, and performance is summarized over $200$ independent replicates. The cumulative/dynamic AUC is computed with \code{timeROC} \citep{blanche2013timeROC} from the marker $1-\widehat S_i(t)$ and the test subjects' latent event times $T_i^*$, which are known in a simulation, so its IPCW weights are identically $1$. At the horizon, it is evaluated at exactly $t=\tau$, with subjects without an event by $\tau$ coded as $T_i^*=\tau+10^{-6}$, so that they count as controls. The markers follow each test subject's covariate history up to $\min(T_i^*,C_i)$, carried forward, so the AUC depends on the censoring fraction. We use it to compare methods within a setting. The integrated Brier score (IBS) is the trapezoidal average over the grid $0,0.1,\ldots,\tau-0.1$ of the IPCW Brier score, computed from the test subjects' observed outcomes with censoring weights from their product-limit estimate. The root mean squared error of the predicted survival curves, $\mathrm{RMSE}_S$, is computed on the grid $0,0.1,\ldots,\tau$ ($31$ points in Simulation~I and $51$ in Simulation~II) against each test subject's true survival curve given its complete generated visit schedule and covariate path, including the visits after $T_i$, with the random effects integrated out.

RMST at $\tau/2$ and median survival are computed from the average predicted curve $\widehat{\bar S}(t)=N_s^{-1}\sum_i\widehat S_i(t)$ over the $N_s$ test subjects, RMST by integration and the median by the first crossing of $0.5$. Their reference values in a replicate come from the uncensored latent event times of its test subjects: $N_s^{-1}\sum_i\min(T_i^*,\tau/2)$ and the sample median of the $T_i^*$. Bias and RMSE are the mean and the root mean square over replicates of the difference between the estimate and its reference value, and the Monte Carlo standard error is that of the mean difference.

Web Tables~\ref{tab:sim1-elpd} and~\ref{tab:sim2-elpd} (Web Appendix~\ref{supp:sim-logscore}) report pairwise differences in subject-level test log scores, computed on the common $0.1$-unit grid with the random effects integrated and the posterior draws averaged before logarithms are taken (Web Appendix~\ref{supp:scores}).

Three AUC references are reported in Simulation~I and two in Simulation~II. ``True model (carried forward)'' applies the same scorer to the same test subjects, latent outcomes and carried-forward paths as the fitted methods, with the marker $1-S_i(t)$ from the true survival probability given the carried-forward path, with the random effects integrated out in Simulation~I, so it also depends on the censoring fraction. In the non-PH models, it lets the coefficient of $x_3$ change after the last observed visit as in the generating model, which with irregular visits uses the subject's later visit times, not observed at the exit. The population references are computed from an independent, uncensored Monte Carlo sample of $10^5$ subjects of each design with complete generated paths and the same coding of the latent event times, with the marker $1-S_i(t)$ from the true survival probability given the complete path. In Simulation~I, the Marginal reference integrates the random effects out and the Oracle reference uses their realized values. Simulation~II has no frailty and a single Reference. The population references do not depend on the censoring fraction and, because the fitted models see stopped, carried-forward paths, they are reference points that need not bound the fitted AUCs.

\subsection{Benchmark implementations}\label{supp:benchmarks}

\paragraph{Time-varying Cox regression.}
Time-varying Cox regression is fitted in counting-process form using \code{survival} version 3.8-6 in \R{} \citep{survival-package}. Covariates are held constant within each observation interval, with $\bm{x}_{i,j-1}$ denoting the measurement at its start. The cumulative hazard is estimated as
\[
\widehat H_i(t)
=
\sum_{j:\,s_{i,j-1}<t}
e^{\widehat{\bm{\beta}}^{\top}\bm{x}_{i,j-1}}
\bigl[
\widehat\Lambda_0\bigl(\min(s_{ij},t)\bigr)
-
\widehat\Lambda_0(s_{i,j-1})
\bigr],
\]
where $\widehat\Lambda_0$ is the Breslow estimator of the cumulative baseline hazard. The predicted survival probability is then $\widehat S_i(t)=\exp\{-\widehat H_i(t)\}$.

\paragraph{Adapted frailty Cox regression.}
An adapted frailty Cox model includes a subject-level log-normal random intercept $b_i\sim\mathcal N(0,\sigma_b^2)$ and is fitted using \code{coxme} version 2.2-22 in \R{} \citep{coxme-package}. To obtain survival predictions, we reconstruct a Breslow-type baseline using a Gaussian approximation to the conditional frailty distribution given the training data. Let $\widehat m_i$ and $\widehat s_i^2$ denote the fitted mode and local variance, respectively. Approximating $E(e^{b_i}\mid\text{training data})$ by $\exp(\widehat m_i+\widehat s_i^2/2)$ gives
\[
d\widehat\Lambda_0(t_k)
=
\frac{d_k}{
\displaystyle
\sum_{(i,j)\in\mathcal R(t_k)}
e^{\widehat{\bm{\beta}}^{\top}\bm{x}_{i,j-1}}
\exp\bigl(\widehat m_i+\tfrac12\widehat s_i^2\bigr)
},
\]
where $d_k$ is the event count at training event time $t_k$ and
$\mathcal R(t_k)=\{(i,j):t_{i,j-1}<t_k\le t_{ij}\}$ is the training risk set. The local Laplace variance is
$\widehat s_i^2=(1/\widehat\sigma_b^2+\widehat\Lambda_i)^{-1}$, where $\widehat\Lambda_i$ is the subject's fitted cumulative hazard computed from an initial baseline constructed by plugging in the frailty modes. This baseline reconstruction uses a one-step Laplace approximation and is not an exact frailty likelihood calculation. For a test subject, $\widehat H_i(t)$ is accumulated as above using the adapted baseline and setting the log-frailty to zero. The unobserved frailty is then integrated out numerically to obtain
\[
\widehat S_i(t)
=
\sum_{g=1}^{G}
w_g\exp\bigl\{-\widehat H_i(t)e^{z_g}\bigr\},
\]
where $(z_g,w_g)_{g=1}^{G}$ are the Gauss--Hermite nodes and normalized weights for $\mathcal N(0,\widehat\sigma_b^2)$, with $G=41$.

\paragraph{Random survival forest.}
The RSF is implemented by using \code{LTRCforests} version $0.7.0$ in \R{} \citep{LTRCforests-package}, with \code{mtry} set to $\lceil\sqrt{p}\rceil$, which is $3$ in both simulations, $200$ trees and the remaining \code{ltrcrrf} defaults. The forest is grown on the interval records, and each subject's curve is obtained from the fitted forest along that subject's record sequence on the common evaluation grid by chaining conditional survival across records \citep{Yao2022}. For subject $i$ with records $(s_{i,j-1},s_{ij}]$, $j=1,\dots,J_i$, let $\widehat S_{bij}$ be the in-bag-weighted product-limit estimate for left-truncated right-censored data, computed from the training records that share the terminal node of record $j$ in tree $b=1,\dots,B$. The forest's conditional survival within record $j$ is the tree average $\bar s_{ij}(t)=B^{-1}\sum_{b=1}^{B}\widehat S_{bij}(t)/\widehat S_{bij}(s_{i,j-1})$, and the predicted curve is
\[
\widehat S_i(t)=\Bigl\{\prod_{k<j}\bar s_{ik}(s_{ik})\Bigr\}\,\bar s_{ij}(t),\qquad s_{i,j-1}\le t<s_{ij},
\]
with the last record extended beyond $s_{iJ_i}$, so that its covariate values are carried forward to the end of the grid as for the other methods. This is the ensemble estimator implemented in \code{predictProb} of \code{LTRCforests}.

\section{Simulation designs}\label{supp:simdesign}

\subsection{Simulation I covariate processes}\label{supp:sim1}
Three continuous covariates follow polynomial mixed models,
\[
x_{kij}=\gamma_{k0}+u_{ki0}+u_{ki1}t_j+u_{ki2}t_j^2+\varepsilon_{kij},
\quad
\bm u_{ki}\sim\Normal(\bm0,D_k),\quad
\varepsilon_{kij}\sim\Normal(0,\sigma_k^2),
\]
with parameters in Web Table~\ref{tab:covariate-dgp}. A binary longitudinal covariate satisfies
$\mathrm{logit}\,P(x_{4ij}=1\mid\bm u_{4i}) =0.2+u_{4i0}+u_{4i1}t_j+u_{4i2}t_j^2,$
where
\[
\bm u_{4i}\sim\Normal\!\left(\bm0,
\begin{pmatrix}
1.000&0.050&0.005\\
0.050&0.300&0.005\\
0.005&0.005&0.020
\end{pmatrix}\right).
\]
The covariate $x_{5i}$ is time-invariant and uniform on $\{1,2,3,4\}$. The hazards of main-text Section~5.2 use the measured values: on $(t_{j-1},t_j]$, $x_k$ is the value measured at $t_{j-1}$, including the measurement errors $\varepsilon_{kij}$ of $x_1,x_2,x_3$.

\begin{table}[H]
\caption{Parameters for the continuous longitudinal-covariate generating processes in Simulation I.}
\label{tab:covariate-dgp}
\centering
\small
\begin{tabular}{cccc}
\toprule
Covariate & $\gamma_{k0}$ & $D_k$ & $\sigma_k$\\
\midrule
$x_1$ & $-0.5$ &
$\left(\begin{smallmatrix}0.250&0.050&0.005\\0.050&0.500&0.005\\0.005&0.005&0.020\end{smallmatrix}\right)$
& $0.30$\\[1.2em]
$x_2$ & $0.5$ &
$\left(\begin{smallmatrix}0.300&-0.050&0.000\\-0.050&0.600&-0.005\\0.000&-0.005&0.030\end{smallmatrix}\right)$
& $0.35$\\[1.2em]
$x_3$ & $0.0$ &
$\left(\begin{smallmatrix}0.200&0.030&0.003\\0.030&0.400&0.003\\0.003&0.003&0.018\end{smallmatrix}\right)$
& $0.25$\\
\bottomrule
\end{tabular}
\end{table}

\subsection{Simulation I with irregular visits}\label{supp:sim1-irr}
The irregular-visit variant of Simulation~I is the non-PH model with subject-specific visit times. Subject $i$ has visits $0=t_{i0}<t_{i1}<\cdots<t_{i,n_i^*}<\tau$, where $t_{ij}=t_{i,j-1}+\Delta_{ij}$, the gaps $\Delta_{ij}$ are independent and uniform on $[0.25,0.75]$, with mean equal to the regular spacing $0.5$, and $n_i^*=\max\{j:t_{ij}<\tau\}$. Every subject has at least $5$ visits before $\tau$ and none in the 200 replicates has more than $10$. The gaps are drawn independently of the covariate processes and of $b_{i0},b_{i1}$, and the event time does not influence them. The covariate models of Web Appendix~\ref{supp:sim1} are used with $t_{ij}$ in place of $t_j$. On $(t_{i,j-1},\min(t_{ij},\tau)]$, $j=1,\ldots,n_i^*+1$, the hazard is that of the non-PH model with $t_{i,j-1}$ in place of $t_{j-1}$ and $\xi=2.84$, so the coefficient of $x_3$ changes at each subject's first visits at or after $t=1$ and $t=2$. Censoring is as in the PH model of main-text Section~5.2, with $r=0.20$ only. 

\subsection{Simulation II baseline distributions and transitions}\label{supp:sim2}
At the visits $t=j$, $j=0,\ldots,4$, the evolving covariates $L_j=(L_{1j},\ldots,L_{6j})$, three binary ($L_1$--$L_3$) and three continuous ($L_4$--$L_6$), and a time-varying binary covariate $X_j$ are measured. At baseline, $L_{k0}\sim\mathrm{Bernoulli}(0.5)$ for $k=1,2,3$, $L_{k0}\sim\Normal(2,1)$ for $k=4,5,6$, and $X_0\sim\mathrm{Bernoulli}\{ \operatorname{expit}(-1+\bar L_0^{\mathrm b}-0.25\bar L_0^{\mathrm c})\}.$ Covariates then change according to
\[
\begin{aligned}
L_{kj}&\sim\mathrm{Bernoulli}\{
\operatorname{expit}(-2-2X_{j-1}-0.5L_{k+3,j-1})\},
&&k=1,2,3,\\
L_{kj}&\sim\Normal(2+X_{j-1}-L_{k-3,j-1}+0.5L_{k,j-1},1),
&&k=4,5,6,\\
X_j&\sim\mathrm{Bernoulli}\{
\operatorname{expit}(-1+\bar L_j^{\mathrm b}-0.25\bar L_j^{\mathrm c})\}.
\end{aligned}
\]
Here $\bar L_j^{\mathrm b}=3^{-1}\sum_{k=1}^3L_{kj}$ and $\bar L_j^{\mathrm c}=3^{-1}\sum_{k=4}^6L_{kj}$ are the averages of the binary and of the continuous covariates $L_j$ thus depends on $(X_{j-1},L_{j-1})$, and $X_j$ on the concurrent $L_j$. The binary covariates are monotone (absorbing at one): each of $L_{1j},L_{2j},L_{3j}$ and $X_j$ is redrawn from its displayed law only while it is still zero and remains one thereafter. The intercepts $(\varphi_1,\ldots,\varphi_5)=(-2.32,\allowbreak-3.52,\allowbreak-3.92,\allowbreak-3.52,\allowbreak-2.32)$ of the event probabilities $p_j$ form a U-shaped sequence, a common level plus the offsets $0.4(j-3)^2-0.8$, with the level chosen so that the marginal event probability by $\tau$ is $85\%$.

\section{Additional simulation results}\label{supp:simresults}

\subsection{Landmark predictions}\label{supp:sim-landmark}

\begin{figure}[tbp]
\centering
\includegraphics[width=\linewidth]{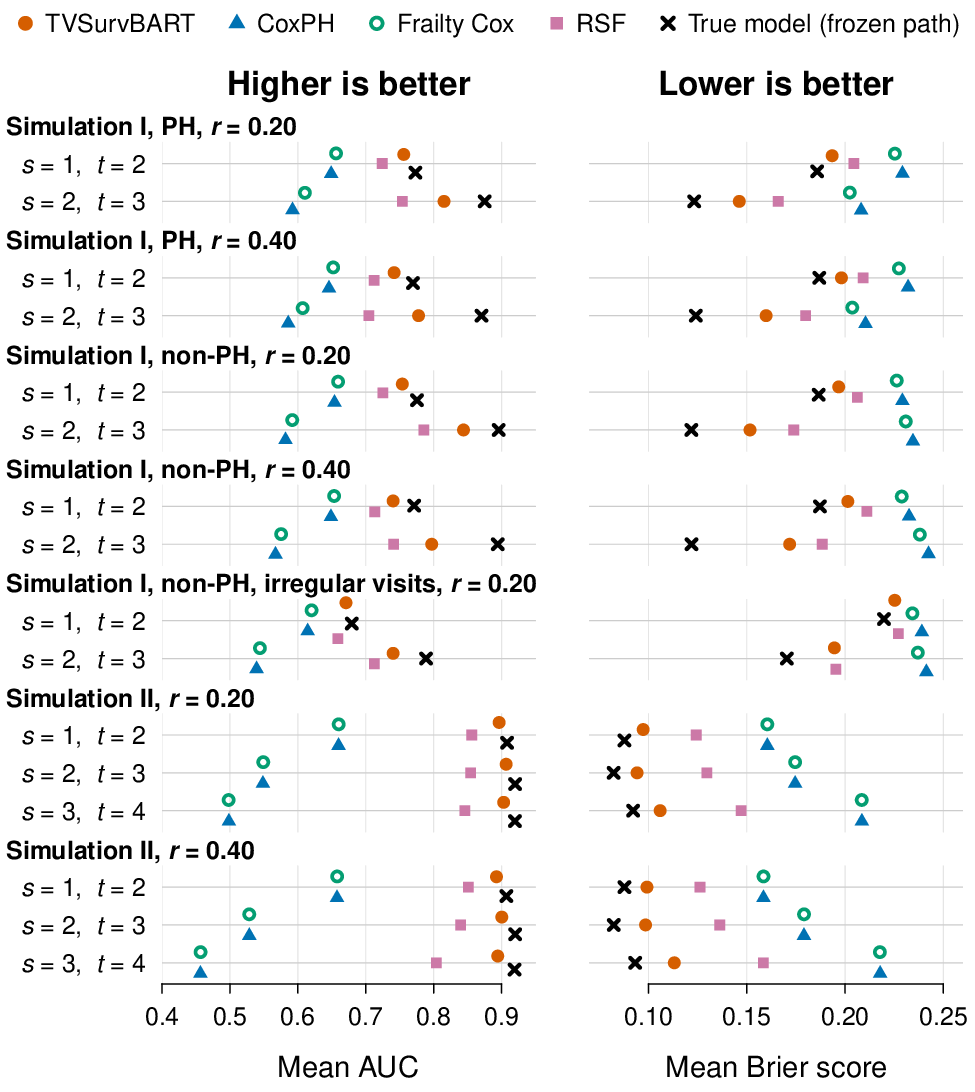}
\caption{Landmark prediction in the simulations, all settings, means over 200 replicates. For test subjects at risk at the landmark time $s$, each method predicts survival to $t=s+1$ given survival to $s$ along the frozen path (Web Appendix~\ref{supp:sim-eval}). True model (frozen path) is the generating model's conditional survival along this path. The cumulative/dynamic AUC and Brier score at $t$ use IPCW estimated within the risk set. Main-text Figure~1 shows the settings with $r=0.40$. Web Tables~\ref{tab:sim1-landmark} and~\ref{tab:sim2-landmark} give the means to three decimals.}
\label{fig:sim-landmark-all}
\end{figure}

Web Appendix~\ref{supp:sim-eval} defines the landmark evaluation summarized in main-text Figure~1, for the settings with $r=0.40$, and in Web Figure~\ref{fig:sim-landmark-all}, for all settings: the landmark times and horizons, the at-risk sets, the frozen paths, the conditional survival predicted by each method, the criteria, and the two true-model references. Web Tables~\ref{tab:sim1-landmark} and~\ref{tab:sim2-landmark} give the means over the 200 replicates for every method.

In the four regular-visit settings of Simulation~I, TVSurvBART along the frozen path has a larger AUC and IPA and a smaller Brier score than CoxPH, frailty Cox, and RSF at both landmarks (Web Table~\ref{tab:sim1-landmark}). Its paired differences with RSF range from $+0.027$ to $+0.073$ in AUC and from $-0.022$ to $-0.010$ in Brier score, each at least $6.9$ Monte Carlo standard errors from zero. The AUC differences against CoxPH and frailty Cox range from $+0.087$ to $+0.262$, each at least $17.9$ standard errors. TVSurvBART's AUC is $0.017$ to $0.031$ below that of the frozen-path truth at $s=1$ and $0.052$ to $0.097$ below it at $s=2$. The complete-path truth also uses the covariates measured after $s$, which no method given the history up to $s$ can use. Its AUC exceeds that of the frozen-path truth by $0.039$ to $0.041$ at $s=1$ and $0.016$ to $0.021$ at $s=2$, the discrimination lost by carrying the last visit forward.

Calibration-in-the-large compares the predictions with outcomes generated along the complete paths. In these four settings, the complete-path truth is calibrated to within $0.004$, whereas the frozen-path truth under-predicts survival by $0.008$ to $0.017$. TVSurvBART's calibration ranges from $-0.012$ to $+0.018$ and is largest in the non-PH settings at $s=1$, $+0.016$ and $+0.018$ at $r=0.20$ and $0.40$. Letting the tree time input follow the scheduled visit times in $(s,t)$, with the covariates still frozen, changes these two values to $-0.003$ and $+0.001$. At both censoring levels and both landmarks, it raises TVSurvBART's AUC by less than $0.001$ and lowers its Brier score by $0.0011$ to $0.0015$.

For irregular visits, the frozen-path truth's AUC is $0.121$ and $0.091$ below that of the complete-path truth at $s=1$ and $s=2$. Along the frozen path, TVSurvBART beats CoxPH and frailty Cox in AUC, Brier score, and IPA at both landmarks, each difference at least $11.8$ standard errors from zero. Its AUC differences with RSF are $+0.012$ ($0.002$) at $s=1$ and $+0.028$ ($0.004$) at $s=2$. Its Brier score difference is $-0.0019$ ($0.0008$) at $s=1$, and at $s=2$, its Brier score and IPA differences, $-0.0007$ ($0.0012$) and $+0.005$ ($0.006$), are smaller than their standard errors. Its calibration at $s=2$ is $+0.076$ ($0.005$), against $+0.032$ for CoxPH, $+0.022$ for frailty Cox, and $+0.035$ for RSF. The two truths differ because the last visit $v$ precedes $s$, by about $0.3$ on average: the frozen-path truth keeps the covariates and the coefficient $\beta_3$ of $x_3$ at their values at $v$, whereas along the complete path, both change at the next visit, after $s$.

Relative to the regular-visit non-PH setting at $r=0.20$, irregular visits lower the AUC of every fitted method along the frozen path by $0.039$ to $0.104$, most for TVSurvBART ($0.083$ at $s=1$ and $0.104$ at $s=2$). TVSurvBART's AUC difference with RSF shrinks from $+0.029$ to $+0.012$ at $s=1$ and from $+0.058$ to $+0.028$ at $s=2$, and its Brier score difference from $-0.0095$ to $-0.0019$ and from $-0.022$ to $-0.0007$.

In Simulation~II, the hazard on $(s,s+1]$ depends only on the covariates measured at $s$, so the frozen-path and complete-path truths coincide for every at-risk subject. TVSurvBART's AUC, $0.893$ to $0.907$, is $0.012$ to $0.025$ below that of the true model (Web Table~\ref{tab:sim2-landmark}). Its paired differences with RSF are $+0.040$ to $+0.090$ in AUC and $-0.045$ to $-0.027$ in Brier score, each at least $19.0$ standard errors from zero. At $s=2$ and $s=3$, CoxPH and frailty Cox have AUCs of $0.457$ to $0.549$ and negative IPA, so they predict worse than the product-limit estimate of survival in the at-risk set. TVSurvBART under-predicts conditional survival at $s=1$ and $s=3$, with calibration from $-0.029$ to $-0.023$ and bias against the true model from $-0.030$ to $-0.024$, and its calibration at $s=2$ is $-0.002$ and $-0.003$ at $r=0.20$ and $0.40$. The true model's calibration lies between $-0.003$ and $+0.004$.

In the four regular-visit settings of Simulation~I, TVSurvBART's $95\%$ intervals cover the frozen-path truth for $0.905$ to $0.941$ of the at-risk subjects at $s=1$ and for only $0.539$ to $0.631$ at $s=2$. In Simulation~II, they cover it for $0.611$ to $0.665$. In the last two cases, most misses come from subjects whose true conditional survival is at least $0.95$. These subjects make up $29\%$ to $30\%$ of the Simulation~I risk sets at $s=2$ and $34\%$ to $44\%$ of the Simulation~II risk sets, against at most $1.6\%$ at $s=1$ in Simulation~I. TVSurvBART's predicted survival for them is too low on average, and their intervals cover the truth for only $0.11$ to $0.34$ of them, against $0.71$ to $0.86$ of the other subjects. Web Appendix~\ref{supp:sim-coverage} reports interval coverage along complete paths (Web Tables~\ref{tab:sim1-coverage} and~\ref{tab:sim-landmark-coverage}).

\begin{table}[H]
\caption{Simulation~I, landmark predictions of the conditional survival probability at $t=s+1$ for the test subjects at risk at $s$, over 200 replicates: mean numbers at risk and of events in $(s,t]$, and means of the IPCW cumulative/dynamic AUC, IPCW Brier score, index of prediction accuracy (IPA), and calibration-in-the-large, which is positive when survival is over-predicted (Web Appendix~\ref{supp:sim-eval}). Irreg.\ is the non-PH model with irregular visits. True model is the true conditional survival along the complete generated path or the frozen path. Monte Carlo standard errors are at most $0.011$ (AUC), $0.004$ (Brier score), $0.016$ (IPA), and $0.008$ (calibration).}
\label{tab:sim1-landmark}
\centering
\footnotesize
\setlength{\tabcolsep}{3pt}
\begin{tabular}{@{}lcccccccccc@{}}
\toprule
 & \multicolumn{5}{c}{$s=1$, $t=2$} & \multicolumn{5}{c}{$s=2$, $t=3$}\\
\cmidrule(lr){2-6}\cmidrule(lr){7-11}
 & \multicolumn{2}{c}{PH} & \multicolumn{2}{c}{Non-PH} & Irreg. & \multicolumn{2}{c}{PH} & \multicolumn{2}{c}{Non-PH} & Irreg.\\
\cmidrule(lr){2-3}\cmidrule(lr){4-5}\cmidrule(lr){6-6}\cmidrule(lr){7-8}\cmidrule(lr){9-10}\cmidrule(lr){11-11}
Method & $r=.20$ & $.40$ & $.20$ & $.40$ & $.20$ & $r=.20$ & $.40$ & $.20$ & $.40$ & $.20$\\
\midrule
At risk & 146 & 108 & 153 & 114 & 154 & 55 & 29 & 59 & 32 & 59\\
Events & 86 & 57 & 88 & 59 & 89 & 14 & 7 & 19 & 9 & 19\\
\midrule
\multicolumn{11}{@{}l}{AUC}\\
True model, complete path & .813 & .809 & .814 & .813 & .800 & .895 & .892 & .912 & .914 & .880\\
True model, frozen path & .773 & .770 & .775 & .771 & .679 & .875 & .870 & .896 & .894 & .789\\
TVSurvBART, frozen path & .756 & .742 & .754 & .740 & .671 & .815 & .778 & .844 & .797 & .740\\
CoxPH & .649 & .646 & .654 & .649 & .615 & .592 & .586 & .582 & .567 & .539\\
Frailty Cox & .656 & .652 & .659 & .653 & .620 & .610 & .607 & .592 & .575 & .544\\
RSF & .724 & .712 & .725 & .713 & .659 & .754 & .705 & .786 & .741 & .713\\
\midrule
\multicolumn{11}{@{}l}{Brier score}\\
True model, complete path & .170 & .171 & .171 & .171 & .177 & .113 & .113 & .112 & .111 & .130\\
True model, frozen path & .186 & .187 & .187 & .187 & .220 & .123 & .124 & .122 & .122 & .170\\
TVSurvBART, frozen path & .193 & .198 & .197 & .201 & .225 & .146 & .160 & .152 & .172 & .195\\
CoxPH & .229 & .232 & .229 & .233 & .239 & .208 & .210 & .235 & .242 & .241\\
Frailty Cox & .225 & .227 & .226 & .229 & .234 & .202 & .204 & .231 & .238 & .237\\
RSF & .204 & .209 & .206 & .211 & .227 & .166 & .180 & .174 & .188 & .195\\
\midrule
\multicolumn{11}{@{}l}{IPA}\\
True model, complete path & .285 & .275 & .289 & .282 & .263 & .406 & .384 & .475 & .467 & .397\\
True model, frozen path & .219 & .208 & .223 & .215 & .084 & .351 & .323 & .427 & .411 & .209\\
TVSurvBART, frozen path & .186 & .160 & .180 & .155 & .061 & .233 & .137 & .289 & .174 & .100\\
CoxPH & .036 & .016 & .046 & .025 & .003 & $-.090$ & $-.135$ & $-.099$ & $-.167$ & $-.118$\\
Frailty Cox & .052 & .036 & .058 & .040 & .023 & $-.061$ & $-.099$ & $-.083$ & $-.148$ & $-.098$\\
RSF & .140 & .113 & .141 & .114 & .053 & .131 & .037 & .185 & .097 & .095\\
\midrule
\multicolumn{11}{@{}l}{Calibration}\\
True model, complete path & $-.001$ & $+.003$ & $-.004$ & $-.001$ & $+.002$ & $-.003$ & $-.003$ & $+.002$ & $+.001$ & $-.003$\\
True model, frozen path & $-.016$ & $-.012$ & $-.017$ & $-.014$ & $-.003$ & $-.013$ & $-.012$ & $-.008$ & $-.010$ & $+.016$\\
TVSurvBART, frozen path & $-.012$ & $-.006$ & $+.016$ & $+.018$ & $+.019$ & $-.001$ & $+.004$ & $+.006$ & $+.013$ & $+.076$\\
CoxPH & $+.017$ & $+.026$ & $+.010$ & $+.018$ & $+.021$ & $+.013$ & $+.014$ & $+.026$ & $+.034$ & $+.032$\\
Frailty Cox & $+.007$ & $+.014$ & .000 & $+.006$ & $+.009$ & $+.007$ & $+.007$ & $+.017$ & $+.024$ & $+.022$\\
RSF & $+.020$ & $+.034$ & $+.016$ & $+.027$ & $+.044$ & $+.022$ & $+.045$ & $+.028$ & $+.055$ & $+.035$\\
\bottomrule
\end{tabular}
\end{table}

\begin{table}[H]
\caption{Simulation~II, landmark predictions over 200 replicates at $s=1,2,3$ with horizon $t=s+1$; entries as in Web Table~\ref{tab:sim1-landmark}. Monte Carlo standard errors are at most $0.009$ (AUC), $0.003$ (Brier score), $0.011$ (IPA), and $0.006$ (calibration).}
\label{tab:sim2-landmark}
\centering
\small
\begin{tabular}{@{}lcccccc@{}}
\toprule
 & \multicolumn{3}{c}{$r=.20$} & \multicolumn{3}{c}{$r=.40$}\\
\cmidrule(lr){2-4}\cmidrule(lr){5-7}
Method & $s=1$ & $s=2$ & $s=3$ & $s=1$ & $s=2$ & $s=3$\\
\midrule
At risk & 171 & 128 & 95 & 137 & 83 & 50\\
Events & 37 & 29 & 26 & 27 & 17 & 13\\
\midrule
\multicolumn{7}{@{}l}{AUC}\\
True model & .908 & .920 & .920 & .907 & .920 & .919\\
TVSurvBART & .896 & .907 & .903 & .893 & .900 & .894\\
CoxPH & .660 & .549 & .499 & .658 & .529 & .457\\
Frailty Cox & .660 & .549 & .498 & .658 & .528 & .457\\
RSF & .856 & .854 & .846 & .851 & .840 & .804\\
\midrule
\multicolumn{7}{@{}l}{Brier score}\\
True model & .088 & .082 & .092 & .088 & .082 & .093\\
TVSurvBART & .097 & .094 & .106 & .099 & .099 & .113\\
CoxPH & .160 & .175 & .209 & .158 & .179 & .218\\
Frailty Cox & .160 & .175 & .208 & .159 & .179 & .218\\
RSF & .124 & .130 & .147 & .126 & .136 & .158\\
\midrule
\multicolumn{7}{@{}l}{IPA}\\
True model & .486 & .525 & .535 & .484 & .523 & .519\\
TVSurvBART & .429 & .456 & .465 & .416 & .430 & .414\\
CoxPH & .060 & $-.005$ & $-.047$ & .068 & $-.033$ & $-.113$\\
Frailty Cox & .060 & $-.005$ & $-.046$ & .068 & $-.033$ & $-.112$\\
RSF & .272 & .253 & .261 & .258 & .215 & .189\\
\midrule
\multicolumn{7}{@{}l}{Calibration}\\
True model & $-.001$ & $+.004$ & $+.003$ & $-.003$ & $+.004$ & $+.001$\\
TVSurvBART & $-.025$ & $-.002$ & $-.023$ & $-.029$ & $-.003$ & $-.029$\\
CoxPH & $-.002$ & $+.008$ & $+.007$ & $-.002$ & $+.012$ & $+.013$\\
Frailty Cox & $-.002$ & $+.007$ & $+.006$ & $-.002$ & $+.012$ & $+.011$\\
RSF & $+.003$ & $+.010$ & $+.011$ & $+.004$ & $+.017$ & $+.019$\\
\bottomrule
\end{tabular}
\end{table}

\subsection{Population summaries along complete paths}\label{supp:sim-complete}
Web Tables~\ref{tab:sim1-complete}--\ref{tab:sim-complete-auc} evaluate every method's predicted survival curves along each test subject's complete generated covariate path: the covariates at every scheduled visit before $\tau$ in Simulation~I, or at every generated visit with irregular visits, and those of all five intervals in Simulation~II. These summaries are well defined because the covariate processes, and with irregular visits the visit times, do not depend on the event or censoring times (Web Appendices~\ref{supp:sim1}--\ref{supp:sim2}), so every test subject has a complete path up to $\tau$ whether or not its event occurs earlier. The reference values are those of Web Appendix~\ref{supp:sim-eval}: RMST at $\tau/2$ and median survival of the average predicted curve are compared with the latent event times of the test sample, and $\mathrm{RMSE}_S$ with each subject's true survival curve along the same complete path. The IBS uses the observed outcomes with IPCW. The True model rows apply the same calculations to the true survival curves along the complete paths, with frailty integrated out in Simulation~I, and so show the errors that remain when the survival curves are known. The complete-path AUC is computed from the latent event times of the test subjects, like the Marginal, Oracle, and Reference rows, which are population values from complete paths. On the test subjects' complete paths, the true model's AUC is within $0.005$ of the Marginal or Reference value at every horizon.

Along complete paths, as along carried-forward paths (Web Appendix~\ref{supp:sim-pred}, Web Tables~\mbox{\ref{tab:sim1-calibration}--\ref{tab:sim2-auc}}), TVSurvBART has the lowest IBS and $\mathrm{RMSE}_S$ in all seven settings and the largest AUC of the fitted methods at every horizon. Here each of its paired differences from the three benchmarks in these criteria exceeds $15$ standard errors, estimated by a paired bootstrap over replicates with $2{,}000$ resamples. Its $\mathrm{RMSE}_S$ is $0.059$ to $0.075$ in Simulation~I and $0.093$ and $0.103$ in Simulation~II, against $0.093$ to $0.102$ and $0.171$ and $0.174$ for RSF.

Differences between methods in the population summaries are small. In Simulation~I, the true model's RMST RMSE is $0.019$, or $0.022$ with irregular visits, and its median RMSE is $0.038$ to $0.048$. The fitted methods' RMST RMSEs range from $0.023$ to $0.029$, and TVSurvBART's RMST biases from $-0.007$ to $-0.004$. TVSurvBART's RMST RMSE differs from RSF's by $-0.0023$ to $+0.0006$ across the seven settings, never by more than $1.7$ bootstrap standard errors, and is smaller than those of CoxPH and frailty Cox in all seven, by $0.0010$ to $0.0135$ ($2.0$ to $4.6$ bootstrap standard errors). Its median RMSE is within $0.8$ bootstrap standard errors of RSF's in Simulation~I and smaller in Simulation~II, by $0.019$ at $r=0.20$ and $0.017$ at $r=0.40$ (bootstrap standard errors $0.007$).

For RMST RMSE, RSF's value is smaller than or equal to TVSurvBART's, at three decimals, in all seven settings along carried-forward paths (Web Appendix~\ref{supp:sim-pred}, Web Tables~\ref{tab:sim1-calibration}, \ref{tab:sim1-irr}, and~\ref{tab:sim2-calibration}), whereas along complete paths, TVSurvBART's is smaller than or equal to RSF's in six of the seven, the exception being the irregular-visit design. In Simulation~II, the true model's RMST bias along complete paths is $-0.002$ ($0.003$) at both censoring levels, against $+0.014$ at $r=0.20$ and $+0.031$ at $r=0.40$ along carried-forward paths, so the latter come from the path extension. Its median bias is $-0.031$ ($0.015$), against $+0.054$ and $+0.155$. Along complete paths, TVSurvBART's RMST biases are $+0.011$ and $+0.006$, against $+0.022$ and $+0.029$ along carried-forward paths, and those of CoxPH are $-0.022$ and $-0.027$, against $-0.032$ and $-0.043$.

\begin{table}[H]
\caption{Simulation~I, out-of-sample results along the complete generated covariate paths over 200 replicates. Entries are bias (Monte Carlo standard error) and RMSE for RMST at half the horizon ($\tau/2=1.5$) and for median survival, against the latent event times of the test subjects, with the mean reference values of Web Tables~\ref{tab:sim1-calibration} and~\ref{tab:sim1-irr}, then IBS, from the observed outcomes, and survival-curve RMSE against the true curve along the same path. ``True model'' is the generating model's survival curve along the same paths, with frailty integrated out.}
\label{tab:sim1-complete}
\small
\centering
\begin{tabular}{@{}lrrrrcc@{}}
\toprule
Method & \multicolumn{2}{c}{RMST($\tau/2$)} & \multicolumn{2}{c}{Median} & IBS & $\mathrm{RMSE}_S$\\
\cmidrule(lr){2-3}\cmidrule(lr){4-5}
& Bias (SE) & RMSE & Bias (SE) & RMSE & &\\
\midrule
\multicolumn{7}{@{}l}{Setting: PH, $r=.20$}\\
TVSurvBART & $-.005$ (.002) & .023 & $+.011$ (.003) & .043 & .127 & .059\\
CoxPH & $-.009$ (.002) & .026 & $-.012$ (.003) & .046 & .140 & .130\\
Frailty Cox & $-.006$ (.002) & .025 & $-.007$ (.003) & .045 & .139 & .126\\
RSF & $-.002$ (.002) & .024 & $+.001$ (.003) & .043 & .132 & .093\\
True model & $+.001$ (.001) & .019 & $+.006$ (.003) & .038 & .123 & .000\\
\midrule
\multicolumn{7}{@{}l}{Setting: PH, $r=.40$}\\
TVSurvBART & $-.004$ (.002) & .023 & $+.010$ (.003) & .043 & .128 & .071\\
CoxPH & $-.007$ (.002) & .025 & $-.010$ (.003) & .045 & .140 & .131\\
Frailty Cox & $-.005$ (.002) & .024 & $-.006$ (.003) & .044 & .139 & .127\\
RSF & $-.001$ (.002) & .023 & $+.001$ (.003) & .042 & .133 & .099\\
True model & $+.001$ (.001) & .019 & $+.006$ (.003) & .038 & .123 & .000\\
\midrule
\multicolumn{7}{@{}l}{Setting: Non-PH, $r=.20$}\\
TVSurvBART & $-.005$ (.002) & .023 & $.000$ (.003) & .045 & .127 & .062\\
CoxPH & $-.008$ (.002) & .025 & $-.010$ (.003) & .049 & .141 & .136\\
Frailty Cox & $-.005$ (.002) & .024 & $-.005$ (.003) & .047 & .140 & .132\\
RSF & $-.002$ (.002) & .023 & $+.001$ (.003) & .046 & .132 & .097\\
True model & $+.002$ (.001) & .019 & $+.007$ (.003) & .040 & .123 & .000\\
\midrule
\multicolumn{7}{@{}l}{Setting: Non-PH, $r=.40$}\\
TVSurvBART & $-.004$ (.002) & .023 & $.000$ (.003) & .046 & .129 & .075\\
CoxPH & $-.007$ (.002) & .025 & $-.008$ (.003) & .049 & .141 & .136\\
Frailty Cox & $-.005$ (.002) & .024 & $-.005$ (.003) & .048 & .140 & .133\\
RSF & $-.001$ (.002) & .023 & $+.003$ (.003) & .046 & .133 & .102\\
True model & $+.002$ (.001) & .019 & $+.007$ (.003) & .040 & .123 & .000\\
\midrule
\multicolumn{7}{@{}l}{Setting: Non-PH, irregular visits, $r=.20$}\\
TVSurvBART & $-.007$ (.002) & .026 & $-.005$ (.004) & .053 & .127 & .065\\
CoxPH & $-.012$ (.002) & .029 & $-.025$ (.004) & .061 & .141 & .136\\
Frailty Cox & $-.008$ (.002) & .027 & $-.017$ (.004) & .058 & .139 & .131\\
RSF & $-.005$ (.002) & .025 & $-.011$ (.004) & .054 & .132 & .099\\
True model & $-.001$ (.002) & .022 & $-.002$ (.003) & .048 & .123 & .000\\
\bottomrule
\end{tabular}
\end{table}

\begin{table}[H]
\caption{Simulation~II, out-of-sample results along the complete generated covariate paths over 200 replicates, as in Web Table~\ref{tab:sim1-complete}. RMST is evaluated at half the horizon ($\tau/2=2.5$); the mean reference values are those of Web Table~\ref{tab:sim2-calibration}.}
\label{tab:sim2-complete}
\small
\centering
\begin{tabular}{@{}lrrrrcc@{}}
\toprule
Method & \multicolumn{2}{c}{RMST($\tau/2$)} & \multicolumn{2}{c}{Median} & IBS & $\mathrm{RMSE}_S$\\
\cmidrule(lr){2-3}\cmidrule(lr){4-5}
& Bias (SE) & RMSE & Bias (SE) & RMSE & &\\
\midrule
\multicolumn{7}{@{}l}{Censoring: $20\%$}\\
TVSurvBART & $+.011$ (.004) & .051 & $-.023$ (.016) & .231 & .132 & .093\\
CoxPH & $-.022$ (.004) & .064 & $-.125$ (.019) & .301 & .185 & .248\\
Frailty Cox & $-.020$ (.004) & .064 & $-.118$ (.019) & .299 & .185 & .248\\
RSF & $+.002$ (.004) & .053 & $-.017$ (.018) & .251 & .152 & .171\\
True model & $-.002$ (.003) & .043 & $-.031$ (.015) & .207 & .123 & .000\\
\midrule
\multicolumn{7}{@{}l}{Censoring: $40\%$}\\
TVSurvBART & $+.006$ (.004) & .050 & $-.049$ (.016) & .235 & .134 & .103\\
CoxPH & $-.027$ (.004) & .064 & $-.146$ (.018) & .299 & .183 & .243\\
Frailty Cox & $-.025$ (.004) & .063 & $-.141$ (.019) & .297 & .183 & .244\\
RSF & $+.004$ (.004) & .053 & $-.009$ (.018) & .252 & .154 & .174\\
True model & $-.002$ (.003) & .043 & $-.031$ (.015) & .207 & .123 & .000\\
\bottomrule
\end{tabular}
\end{table}

\begin{table}[H]
\caption{Mean test-sample cumulative/dynamic AUC over 200 replicates, every method evaluated along the complete generated covariate paths with the latent event times of the test subjects. The population references are those of Web Tables~\ref{tab:sim1-auc}, \ref{tab:sim1-irr}, and~\ref{tab:sim2-auc}, also computed from complete paths and latent event times: ``Marginal'' integrates out frailty, ``Oracle'' uses realized frailty, and Simulation~II, which has no frailty, has one ``Reference''. (a) Simulation~I; the irregular-visit design was run at $r=0.20$ only. (b) Simulation~II. Monte Carlo standard errors are at most $0.0033$.}
\label{tab:sim-complete-auc}
\centering
\small
(a) Simulation~I\\[2pt]
\begin{tabular}{@{}llcccccc@{}}
\toprule
Setting & Method & \multicolumn{3}{c}{$r=.20$} & \multicolumn{3}{c}{$r=.40$}\\
\cmidrule(lr){3-5}\cmidrule(lr){6-8}
& & $t=1$ & $t=2$ & $t=3$ & $t=1$ & $t=2$ & $t=3$\\
\midrule
PH & Marginal & .758 & .827 & .870 & .758 & .827 & .870\\
 & Oracle & .880 & .946 & .964 & .880 & .946 & .964\\
 & TVSurvBART & .755 & .812 & .848 & .753 & .804 & .832\\
 & CoxPH & .731 & .730 & .722 & .731 & .729 & .721\\
 & Frailty Cox & .732 & .733 & .725 & .732 & .733 & .724\\
 & RSF & .743 & .783 & .811 & .740 & .776 & .800\\
\midrule
Non-PH & Marginal & .777 & .823 & .869 & .777 & .823 & .869\\
 & Oracle & .887 & .944 & .965 & .887 & .944 & .965\\
 & TVSurvBART & .770 & .808 & .846 & .768 & .799 & .827\\
 & CoxPH & .727 & .730 & .720 & .732 & .730 & .714\\
 & Frailty Cox & .731 & .734 & .719 & .736 & .733 & .712\\
 & RSF & .751 & .781 & .813 & .750 & .774 & .801\\
\midrule
Non-PH, irregular & Marginal & .778 & .828 & .872 & -- & -- & --\\
 & Oracle & .886 & .944 & .965 & -- & -- & --\\
 & TVSurvBART & .767 & .808 & .841 & -- & -- & --\\
 & CoxPH & .730 & .731 & .717 & -- & -- & --\\
 & Frailty Cox & .734 & .735 & .716 & -- & -- & --\\
 & RSF & .750 & .778 & .806 & -- & -- & --\\
\bottomrule
\end{tabular}

\vspace{8pt}
(b) Simulation~II\\[2pt]
\begin{tabular}{@{}llccccc@{}}
\toprule
Censoring & Method & $t=1$ & $t=2$ & $t=3$ & $t=4$ & $t=5$\\
\midrule
$20\%$ & Reference & .871 & .878 & .888 & .899 & .914\\
 & TVSurvBART & .858 & .865 & .871 & .883 & .897\\
 & CoxPH & .734 & .718 & .695 & .665 & .640\\
 & Frailty Cox & .734 & .718 & .694 & .664 & .637\\
 & RSF & .829 & .824 & .820 & .825 & .847\\
\midrule
$40\%$ & Reference & .871 & .878 & .888 & .899 & .914\\
 & TVSurvBART & .857 & .863 & .867 & .877 & .891\\
 & CoxPH & .760 & .740 & .708 & .668 & .626\\
 & Frailty Cox & .760 & .740 & .707 & .667 & .624\\
 & RSF & .831 & .823 & .816 & .816 & .835\\
\bottomrule
\end{tabular}
\end{table}

\subsection{Predictions along carried-forward paths}\label{supp:sim-pred}
Web Tables~\ref{tab:sim1-calibration}--\ref{tab:sim2-auc} report the retrospective comparison along carried-forward paths (Web Appendix~\ref{supp:sim-eval}): IBS, $\mathrm{RMSE}_S$, AUC at $t=1$ and $t=\tau$, and the RMSEs of RMST and median survival, together with the biases and their Monte Carlo standard errors, the mean reference values of RMST and median survival, the population AUC references, and the AUC at the other horizons; main-text Figures~2 and~3 show the mean AUC$(t)$ curves. In every setting, TVSurvBART has the lowest IBS and $\mathrm{RMSE}_S$, the largest AUC of the fitted methods at every tabulated horizon, and the highest test log score (Web Appendix~\ref{supp:sim-logscore}, Web Tables~\ref{tab:sim1-elpd} and~\ref{tab:sim2-elpd}). No fitted method's mean AUC exceeds the True model (carried forward) rows of Web Tables~\ref{tab:sim1-auc}, \ref{tab:sim1-irr}, and~\ref{tab:sim2-auc} at any tabulated horizon. These rows differ from the frozen-path True model rows of the landmark comparison (Web Appendix~\ref{supp:sim-landmark}, Web Tables~\ref{tab:sim1-landmark} and~\ref{tab:sim2-landmark}).

In the four Simulation~I settings (PH and non-PH, each at $r=0.20$ and $r=0.40$), TVSurvBART's median bias ranges from $0.000$ to $0.010$, while its RMST bias ranges from $-0.008$ to $-0.001$ (Web Table~\ref{tab:sim1-calibration}). Its RMST RMSE lies between those of RSF and the Cox models, and its median RMSE is no larger than any benchmark's except under non-PH at $r=0.40$, where that of RSF is $0.001$ smaller. The marginal reference integrates out frailty and the oracle uses its realized value (Web Table~\ref{tab:sim1-auc}). Their difference measures the information that the latent effects supply when the complete paths are known.

With irregular visits (Web Table~\ref{tab:sim1-irr}), TVSurvBART again has the lowest IBS ($0.126$) and $\mathrm{RMSE}_S$ ($0.105$) and, among the fitted methods, the largest AUC at $t=1,2,3$ ($0.769$, $0.810$ and $0.819$). On the regular grid in the same setting, its IBS and $\mathrm{RMSE}_S$ are the same to three decimals and its AUC is $0.769$, $0.816$ and $0.826$ (Web Tables~\ref{tab:sim1-calibration}--\ref{tab:sim1-auc}). On each of these criteria, the four methods rank in the same order under both visit processes. TVSurvBART has the smallest absolute median bias ($-0.010$) and, with RSF, the smallest median RMSE ($0.054$), and RSF has the smallest absolute RMST bias ($-0.006$) and RMST RMSE ($0.025$).

In Simulation~II, TVSurvBART's rows in Web Tables~\ref{tab:sim2-calibration}--\ref{tab:sim2-auc} and~\ref{tab:sim2-elpd} use the random-intercept-and-slope fit. RSF has the smallest absolute RMST biases and RMST RMSE, and TVSurvBART the smallest median RMSE, at both censoring levels (Web Table~\ref{tab:sim2-calibration}). The reported true-model calculations on the same carried-forward paths give RMST biases of $+0.014$ and $+0.031$ and median biases of $+0.054$ and $+0.155$ at $20\%$ and $40\%$ censoring. These show that path extension itself can distort population summaries. TVSurvBART's RMST biases, $+0.022$ and $+0.029$ (Web Table~\ref{tab:sim2-calibration}), are $0.008$ above and $0.002$ below those of the true model. TVSurvBART's AUC ranges from $0.848$ to $0.858$ under $20\%$ censoring and declines from $0.857$ to $0.787$ under $40\%$ censoring (Web Table~\ref{tab:sim2-auc}). Fitting at $r=0.20$ with a random intercept only or without random effects (Web Appendix~\ref{supp:sim-ablation}) change TVSurvBART's IBS, $\mathrm{RMSE}_S$, and AUC by at most $0.0014$ (Web Table~\ref{tab:sim-ablation}) and its RMST and median RMSEs by at most $0.002$.%

\begin{table}[H]
\caption{Simulation I, out-of-sample results along carried-forward paths over 200 replicates. Entries are bias (Monte Carlo standard error) and RMSE for RMST at half the horizon ($\tau/2=1.5$; mean test-sample reference values $1.015$ under PH, $1.034$ under non-PH), then bias and RMSE for median survival (mean test-sample reference values $1.022$ and $1.054$), IBS, and survival-curve RMSE. The RMST and median reference values of a replicate are computed from the latent event times of its test subjects (Web Appendix~\ref{supp:sim-eval}) and are the same at both censoring fractions.}
\label{tab:sim1-calibration}
\small
\centering
\begin{tabular}{@{}lrrrrcc@{}}
\toprule
Method & \multicolumn{2}{c}{RMST($\tau/2$)}
& \multicolumn{2}{c}{Median}
& IBS & $\mathrm{RMSE}_S$\\
\cmidrule(lr){2-3}\cmidrule(lr){4-5}
& Bias (SE) & RMSE & Bias (SE) & RMSE & &\\
\midrule
\multicolumn{7}{@{}l}{Setting: PH, $r=.20$}\\
TVSurvBART & $-.008$ (.002) & .024 & $.004$ (.003) & .042 & .123 & .098\\%
CoxPH & $-.011$ (.002) & .027 & $-.018$ (.003) & .048 & .139 & .140\\%
Frailty Cox & $-.008$ (.002) & .025 & $-.012$ (.003) & .046 & .138 & .136\\%
RSF & $-.003$ (.002) & .024 & $-.004$ (.003) & .044 & .133 & .118\\%
\midrule
\multicolumn{7}{@{}l}{Setting: PH, $r=.40$}\\
TVSurvBART & $-.005$ (.002) & .023 & $.007$ (.003) & .043 & .125 & .117\\%
CoxPH & $-.008$ (.002) & .025 & $-.014$ (.003) & .046 & .139 & .148\\%
Frailty Cox & $-.005$ (.002) & .024 & $-.009$ (.003) & .045 & .138 & .143\\%
RSF & $.002$ (.002) & .023 & $.002$ (.003) & .043 & .136 & .137\\%
\midrule
\multicolumn{7}{@{}l}{Setting: Non-PH, $r=.20$}\\
TVSurvBART & $-.007$ (.002) & .024 & $.000$ (.003) & .046 & .126 & .105\\%
CoxPH & $-.010$ (.002) & .026 & $-.014$ (.003) & .050 & .140 & .145\\%
Frailty Cox & $-.007$ (.002) & .025 & $-.008$ (.003) & .048 & .139 & .140\\%
RSF & $-.003$ (.002) & .023 & $-.002$ (.003) & .047 & .133 & .121\\%
\midrule
\multicolumn{7}{@{}l}{Setting: Non-PH, $r=.40$}\\
TVSurvBART & $-.001$ (.002) & .024 & $.010$ (.003) & .049 & .128 & .127\\%
CoxPH & $-.008$ (.002) & .025 & $-.011$ (.003) & .050 & .140 & .151\\%
Frailty Cox & $-.005$ (.002) & .024 & $-.006$ (.003) & .049 & .139 & .146\\%
RSF & $.001$ (.002) & .023 & $.006$ (.003) & .048 & .136 & .138\\%
\bottomrule
\end{tabular}
\end{table}

\begin{table}[H]
\caption{Simulation I, mean test-sample cumulative/dynamic AUC along carried-forward paths over 200 replicates. ``True model (carried forward)'' is the AUC of the true survival probabilities, with frailty integrated out, along the same carried-forward paths of the same test subjects; like the fitted methods, it depends on the censoring fraction. ``Marginal'' integrates out frailty and ``Oracle'' uses realized frailty. Both are population values computed from complete paths (Web Appendix~\ref{supp:sim-eval}).}
\label{tab:sim1-auc}
\centering
\small
\begin{tabular}{@{}llcccccc@{}}
\toprule
Setting & Method & \multicolumn{3}{c}{$r=.20$} & \multicolumn{3}{c}{$r=.40$}\\
\cmidrule(lr){3-5}\cmidrule(lr){6-8}
& & $t=1$ & $t=2$ & $t=3$ & $t=1$ & $t=2$ & $t=3$\\
\midrule
PH & Marginal & .758 & .827 & .870 & .758 & .827 & .870\\
 & Oracle & .880 & .946 & .964 & .880 & .946 & .964\\
 & True model (carried forward) & .776 & .854 & .913 & .763 & .815 & .853\\%
 & TVSurvBART & .765 & .826 & .866 & .750 & .783 & .803\\
 & CoxPH & .738 & .730 & .721 & .728 & .714 & .704\\
 & Frailty Cox & .740 & .736 & .725 & .731 & .719 & .709\\
 & RSF & .746 & .765 & .775 & .731 & .730 & .729\\
\midrule
Non-PH & Marginal & .777 & .823 & .869 & .777 & .823 & .869\\
 & Oracle & .887 & .944 & .965 & .887 & .944 & .965\\
 & True model (carried forward) & .791 & .849 & .906 & .779 & .811 & .846\\%
 & TVSurvBART & .769 & .816 & .826 & .756 & .775 & .770\\
 & CoxPH & .733 & .731 & .720 & .729 & .715 & .700\\
 & Frailty Cox & .738 & .736 & .721 & .734 & .720 & .703\\
 & RSF & .754 & .763 & .776 & .742 & .729 & .728\\
\bottomrule
\end{tabular}
\end{table}

\begin{table}[H]
\caption{Simulation I with irregular visit times (non-PH, $20\%$ censoring), out-of-sample results along carried-forward paths over 200 replicates. Entries are bias (Monte Carlo standard error) and RMSE for RMST at half the horizon ($\tau/2=1.5$; mean test-sample reference value $1.045$) and for median survival (mean test-sample reference value $1.078$), IBS, survival-curve RMSE, and mean cumulative/dynamic AUC at $t=1,2,3$ (Monte Carlo standard errors at most $0.0033$). The ``True model (carried forward)'' row is the AUC of the true survival probabilities along the same carried-forward paths of the same test subjects (Web Appendix~\ref{supp:sim-eval}). The population marginal and oracle AUC references, computed from complete paths, are $0.778$, $0.828$, $0.872$ and $0.886$, $0.944$, $0.965$.}
\label{tab:sim1-irr}
\centering
\small
\setlength{\tabcolsep}{3pt}
\begin{tabular}{@{}lrrrrccccc@{}}
\toprule
Method & \multicolumn{2}{c}{RMST($\tau/2$)} & \multicolumn{2}{c}{Median} & IBS & $\mathrm{RMSE}_S$ & \multicolumn{3}{c}{AUC}\\
\cmidrule(lr){2-3}\cmidrule(lr){4-5}\cmidrule(lr){8-10}
& Bias (SE) & RMSE & Bias (SE) & RMSE & & & $t=1$ & $t=2$ & $t=3$\\
\midrule
\multicolumn{7}{@{}l}{True model (carried forward)} & .784 & .847 & .902\\%
TVSurvBART & $-.009$ (.002) & .027 & $-.010$ (.004) & .054 & .126 & .105 & .769 & .810 & .819\\%
CoxPH & $-.014$ (.002) & .029 & $-.030$ (.004) & .062 & .140 & .145 & .735 & .733 & .717\\%
Frailty Cox & $-.010$ (.002) & .027 & $-.022$ (.004) & .059 & .138 & .139 & .739 & .738 & .719\\%
RSF & $-.006$ (.002) & .025 & $-.014$ (.004) & .054 & .133 & .122 & .750 & .764 & .761\\%
\bottomrule
\end{tabular}
\end{table}

\begin{table}[H]
\caption{Simulation II, out-of-sample results along carried-forward paths over 200 replicates. RMST is evaluated at half the horizon ($\tau/2=2.5$; mean test-sample reference value $1.461$); the mean test-sample median is $1.564$. The reference values of a replicate are computed from the latent event times of its test subjects (Web Appendix~\ref{supp:sim-eval}).}
\label{tab:sim2-calibration}
\centering
\small
\setlength{\tabcolsep}{2.5pt}
\begin{tabular}{@{}lrrrrcc@{}}
\toprule
Method & \multicolumn{2}{c}{RMST($\tau/2$)}
& \multicolumn{2}{c}{Median}
& IBS & $\mathrm{RMSE}_S$\\
\cmidrule(lr){2-3}\cmidrule(lr){4-5}
& Bias (SE) & RMSE & Bias (SE) & RMSE & &\\
\midrule
\multicolumn{7}{@{}l}{Censoring: $20\%$}\\
TVSurvBART & $+.022$ (.004) & .055 & $+.028$ (.017) & .236
& .139 & .143\\%
CoxPH & $-.032$ (.004) & .068 & $-.162$ (.019) & .314
& .182 & .248\\%
Frailty Cox & $-.031$ (.004) & .068 & $-.156$ (.019) & .312
& .182 & .248\\%
RSF & $-.002$ (.004) & .053 & $-.046$ (.018) & .253
& .159 & .190\\%
\midrule
\multicolumn{7}{@{}l}{Censoring: $40\%$}\\
TVSurvBART & $+.029$ (.004) & .058 & $+.062$ (.017) & .245
& .141 & .176\\%
CoxPH & $-.043$ (.004) & .072 & $-.199$ (.018) & .324
& .177 & .244\\%
Frailty Cox & $-.042$ (.004) & .072 & $-.195$ (.018) & .322
& .177 & .244\\%
RSF & $+.009$ (.004) & .054 & $+.007$ (.018) & .253
& .163 & .208\\%
\bottomrule
\end{tabular}
\end{table}

\begin{table}[H]
\caption{Simulation II, mean test-sample cumulative/dynamic AUC along carried-forward paths over 200 replicates (Monte Carlo standard errors at most 0.003). ``True model (carried forward)'' is the AUC of the true survival probabilities along the same carried-forward paths of the same test subjects and, like the fitted methods, depends on the censoring fraction. ``Reference'' is a population value computed from complete generated paths (Web Appendix~\ref{supp:sim-eval}).}
\label{tab:sim2-auc}
\centering
\small
\setlength{\tabcolsep}{4pt}
\begin{tabular}{llccccc}
\toprule
Censoring & Method & $t=1$ & $t=2$ & $t=3$ & $t=4$ & $t=5$\\
\midrule
$20\%$ & Reference & .871 & .878 & .888 & .899 & .914\\
& True model (carried forward) & .872 & .865 & .856 & .854 & .867\\%
& TVSurvBART & .858 & .855 & .848 & .848 & .854\\
& CoxPH & .734 & .731 & .721 & .709 & .700\\
& Frailty Cox & .734 & .732 & .723 & .711 & .703\\
& RSF & .826 & .814 & .793 & .771 & .766\\
\midrule
$40\%$ & Reference & .871 & .878 & .888 & .899 & .914\\
& True model (carried forward) & .872 & .853 & .829 & .809 & .800\\%
& TVSurvBART & .857 & .841 & .819 & .801 & .787\\
& CoxPH & .760 & .751 & .734 & .714 & .696\\
& Frailty Cox & .760 & .752 & .735 & .715 & .697\\
& RSF & .827 & .803 & .768 & .732 & .701\\
\bottomrule
\end{tabular}
\end{table}

\subsection{Coverage of posterior intervals}\label{supp:sim-coverage}
This appendix assesses the frequentist coverage of TVSurvBART's $95\%$ posterior intervals for subject-level survival probabilities along each test subject's complete generated covariate path, which separates posterior uncertainty from the carried-forward approximation. Each retained draw $d$ gives the survival probability $\bar S_i^{(d)}(t)$ with the random effect integrated out against $\mathcal N(\bm 0,\Sigma_b^{(d)})$, a single random effect for the whole path, as in Web Appendix~\ref{supp:sim-eval}, and the interval is formed by the $2.5\%$ and $97.5\%$ quantiles of the $2{,}000$ draws. The target is the true survival probability along the same path, with the random effects integrated out over their generating distribution in Simulation~I. The subjects' realized random effects are not used. TVSurvBART only approximates the generating baseline and covariate effects, so exact nominal coverage is not expected.

Web Table~\ref{tab:sim1-coverage} gives the proportion of intervals that contain the target at $t=1,2,3$ in Simulation~I, over the $300$ test subjects of each of the $200$ replicates. Coverage is close to nominal at $t=1$, from $0.940$ to $0.971$, and falls with the horizon, to $0.856$--$0.892$ at $t=2$ and $0.676$--$0.763$ at $t=3$, while the mean widths decrease from $0.240$--$0.269$ to $0.136$--$0.166$. The irregular-visit design gives values close to those of the non-PH model at $r=0.20$. In Simulation~II, where the intervals were stored at $t=1,3,5$ only, coverage is $0.893$, $0.817$, and $0.854$ at $r=0.20$ and $0.896$, $0.805$, and $0.827$ at $r=0.40$.

For the landmarks of Web Appendix~\ref{supp:sim-landmark}, the target is the conditional survival probability $\Pr(T_i^*>t\mid T_i^*>s)$, $t=s+1$, along the complete generated path, which depends only on the path and the parameters and is therefore evaluated for every test subject. In draw $d$, the prediction is the ratio $\bar S_i^{(d)}(t)/\bar S_i^{(d)}(s)$, and the interval is formed by the $2.5\%$ and $97.5\%$ quantiles of these ratios with draw $d$ weighted by $\bar S_i^{(d)}(s)$, the survival weighting of Web Appendix~\ref{supp:prediction}. The weighted mean of the ratios is then the point prediction, the ratio of the posterior means of $\bar S_i(t)$ and $\bar S_i(s)$. The target is the corresponding ratio of true survival probabilities along the same path.

Along complete paths (Web Table~\ref{tab:sim-landmark-coverage}), coverage is $0.835$ to $0.875$ at $s=1$ and $0.491$ to $0.608$ at $s=2$ in the four regular Simulation~I settings, $0.850$ and $0.605$ with irregular visits, and $0.642$ to $0.707$ in Simulation~II. Among the subjects at risk at $s$, it is similar at $s=1$ and lower at $s=2$, $0.447$ to $0.575$, and in Simulation~II, $0.612$ to $0.666$.

The undercoverage is concentrated where the target is extreme. Within quartiles of the target in each replicate, coverage at $s=1$ in the regular Simulation~I settings is $0.941$ to $0.957$ in the two middle quartiles, against $0.743$ to $0.820$ in the lowest and $0.698$ to $0.774$ in the highest, where the point predictions lie on average $0.027$ to $0.067$ above and $0.079$ to $0.117$ below the target. At $s=2$, these mean errors grow to $0.230$ to $0.400$ above and $0.119$ to $0.183$ below the target, and coverage in the two outer quartiles falls to $0.137$ to $0.445$. In Simulation~II, the highest quartile, with a mean target of $0.978$ to $0.987$, is predicted on average $0.031$ to $0.064$ too low and covered in $0.189$ to $0.292$ of cases, and the second highest in $0.636$ to $0.760$. The predicted risks are thus attenuated toward the average risk. The fixed-horizon predictions show the same pattern, growing with the horizon: in the quartile with the largest true $S_i(t)$, the posterior mean of $\bar S_i(t)$ in the regular Simulation~I settings is on average $0.014$ to $0.036$ below the truth at $t=1$, $0.046$ to $0.071$ at $t=2$, and $0.073$ to $0.106$ at $t=3$. The posterior intervals describe uncertainty under the fitted model and do not account for this systematic error. Correspondingly, along the frozen paths of Web Appendix~\ref{supp:sim-landmark} the RMSE of the landmark predictions is $1.1$ to $1.2$ times the standard deviation implied by the interval width, width$/3.92$, at $s=1$ in the regular Simulation~I settings, where the intervals cover the frozen-path truth for $0.905$ to $0.941$ of the subjects at risk, and $1.5$ to $2.0$ times at $s=2$ and in Simulation~II, where they cover it for $0.539$ to $0.665$.

\begin{table}[H]
\caption{Simulation~I, coverage of TVSurvBART's $95\%$ posterior intervals for the survival probability $S_i(t)$ along the complete generated covariate paths, over 200 replicates. Monte Carlo standard errors are at most $0.007$ for coverage and $0.001$ for the mean width.}
\label{tab:sim1-coverage}
\centering
\small
\begin{tabular}{@{}lcccccc@{}}
\toprule
Setting & \multicolumn{3}{c}{Coverage} & \multicolumn{3}{c}{Mean width}\\
\cmidrule(lr){2-4}\cmidrule(lr){5-7}
 & $t=1$ & $t=2$ & $t=3$ & $t=1$ & $t=2$ & $t=3$\\
\midrule
PH, $r=.20$ & .971 & .891 & .745 & .241 & .168 & .136\\
PH, $r=.40$ & .970 & .856 & .677 & .262 & .185 & .155\\
Non-PH, $r=.20$ & .944 & .892 & .763 & .250 & .185 & .147\\
Non-PH, $r=.40$ & .946 & .859 & .676 & .269 & .203 & .166\\
Non-PH, irregular visits, $r=.20$ & .940 & .876 & .747 & .240 & .190 & .151\\
\bottomrule
\end{tabular}
\end{table}

\begin{table}[H]
\caption{Coverage of TVSurvBART's $95\%$ posterior intervals for the landmark conditional survival probability $\Pr(T_i^*>t\mid T_i^*>s)$, $t=s+1$, along the complete generated covariate paths, over 200 replicates. Monte Carlo standard errors are at most $0.007$ for All, $0.009$ for At risk, and $0.016$ within quartiles.}
\label{tab:sim-landmark-coverage}
\centering
\small
\begin{tabular}{@{}lccccccc@{}}
\toprule
 & & & & \multicolumn{4}{c}{Quartile of the target}\\
\cmidrule(lr){5-8}
Landmark & All & Width & At risk & Q1 & Q2 & Q3 & Q4\\
\midrule
\multicolumn{8}{@{}l}{Simulation~I, PH, $r=.20$}\\
$s=1$, $t=2$ & .869 & .263 & .871 & .805 & .953 & .945 & .771\\
$s=2$, $t=3$ & .608 & .375 & .543 & .370 & .835 & .883 & .346\\
\midrule
\multicolumn{8}{@{}l}{Simulation~I, PH, $r=.40$}\\
$s=1$, $t=2$ & .836 & .293 & .833 & .743 & .946 & .942 & .712\\
$s=2$, $t=3$ & .533 & .410 & .486 & .227 & .704 & .877 & .324\\
\midrule
\multicolumn{8}{@{}l}{Simulation~I, non-PH, $r=.20$}\\
$s=1$, $t=2$ & .875 & .291 & .875 & .820 & .957 & .947 & .774\\
$s=2$, $t=3$ & .583 & .416 & .531 & .445 & .875 & .842 & .169\\
\midrule
\multicolumn{8}{@{}l}{Simulation~I, non-PH, $r=.40$}\\
$s=1$, $t=2$ & .835 & .322 & .833 & .753 & .949 & .941 & .698\\
$s=2$, $t=3$ & .491 & .463 & .447 & .261 & .736 & .830 & .137\\
\midrule
\multicolumn{8}{@{}l}{Simulation~I, non-PH, irregular visits, $r=.20$}\\
$s=1$, $t=2$ & .850 & .275 & .855 & -- & -- & -- & --\\
$s=2$, $t=3$ & .605 & .408 & .575 & -- & -- & -- & --\\
\midrule
\multicolumn{8}{@{}l}{Simulation~II, $r=.20$}\\
$s=1$, $t=2$ & .660 & .255 & .624 & .879 & .922 & .636 & .203\\
$s=2$, $t=3$ & .657 & .256 & .628 & .783 & .940 & .673 & .230\\
$s=3$, $t=4$ & .707 & .307 & .666 & .826 & .954 & .756 & .292\\
\midrule
\multicolumn{8}{@{}l}{Simulation~II, $r=.40$}\\
$s=1$, $t=2$ & .659 & .282 & .622 & .869 & .929 & .648 & .189\\
$s=2$, $t=3$ & .642 & .290 & .612 & .750 & .945 & .670 & .205\\
$s=3$, $t=4$ & .690 & .358 & .653 & .773 & .954 & .760 & .271\\
\bottomrule
\end{tabular}
\end{table}

\subsection{Test log scores}\label{supp:sim-logscore}
Web Tables~\ref{tab:sim1-elpd}--\ref{tab:sim2-elpd} report pairwise differences in the subject-level test log score of Web Appendix~\ref{supp:scores}. TVSurvBART's score exceeds that of CoxPH by $23.8$--$51.8$, that of frailty Cox by $20.0$--$46.6$, and that of RSF by $34.6$--$48.0$ across the four Simulation~I settings. With irregular visits, TVSurvBART's test log score exceeds those of CoxPH, frailty Cox and RSF by $47.9$, $42.3$ and $43.0$ (Monte Carlo standard errors $0.8$, $0.7$ and $0.8$). In Simulation~II, the log-score advantages for TVSurvBART are $162.2$ and $106.0$ over CoxPH ($162.1$ and $105.9$ over frailty Cox) and $111.6$ and $78.0$ over RSF at $20\%$ and $40\%$ censoring, respectively. RSF in turn exceeds CoxPH and frailty Cox by $50.6$ and $50.5$ at $20\%$ censoring and both by $27.9$ at $40\%$.

\begin{table}[H]
\caption{Simulation I, pairwise differences in the test log score over 200 replicates, summed over all 300 test subjects: the log predicted density of the observed event time for subjects with an event and the log predicted survival probability at the censoring time for censored subjects (method in the left column minus method in the top row; Monte Carlo standard error of the mean paired difference in parentheses; positive values favor the row method).}
\label{tab:sim1-elpd}
\centering
\small
\begin{tabular}{@{}lrrr@{}}
\toprule
 & TVSurvBART & RSF & CoxPH\\
\midrule
\multicolumn{4}{@{}l}{Setting: PH, $r=.20$}\\
RSF & $-45.9$ (0.8) & -- & --\\
CoxPH & $-42.0$ (0.7) & $+3.9$ (0.8) & --\\
Frailty Cox & $-36.7$ (0.6) & $+9.2$ (0.8) & $+5.3$ (0.2)\\
\midrule
\multicolumn{4}{@{}l}{Setting: PH, $r=.40$}\\
RSF & $-34.6$ (0.8) & -- & --\\
CoxPH & $-23.8$ (0.6) & $+10.9$ (0.7) & --\\
Frailty Cox & $-20.0$ (0.5) & $+14.6$ (0.7) & $+3.8$ (0.1)\\
\midrule
\multicolumn{4}{@{}l}{Setting: Non-PH, $r=.20$}\\
RSF & $-48.0$ (0.8) & -- & --\\
CoxPH & $-51.8$ (0.7) & $-3.8$ (0.8) & --\\
Frailty Cox & $-46.6$ (0.7) & $+1.5$ (0.8) & $+5.3$ (0.2)\\
\midrule
\multicolumn{4}{@{}l}{Setting: Non-PH, $r=.40$}\\
RSF & $-35.8$ (0.7) & -- & --\\
CoxPH & $-30.0$ (0.6) & $+5.8$ (0.7) & --\\
Frailty Cox & $-26.0$ (0.5) & $+9.9$ (0.7) & $+4.0$ (0.1)\\
\bottomrule
\end{tabular}
\end{table}

\begin{table}[H]
\caption{Simulation II, pairwise differences in the test log score over 200 replicates, summed over all 300 test subjects: the log predicted density of the observed event time for subjects with an event and the log predicted survival probability at the censoring time for censored subjects (method in the left column minus method in the top row; Monte Carlo standard error of the mean paired difference in parentheses; positive values favor the row method).}
\label{tab:sim2-elpd}
\centering
\small
\begin{tabular}{@{}lrrr@{}}
\toprule
 & TVSurvBART & RSF & CoxPH\\
\midrule
\multicolumn{4}{@{}l}{Censoring: $20\%$}\\
RSF & $-111.6$ (1.1) & -- & --\\
CoxPH & $-162.2$ (1.1) & $-50.6$ (1.1) & --\\
Frailty Cox & $-162.1$ (1.1) & $-50.5$ (1.1) & $+0.08$ (0.02)\\
\midrule
\multicolumn{4}{@{}l}{Censoring: $40\%$}\\
RSF & $-78.0$ (1.0) & -- & --\\
CoxPH & $-106.0$ (1.0) & $-27.9$ (0.9) & --\\
Frailty Cox & $-105.9$ (1.0) & $-27.9$ (0.9) & $+0.03$ (0.01)\\
\bottomrule
\end{tabular}
\end{table}

\subsection{Random-effect ablation}\label{supp:sim-ablation}
To isolate the contribution of the random effects, we refit two reduced versions of TVSurvBART: TVSurvBART with a random intercept only, which has $\bm z_i(t)=1$, so that $q=1$, and a half-$t_2(5)$ prior on the intercept standard deviation (Web Appendix~\ref{supp:leafprior}), and TVSurvBART without random effects, which omits $\bm b_i$ from the hazard. Both were fitted in two settings with $r=0.20$: Simulation~I under PH, whose data have a random intercept and slope with $\sigma_0=1$ and $\sigma_1=2$ (the slope multiplies $t/\tau$), and Simulation~II, whose data have no frailty. In each setting, the reduced models are fitted to the same $200$ replicates and training--test splits as the main analysis, with the same remaining priors and MCMC settings ($m=100$ trees, $20{,}000$ burn-in iterations, and $2{,}000$ retained draws, every $50$th). Web Table~\ref{tab:sim-ablation} reports the three fits and the benchmarks of the main analysis on the landmark criteria, the test log score, and the criteria along carried-forward paths.

In Simulation~I, removing all random effects makes the predictions slightly worse at $s=1$ and along carried-forward paths. At $s=1$, the paired differences from the full model are $+0.0017$ ($0.0003$) in the landmark Brier score and $-0.0011$ ($0.0006$) in AUC, and the coverage of the frozen-path truth falls from $0.941$ to $0.897$ although the intervals are wider (mean width $0.334$ against $0.308$). Along carried-forward paths, the differences are $+0.0010$ ($0.0001$) in IBS, $+0.0046$ ($0.0001$) in $\mathrm{RMSE}_S$, $-0.0061$ ($0.0004$) in AUC at $t=\tau$, and $-4.46$ ($0.21$) in the test log score, a loss that comes almost entirely from the subjects with events. TVSurvBART with a random intercept only matches the full model at $s=1$, with differences of $+0.0004$ ($0.0003$) in AUC and $+0.0001$ ($0.0001$) in Brier score and the same coverage, $0.941$. Along carried-forward paths, its differences are $-0.0001$ in IBS, $+0.0002$ in $\mathrm{RMSE}_S$, and $+0.0011$ in AUC at $t=\tau$ (standard errors at most $0.0002$), and $+0.27$ ($0.06$) in the test log score. At $s=2$, both reduced models are slightly better than the full model: their Brier scores differ from it by $-0.0018$ ($0.0002$) with a random intercept only and by $-0.0019$ ($0.0006$) without random effects, and from each other by $-0.0002$ ($0.0004$). In this setting, the predictive gain from the random effects therefore comes from the random intercept, and the random slope adds little to it.

We did not establish why the slope does not help, although the data were generated with $\sigma_1=2$. The full model underestimates $\sigma_1$, with a mean posterior median of $0.800$ (Web Table~\ref{tab:sim1-re}), and without the slope, the mean posterior median of $\sigma_0$ rises from $0.741$ to $0.866$. At $s=2$, the full model's predicted conditional survival probabilities vary less across the at-risk subjects of a replicate than those of the reduced models: the square root of the mean within-replicate variance is $0.188$, against $0.204$ with a random intercept only, $0.216$ without random effects, and $0.267$ for the frozen-path truth. The gain from dropping the slope at $s=2$ also increases with the full model's posterior mean of $\sigma_1$ (Spearman correlation $-0.29$ between this posterior mean and the paired Brier-score difference, random intercept only minus full model).

In Simulation~II, which has no frailty, neither reduced model changes the predictions materially. At the three landmarks, the paired differences from the full model are at most $0.0008$ in AUC, $0.0003$ in Brier score, and $0.0035$ in coverage in absolute value, so the low coverage of the frozen-path truth, $0.620$ to $0.668$ across the three fits and landmarks, does not come from the random effects. Along carried-forward paths, IBS, $\mathrm{RMSE}_S$, and AUC change by at most $0.0014$, and the test log score changes by $+0.02$ ($0.12$) with a random intercept only and by $+0.23$ ($0.14$) without random effects.

Both reduced models remain better than CoxPH, frailty Cox, and RSF in both settings: each of their paired differences from these benchmarks in landmark AUC and Brier score at every landmark, and in IBS, $\mathrm{RMSE}_S$, AUC at each horizon, and test log score along carried-forward paths, favors them by at least $12.9$ Monte Carlo standard errors. The smallest margin is in AUC over RSF at $s=2$ in Simulation~I, $+0.063$ ($0.005$) for both. In Simulation~I, removing the random effects gives up $4.6\%$ of the full model's landmark Brier-score advantage over CoxPH at $s=1$ and $15\%$ of that over RSF, none at $s=2$, and $4\%$ to $11\%$ of its carried-forward advantages over CoxPH in IBS, $\mathrm{RMSE}_S$, AUC at $t=\tau$, and test log score. Most of TVSurvBART's advantage over the benchmarks is therefore retained without random effects.

\begin{table}[H]
\caption{Random-effect ablation in Simulation~I (PH, $r=0.20$) and Simulation~II ($r=0.20$): means over 200 replicates. TVSurvBART is the full model. The next two rows refit it with a random intercept only and without random effects on the same data. (a) Landmark AUC and Brier score at the horizon $t=s+1$ for the test subjects at risk at the landmark time $s$, with histories frozen at the last visit at or before $s$ (Web Appendix~\ref{supp:sim-eval}), and coverage of the frozen-path truth by the $95\%$ intervals at $s=1$. (b) Mean paired difference from the full model in the test log score summed over the 300 test subjects (Monte Carlo standard error), positive values favoring the row method, and IBS, $\mathrm{RMSE}_S$, and AUC at $t=\tau$. Other Monte Carlo standard errors are at most $0.007$.}
\label{tab:sim-ablation}
\centering
\small
\setlength{\tabcolsep}{4pt}
\begin{tabular}{@{}lccccccc@{}}
\toprule
\multicolumn{8}{@{}l}{(a) Landmark predictions along frozen paths}\\
 & \multicolumn{3}{c}{AUC} & \multicolumn{3}{c}{Brier score} & Coverage\\
\cmidrule(lr){2-4}\cmidrule(lr){5-7}\cmidrule(lr){8-8}
 & $s=1$ & $s=2$ & $s=3$ & $s=1$ & $s=2$ & $s=3$ & $s=1$\\
\midrule
\multicolumn{8}{@{}l}{Simulation~I, PH, $r=.20$}\\
TVSurvBART & .756 & .815 & -- & .193 & .146 & -- & .941\\
Random intercept only & .756 & .817 & -- & .194 & .144 & -- & .941\\
No random effects & .755 & .817 & -- & .195 & .144 & -- & .897\\
CoxPH & .649 & .592 & -- & .229 & .208 & -- & --\\
Frailty Cox & .656 & .610 & -- & .225 & .202 & -- & --\\
RSF & .724 & .754 & -- & .204 & .166 & -- & --\\
\multicolumn{8}{@{}l}{Simulation~II, $r=.20$}\\
TVSurvBART & .896 & .907 & .903 & .097 & .094 & .106 & .624\\
Random intercept only & .897 & .907 & .903 & .097 & .094 & .106 & .620\\
No random effects & .897 & .907 & .903 & .097 & .094 & .106 & .621\\
CoxPH & .660 & .549 & .499 & .160 & .175 & .209 & --\\
Frailty Cox & .660 & .549 & .498 & .160 & .175 & .208 & --\\
RSF & .856 & .854 & .846 & .124 & .130 & .147 & --\\
\midrule
\multicolumn{8}{@{}l}{(b) Predictions along carried-forward paths}\\
 & \multicolumn{2}{c}{\makecell{Log score\\difference}} & \multicolumn{2}{c}{IBS} & \multicolumn{2}{c}{$\mathrm{RMSE}_S$} & \makecell{AUC,\\$t=\tau$}\\
\midrule
\multicolumn{8}{@{}l}{Simulation~I, PH, $r=.20$}\\
TVSurvBART & \multicolumn{2}{c}{--} & \multicolumn{2}{c}{.123} & \multicolumn{2}{c}{.098} & .866\\
Random intercept only & \multicolumn{2}{c}{$+0.27$ (0.06)} & \multicolumn{2}{c}{.123} & \multicolumn{2}{c}{.098} & .867\\
No random effects & \multicolumn{2}{c}{$-4.46$ (0.21)} & \multicolumn{2}{c}{.124} & \multicolumn{2}{c}{.102} & .860\\
CoxPH & \multicolumn{2}{c}{$-42.0$ (0.7)} & \multicolumn{2}{c}{.139} & \multicolumn{2}{c}{.140} & .721\\
Frailty Cox & \multicolumn{2}{c}{$-36.7$ (0.6)} & \multicolumn{2}{c}{.138} & \multicolumn{2}{c}{.136} & .725\\
RSF & \multicolumn{2}{c}{$-45.9$ (0.8)} & \multicolumn{2}{c}{.133} & \multicolumn{2}{c}{.118} & .775\\
\multicolumn{8}{@{}l}{Simulation~II, $r=.20$}\\
TVSurvBART & \multicolumn{2}{c}{--} & \multicolumn{2}{c}{.139} & \multicolumn{2}{c}{.143} & .854\\
Random intercept only & \multicolumn{2}{c}{$+0.02$ (0.12)} & \multicolumn{2}{c}{.139} & \multicolumn{2}{c}{.143} & .854\\
No random effects & \multicolumn{2}{c}{$+0.23$ (0.14)} & \multicolumn{2}{c}{.139} & \multicolumn{2}{c}{.143} & .853\\
CoxPH & \multicolumn{2}{c}{$-162.2$ (1.1)} & \multicolumn{2}{c}{.182} & \multicolumn{2}{c}{.248} & .700\\
Frailty Cox & \multicolumn{2}{c}{$-162.1$ (1.1)} & \multicolumn{2}{c}{.182} & \multicolumn{2}{c}{.248} & .703\\
RSF & \multicolumn{2}{c}{$-111.6$ (1.1)} & \multicolumn{2}{c}{.159} & \multicolumn{2}{c}{.190} & .766\\
\bottomrule
\end{tabular}
\end{table}

\subsection{Random-effect covariance and prior sensitivity}\label{supp:sim-re}
Web Table~\ref{tab:sim1-re} summarizes the posterior of the random-effect covariance, including the bias and RMSE of the posterior medians. Across the four Simulation~I settings, the mean posterior median of the intercept standard deviation ranges from $0.605$ to $0.741$ (true value $1$), and that of the slope standard deviation from $0.595$ to $1.322$ (true value $2$). For the slope standard deviation, the coverage of the $95\%$ credible intervals is $0.270$ under PH and $0.120$ under non-PH at $20\%$ censoring. At $40\%$ censoring, the posterior medians are larger and the intervals wider, and the coverage is $0.810$ and $0.725$. For the intercept standard deviation, the coverage ranges from $0.485$ to $0.755$. The mean posterior median of the correlation is positive in every setting, $0.190$ to $0.384$. Its intervals are wide, with mean width $1.313$ to $1.502$, and contain the true value $0$ in a proportion of at least $0.980$ of replicates. Under the default scale, the half-$t_2(5)$ prior density of each standard deviation at $2$ is $0.89$ times its value at $0$, so the prior penalizes the true values only mildly. The posteriors nonetheless move with the prior scales (Web Table~\ref{tab:sim1-prior}), and with at most one event per subject, the data carry limited information about the random-effect covariance. With irregular visits, the random-effect posterior shows similar shrinkage (Web Table~\ref{tab:sim1-re}, last column).

To assess prior sensitivity, we refit TVSurvBART in the PH setting with $20\%$ censoring on the first 100 replicates, with the data and chain seeds of the main analysis. Two alternative priors reduce the half-$t$ scales of both random-effect standard deviations from $A_1=A_2=5$ to $A_1=A_2=2$ or $A_1=A_2=1$. A third reduces the ensemble scale from $s_F\approx2.492$ to $s_F\approx1.661$, which narrows the central $95\%$ prior hazard-ratio interval for two inputs with uncorrelated tree values from $[0.001,1000]$ to $[0.01,100]$. Under the default prior, the refit reproduces the main-analysis fits of these replicates exactly. The smaller half-$t$ scales leave the predictions essentially unchanged (Web Table~\ref{tab:sim1-prior}): relative to the default, the mean paired differences in IBS, $\mathrm{RMSE}_S$ and AUC at $t=1,2,3$ are below $0.0005$ in absolute value, and those in the log score are $+0.02$ (Monte Carlo standard error $0.08$) for $A_1=A_2=2$ and $+0.14$ ($0.06$) for $A_1=A_2=1$. These scales lower the mean posterior median of the slope standard deviation from $0.800$ to $0.760$ and $0.681$ and the coverage of its $95\%$ intervals from $0.260$ to $0.210$ and $0.150$, so smaller scales shrink the slope standard deviation further. The smaller ensemble scale lowers the AUC at $t=2$ and $t=3$ by $0.005$ and $0.011$, raises IBS and $\mathrm{RMSE}_S$ by $0.001$ and $0.002$, and lowers the log score by $2.8$ ($0.1$). It also lowers the mean posterior median of the intercept standard deviation from $0.738$ to $0.639$. The default $s_F\approx2.492$, which corresponds to the interval $[0.001,1000]$ (Web Appendix~\ref{supp:leafprior}), was fixed before the analyses reported here and was not chosen by their held-out performance. All other simulation fits use the default prior. Because only replicates 1--100 enter, the default column of Web Table~\ref{tab:sim1-prior} differs slightly from the PH, $r=.20$ columns of Web Tables~\ref{tab:sim1-calibration}, \ref{tab:sim1-auc} and~\ref{tab:sim1-re}.

\begin{table}[H]
\caption{Simulation I, posterior summaries of the TVSurvBART random-effect covariance $\Sigma_b$ over 200 replicates per setting: intercept standard deviation $\sigma_0$ and slope standard deviation $\sigma_1$, the square roots of the diagonal entries of $\Sigma_b$ (the slope multiplies $t/\tau$), and correlation $\rho$. For each parameter, entries are the mean over replicates of the posterior median (Monte Carlo standard error), its bias (the mean minus the true value, with the same standard error), its RMSE (the root mean square over replicates of the posterior median minus the true value), the proportion of replicates whose $95\%$ credible interval contains the true value, and the mean width of that interval. The last column is the irregular-visit variant (Web Table~\ref{tab:sim1-irr}).}
\label{tab:sim1-re}
\centering
\small
\setlength{\tabcolsep}{3pt}
\begin{tabular}{@{}lccccc@{}}
\toprule
 & \multicolumn{2}{c}{PH} & \multicolumn{2}{c}{Non-PH} & Non-PH, irregular\\
\cmidrule(lr){2-3}\cmidrule(lr){4-5}\cmidrule(lr){6-6}
 & $r=.20$ & $r=.40$ & $r=.20$ & $r=.40$ & $r=.20$\\
\midrule
\multicolumn{6}{@{}l}{$\sigma_0$ (true value $1$)}\\
Posterior median (SE) & .741 (.008) & .640 (.008) & .720 (.008) & .605 (.009) & .710 (.009)\\
Bias & $-.259$ & $-.360$ & $-.280$ & $-.395$ & $-.290$\\%
RMSE & .280 & .377 & .300 & .415 & .314\\%
Coverage & .755 & .605 & .635 & .485 & .645\\
Width & .695 & .822 & .702 & .833 & .716\\
\midrule
\multicolumn{6}{@{}l}{$\sigma_1$ (true value $2$)}\\
Posterior median (SE) & .800 (.027) & 1.322 (.041) & .595 (.019) & 1.174 (.043) & .708 (.023)\\
Bias & $-1.200$ & $-.678$ & $-1.405$ & $-.826$ & $-1.292$\\%
RMSE & 1.260 & .887 & 1.429 & 1.022 & 1.333\\%
Coverage & .270 & .810 & .120 & .725 & .205\\
Width & 1.650 & 2.458 & 1.486 & 2.372 & 1.600\\
\midrule
\multicolumn{6}{@{}l}{$\rho$ (true value $0$)}\\
Posterior median (SE) & .322 (.012) & .384 (.010) & .190 (.011) & .287 (.009) & .244 (.010)\\
Bias & $.322$ & $.384$ & $.190$ & $.287$ & $.244$\\%
RMSE & .361 & .408 & .244 & .316 & .284\\%
Coverage & .980 & .990 & 1.000 & 1.000 & 1.000\\
Width & 1.380 & 1.313 & 1.502 & 1.417 & 1.455\\
\bottomrule
\end{tabular}
\end{table}

\begin{table}[H]
\caption{Simulation I, prior sensitivity of TVSurvBART in the PH setting with $20\%$ censoring, over replicates 1--100 with the data and chain seeds of the main analysis. The default prior has half-$t_2(5)$ priors on both random-effect standard deviations ($A_1=A_2=5$) and ensemble scale $s_F\approx2.492$. The next two columns change the common half-$t$ scale, and the last column changes $s_F$ only. Prediction entries are means over replicates, rounded to three decimals. The log score is the mean paired difference from the default (Monte Carlo standard error), positive values favoring the column prior. RMST and median biases are relative to the latent event times of the test subjects, with mean reference values $1.012$ and $1.016$ over these replicates. Random-effect entries are the mean posterior median followed by the coverage of the $95\%$ credible interval in brackets (true values $\sigma_0=1$, $\sigma_1=2$, $\rho=0$).}
\label{tab:sim1-prior}
\centering
\small
\setlength{\tabcolsep}{4pt}
\begin{tabular}{@{}lcccc@{}}
\toprule
 & Default & $A_1=A_2=2$ & $A_1=A_2=1$ & $s_F\approx1.661$\\
\midrule
\multicolumn{5}{@{}l}{Prediction}\\
RMST($\tau/2$) bias (SE) & $-.008$ (.002) & $-.008$ (.002) & $-.009$ (.002) & $-.011$ (.002)\\%
Median bias (SE) & $.006$ (.004) & $.006$ (.004) & $.006$ (.004) & $.003$ (.004)\\%
IBS & .123 & .123 & .123 & .124\\
$\mathrm{RMSE}_S$ & .098 & .098 & .098 & .100\\
AUC, $t=1$ & .765 & .765 & .765 & .765\\
AUC, $t=2$ & .826 & .826 & .826 & .821\\
AUC, $t=3$ & .867 & .867 & .868 & .856\\
Log score difference & -- & $+0.02$ (0.08) & $+0.14$ (0.06) & $-2.8$ (0.1)\\
\midrule
\multicolumn{5}{@{}l}{Random-effect posterior}\\
$\sigma_0$ & .738 [.740] & .737 [.710] & .734 [.680] & .639 [.350]\\
$\sigma_1$ & .800 [.260] & .760 [.210] & .681 [.150] & .731 [.220]\\
$\rho$ & .326 [1.000] & .353 [.980] & .412 [.970] & .270 [1.000]\\
\bottomrule
\end{tabular}
\end{table}

In Simulation~II, no frailty is generated and the fitted model is not correctly specified, so the random effects may absorb part of the misspecification. With $r=0.20$ and $r=0.40$, respectively, the mean posterior medians over the $200$ replicates are $0.129$ and $0.139$ for the intercept standard deviation $\sigma_0$ and $0.265$ and $0.319$ for the slope standard deviation $\sigma_1$, where the slope multiplies $t/\tau$ with $\tau=5$. The mean upper $95\%$ credible limits are $0.399$ and $0.422$ for $\sigma_0$ and $0.801$ and $0.991$ for $\sigma_1$, far below $4.08$, the prior median of each standard deviation. The mean posterior median of the correlation $\rho$ is $-0.195$ and $-0.176$, and the mean width of its $95\%$ credible intervals, $1.638$ and $1.633$, is somewhat below the width $1.9$ of the central $95\%$ interval of its uniform prior.

\section{Case study details}\label{supp:app}

\subsection{Cohort, endpoint and exclusions}\label{supp:app-cohort}
Eligibility for the case study of main-text Section~6 required at least 6 months of adalimumab use and clinically inactive disease at two consecutive visits at least 5 months apart. Inactive disease was defined as a joint count of $0$, physician global assessment $\leq 1$, and no active uveitis. Subjects with systemic juvenile idiopathic arthritis (JIA), other systemic rheumatic diseases (e.g., lupus), or inflammatory bowel disease were excluded.

The endpoint is time to the first recorded event satisfying the registry's composite disease-flare definition (\texttt{flare\_12345}). Its date is the first on which either (a) a change in therapy (an adalimumab dose change, discontinuation or restart, the start of a new disease-modifying antirheumatic drug (DMARD), or glucocorticoid use) was recorded for disease flare, active or uncontrolled disease, active uveitis, or lack of primary response, or (b) a clinic visit recorded an active joint count $\ge 1$, a physician global assessment $>1$ (0--10 scale), or new uveitis activity. Follow-up for each subject runs from the index date to the flare date for subjects who flare, and to the last recorded visit otherwise. Visit records after the flare date are discarded. The index date marks cohort entry based on inactive disease at two consecutive visits and anchors the recorded adalimumab continuation, tapering, and discontinuation patterns. No earlier visits are available. Fitting uses the recorded event and censoring dates, measured in years from index.

Starting from $862$ subjects ($3{,}830$ visit records), we excluded: (i) five subjects with internally inconsistent outcome records: one with a flare date recorded despite a negative flare indicator, and four whose flare date in the visit file preceded that in the registry outcome file by $54$ to $197$ days; and (ii) $126$ subjects with zero at-risk follow-up, i.e.\ whose event or censoring date coincided with the index date. Of the $126$ subjects in (ii), $125$ were censored at the index date with the index visit as their only record, and one had a flare recorded on the index date. Subjects without follow-up carry no information about the time to flare after index, and a flare recorded on the index date is incompatible with the entry criterion of inactive disease at that visit, so we treat it as a recording inconsistency. The analytic cohort is therefore defined by an event or censoring date after the index date. This is a weaker requirement than a post-index visit, because a flare between the index visit and the next visit gives positive time at risk and a constructed terminal record. The excluded subjects entered late: the extract ends on 26 August 2022, and $82.5\%$ of them have index dates in 2021 or 2022, against $19.6\%$ of the analytic cohort.

\subsection{Predictors and missing values}\label{supp:app-missing}
Predictors include lagged weekly adalimumab dose and off-drug status; visit-level disease activity, pain, body mass index, joint findings, JIA category, and uveitis history; concurrent use of methotrexate and other recorded DMARDs; and baseline age, sex, race and ethnicity, antinuclear antibody status, and human leukocyte antigen B27 (HLA-B27) status. Dose is taken from the latest dosing period starting before the visit that opens the at-risk interval.

Missing visit measurements were carried forward within subject, which filled at least one value in $872$ of the $2{,}706$ visit records ($32.2\%$). The values still missing after this step are leading, i.e.\ no value of the predictor had yet been recorded for the subject. For pain intensity, $141$ of the $289$ affected subjects have no recorded value during follow-up. Leading missing values occur in $791$ of the $2{,}765$ records ($28.6\%$) and in $349$ subjects ($47.7\%$) at entry, mostly for pain intensity ($668$ records, $24.2\%$), patient global assessment ($157$, $5.7\%$), active enthesitis ($154$, $5.6\%$) and BMI ($106$, $3.8\%$). In the main analysis, leading missing numeric values were replaced by training-fold medians with indicators of a not-yet-observed value. A missing categorical value formed a separate level. Web Appendix~\ref{supp:app-mice} compares it with multiple imputation (Web Table~\ref{tab:app-mice}).

\subsection{Model settings and evaluation}\label{supp:app-eval}
In the application, the trees receive the covariates only, without the scaled time input, so time enters through the baseline hazard and the random slope. TVSurvBART uses $m=100$ trees with the default priors: a half-$t_2(5)$ prior on each random-effect standard deviation and ensemble scale $s_F\approx2.492$, which with $m=100$ trees gives leaf hyperparameters $c\approx16.60$ and $d\approx16.10$. The random slope multiplies $t/\tau$ with $\tau=5$ years. The piecewise cut points are chosen from each training set: scanning candidate points every $91$ days from the index date, a cut point is placed once the current piece contains $50$ training flares, and a last piece with fewer flares is merged with the previous one. In most folds, this places cut points near $0.5$, $1$, $1.5$, and $2$ years. The spline baseline has boundary knots at $0$ and at the maximum follow-up time and two equally spaced interior knots. Each fit runs one chain of $20{,}000$ burn-in iterations followed by $100{,}000$ iterations, of which every $50$th is kept.

In the cross-validation of main-text Section~6, all records of a subject are held out together. As in the simulations, the held-out survival curves of TVSurvBART average over the retained posterior draws and integrate out one random effect along each subject's whole path by Gauss--Hermite quadrature with $16$ nodes per dimension (Web Appendix~\ref{supp:sim-eval}). The cross-validated sensitivity analyses of Web Appendices~\ref{supp:app-prior} and~\ref{supp:app-mice} use the folds of the first repetition. The default and prior-sensitivity fits also use its preprocessing and chain seeds.

\paragraph{Landmark evaluation.} The landmark evaluation of main-text Section~6.3 follows Web Appendix~\ref{supp:sim-eval} with landmarks $s=1$ and $2$ years and horizons $t=s+1$, plus $t=3$ from $s=1$. Each subject is held out once in every repetition, so the pooled held-out subjects at risk are the same in all repetitions: $438$ at $s=1$ and $194$ at $s=2$. Their frozen path consists of the at-risk intervals that start at or before $s$, the last of which is carried forward beyond $s$ as one interval. Visits fall on whole days, so no subject has a visit at $s$. The last visit precedes $s$ by a median of $0.33$ years at $s=1$ and $0.35$ years at $s=2$. The trees receive no time input, so no time input is held fixed. TVSurvBART predicts $\bar S_i(t)/\bar S_i(s)$, the ratio of posterior means, along this path (Web Appendix~\ref{supp:prediction}). CoxPH, frailty Cox, with $41$ Gauss--Hermite nodes for the frailty along the whole path, and RSF predict along the same path as in Web Appendix~\ref{supp:sim-eval}. Because $s$ falls inside the last interval of every frozen path, the forest's ratio, which defines its prediction, need not equal its chained conditional survival from $s$, as with irregular visits in the simulations. Within the risk set, $26.7\%$, $42.2\%$, and $35.1\%$ of the subjects are censored before $t$ for $(s,t)=(1,2)$, $(1,3)$, and $(2,3)$, so the Kaplan--Meier censoring weights, which ignore the covariates, carry more weight than in most simulation settings, where on average $3\%$ to $27\%$ of the subjects at risk are censored before $t$. The criteria are computed in each repetition on the pooled held-out subjects at risk and averaged over the $100$ repetitions. Standard errors come from $1{,}000$ bootstrap resamples of the subjects at risk, the same resample for every repetition and method, with the fitted models held fixed, so they omit the variation from refitting. Across repetitions, the standard deviation of a method's AUC is $0.009$ to $0.026$. Web Table~\ref{tab:app-landmark-full} adds the index of prediction accuracy, the calibration-in-the-large, and the paired differences from piecewise TVSurvBART.

\begin{table}[H]
\caption{Case study, landmark prediction averaged over $100$ repetitions of $10$-fold cross-validation (main-text Table~2), with the index of prediction accuracy (IPA), the calibration-in-the-large (mean predicted minus risk-set Kaplan--Meier survival at $t$, positive when survival is over-predicted), and the differences piecewise TVSurvBART (PW) minus each method. Bootstrap standard errors in parentheses ($1{,}000$ resamples of the subjects at risk, fitted models held fixed). Differences are computed from unrounded values and can differ by $.001$ from the difference of the displayed entries. NCS: natural-cubic-spline baseline; RSF: random survival forest.}
\label{tab:app-landmark-full}
\centering
\small
\setlength{\tabcolsep}{2pt}
\begin{tabular}{@{}lcccccc@{}}
\toprule
 & AUC & Brier score & IPA & Calibration & $\Delta$AUC & $\Delta$Brier\\
\midrule
\multicolumn{7}{@{}l}{$s=1$, $t=2$: $438$ at risk, $127$ flares by $t$, $194$ flare-free at $t$}\\
TVSurvBART (PW) & .572 (.033) & .219 (.010) & $+.023$ (.026) & $+.017$ & -- & --\\
TVSurvBART (NCS) & .567 (.032) & .220 (.008) & $+.020$ (.023) & $-.008$ & $+.005$ (.004) & $-.001$ (.002)\\
CoxPH & .572 (.033) & .224 (.011) & $+.004$ (.035) & $+.007$ & $.000$ (.012) & $-.004$ (.003)\\
Frailty Cox & .572 (.033) & .224 (.011) & $+.004$ (.035) & $+.008$ & $.000$ (.012) & $-.004$ (.003)\\
RSF & .617 (.030) & .211 (.010) & $+.061$ (.024) & $+.024$ & $-.045$ (.020) & $+.008$ (.003)\\
\midrule
\multicolumn{7}{@{}l}{$s=1$, $t=3$: $438$ at risk, $178$ flares by $t$, $75$ flare-free at $t$}\\
TVSurvBART (PW) & .577 (.037) & .246 (.007) & $+.005$ (.030) & $+.031$ & -- & --\\
TVSurvBART (NCS) & .568 (.037) & .246 (.006) & $+.006$ (.025) & $+.019$ & $+.010$ (.005) & $.000$ (.002)\\
CoxPH & .585 (.037) & .250 (.010) & $-.010$ (.040) & $+.024$ & $-.008$ (.014) & $-.004$ (.004)\\
Frailty Cox & .585 (.037) & .250 (.010) & $-.010$ (.040) & $+.024$ & $-.008$ (.014) & $-.004$ (.004)\\
RSF & .544 (.037) & .252 (.007) & $-.019$ (.033) & $+.061$ & $+.033$ (.022) & $-.006$ (.005)\\
\midrule
\multicolumn{7}{@{}l}{$s=2$, $t=3$: $194$ at risk, $51$ flares by $t$, $75$ flare-free at $t$}\\
TVSurvBART (PW) & .570 (.049) & .214 (.016) & $+.014$ (.039) & $+.011$ & -- & --\\
TVSurvBART (NCS) & .566 (.048) & .215 (.017) & $+.006$ (.033) & $+.027$ & $+.004$ (.006) & $-.002$ (.002)\\
CoxPH & .583 (.048) & .216 (.017) & $+.004$ (.051) & $+.002$ & $-.012$ (.019) & $-.002$ (.004)\\
Frailty Cox & .583 (.048) & .216 (.017) & $+.004$ (.051) & $+.002$ & $-.012$ (.019) & $-.002$ (.004)\\
RSF & .455 (.050) & .228 (.019) & $-.052$ (.051) & $+.031$ & $+.116$ (.038) & $-.014$ (.006)\\
\bottomrule
\end{tabular}
\end{table}

\paragraph{Retrospective comparison.} The retrospective comparison of main-text Section~6.4 evaluates each method along the held-out subjects' observed, carried-forward histories (Web Appendix~\ref{supp:sim-eval}). Its cumulative/dynamic AUC, the path AUC, uses IPCW from the observed outcomes with the marker $1-\widehat S_i(t)$ at $t$ from $0.3$ to $3$ years and is computed in each held-out fold and averaged over the folds and repetitions (main-text Figure~4). The secondary interval-probability score (Web Appendix~\ref{supp:scores}) summarizes the probability assigned between the last visit without a recorded flare, $t_{i,\mathrm{last}}$, and the recorded date $T_i$. Of $361$ flares, $302$ are recorded at visits (median interval $184$ days). The other $59$ are dated between visits (median $126$ days). Every method uses the same intervals. This score summarizes probabilities over broader time windows than the recorded-date fitting likelihood. We linearly interpolate survival probabilities on a common grid consisting of $0$ and the points from $0.3$ to $3$ years in $0.1$-year steps, so the score depends on each method's curve only through its values on this grid. Fold scores are summed within each repetition and averaged over repetitions. Standard errors for paired contrasts use the Nadeau--Bengio correction \citep{NadeauBengio2003} and are approximate. Web Table~\ref{tab:app-cv} gives the score contrasts.

\begin{table}[H]
\centering
\caption{Case study, secondary interval-probability score differences over the first three years, row model minus column model, as cohort-total differences averaged over $100$ cross-validation repetitions (negative values favor the column model); approximate Nadeau--Bengio standard errors in parentheses. PW: piecewise baseline; NCS: natural-cubic-spline baseline; RSF: random survival forest.}
\label{tab:app-cv}
\small
\begin{tabular}{@{}lcccc@{}}
\toprule
 & \multicolumn{2}{c}{TVSurvBART} & RSF & CoxPH \\
\cmidrule(lr){2-3}
 & PW & NCS & & \\
\midrule
TVSurvBART (NCS) & $-8.8$ (6.0) & -- & -- & -- \\
RSF & $-34.1$ (9.8) & $-25.3$ (10.5) & -- & -- \\
CoxPH & $-35.8$ (7.6) & $-27.0$ (9.8) & $-1.7$ (11.7) & -- \\
Frailty Cox & $-35.9$ (7.6) & $-27.1$ (9.7) & $-1.8$ (11.7) & $-0.1$ (0.2) \\
\bottomrule
\end{tabular}
\end{table}

\subsection{Random-effect covariance}\label{supp:app-re}
To summarize the random-effect covariance, we fit TVSurvBART to the full analytic cohort. Each chain has the same length as the cross-validation and simulation fits, $20{,}000$ burn-in iterations followed by $100{,}000$ iterations of which every $50$th is kept. With the piecewise baseline (Web Appendix~\ref{supp:app-prior}, Web Table~\ref{tab:app-prior}, Default column), the pooled chains give a posterior median of the intercept standard deviation $\sigma_0$ of $0.372$ ($95\%$ credible interval $0.024$ to $0.887$), of the slope standard deviation $\sigma_1$ of $0.650$ ($0.038$ to $2.236$), where the slope multiplies $t/5$, and of the correlation $\rho$ of $0.059$ ($-0.779$ to $0.831$). Under the piecewise baseline, the data therefore rule out the very large standard deviations that the prior allows, but they do not separate small values of either standard deviation from zero, and they carry little information about $\rho$. With the spline baseline, the posterior medians are $0.720$ ($0.062$ to $1.399$) for $\sigma_0$, $1.095$ ($0.052$ to $4.259$) for $\sigma_1$, and $0.157$ ($-0.749$ to $0.864$) for $\rho$.

\subsection{Prior sensitivity}\label{supp:app-prior}
The prior sensitivity analysis follows that of the simulations (Web Appendix~\ref{supp:sim-re}). We refit TVSurvBART with the piecewise baseline on the ten folds of the first repetition and on the full cohort under three alternative priors: the half-$t$ scales of both random-effect standard deviations reduced from $A_1=A_2=5$ to $A_1=A_2=2$ or $A_1=A_2=1$, and, separately, the ensemble scale reduced from $s_F\approx2.492$ to $s_F\approx1.661$, which narrows the central $95\%$ prior hazard-ratio interval for two inputs with uncorrelated tree values from $[0.001,1000]$ to $[0.01,100]$. Prediction changes little (Web Table~\ref{tab:app-prior}). Relative to the default, the path AUC at $t=1,2,3$ changes by at most $0.003$ under the smaller half-$t$ scales and at most $0.004$ under the smaller ensemble scale. In the three landmark windows, the landmark AUC changes by at most $0.004$ and the landmark Brier score by at most $0.002$ (paired bootstrap standard errors at most $0.007$ and $0.001$). The interval-probability score changes by $-0.2$ (SE $0.6$) for $A_1=A_2=2$, $+0.2$ (SE $0.4$) for $A_1=A_2=1$, and $+2.2$ (SE $1.4$) for $s_F\approx1.661$, with standard errors over the ten folds and the Nadeau--Bengio correction. All other case-study fits use the default prior. The posterior of $\sigma_0$ is nearly unchanged across the half-$t$ scales, with medians from $0.345$ to $0.372$ and upper limits from $0.844$ to $0.887$, although the prior medians range from $0.82$ to $4.08$. The posterior of $\sigma_1$ moves with the prior, with medians $0.484$, $0.618$, and $0.650$ and upper limits $1.624$, $1.955$, and $2.236$ for half-$t$ scales $1$, $2$, and $5$.

\begin{table}[H]
\caption{Prior sensitivity of TVSurvBART with the piecewise baseline. The default prior has half-$t_2$ priors with scale $A_1=A_2=5$ on the two random-effect standard deviations and ensemble scale $s_F\approx2.492$; each other column changes one of these. Prediction entries come from the ten folds of the first cross-validation repetition, with the folds, preprocessing, and chain seeds of the default fits: path AUC is the mean over folds; the landmark AUC and Brier score are computed as for main-text Table~2 (Sections~6.2 and~6.3) on the pooled held-out subjects at risk at $s$ ($438$ at $s=1$, $194$ at $s=2$), with IPCW estimated within the risk set, and the Default landmark entries are those of the first of the $100$ repetitions averaged in that table; and the score difference is the cohort-total interval-probability score minus that of the default (Nadeau--Bengio standard error over folds). The landmark differences from the default have paired bootstrap standard errors ($2{,}000$ resamples of the subjects at risk, fitted models held fixed) of at most $0.007$ for AUC and $0.001$ for Brier score. Random-effect entries are posterior medians and $95\%$ credible intervals from four pooled full-cohort chains; $\sigma_1$ multiplies $t/5$. The last two rows are taken over $\sigma_0$, $\sigma_1$, and $\rho$.}
\label{tab:app-prior}
\centering
\small
\setlength{\tabcolsep}{4pt}
\begin{tabular}{@{}lcccc@{}}
\toprule
 & Default & $A_1=A_2=2$ & $A_1=A_2=1$ & $s_F\approx1.661$\\
\midrule
\multicolumn{5}{@{}l}{Prediction}\\
Path AUC, $t=1$ & .625 & .625 & .625 & .627\\
Path AUC, $t=2$ & .608 & .605 & .608 & .604\\
Path AUC, $t=3$ & .602 & .600 & .603 & .605\\
Landmark AUC, $s=1$, $t=2$ & .575 & .576 & .576 & .575\\
Landmark AUC, $s=1$, $t=3$ & .575 & .575 & .578 & .579\\
Landmark AUC, $s=2$, $t=3$ & .574 & .576 & .574 & .577\\
Landmark Brier score, $s=1$, $t=2$ & .219 & .219 & .219 & .218\\
Landmark Brier score, $s=1$, $t=3$ & .245 & .246 & .246 & .244\\
Landmark Brier score, $s=2$, $t=3$ & .213 & .213 & .212 & .212\\
Score difference & -- & $-0.2$ (0.6) & $+0.2$ (0.4) & $+2.2$ (1.4)\\
\midrule
\multicolumn{5}{@{}l}{Random-effect posterior, full cohort}\\
$\sigma_0$ & .372 & .345 & .353 & .279\\
\quad $95\%$ interval & (.024, .887) & (.015, .848) & (.019, .844) & (.013, .771)\\
$\sigma_1$ & .650 & .618 & .484 & .591\\
\quad $95\%$ interval & (.038, 2.236) & (.031, 1.955) & (.020, 1.624) & (.028, 1.976)\\
$\rho$ & .059 & .053 & .052 & $-.010$\\
\quad $95\%$ interval & ($-.779$, .831) & ($-.786$, .827) & ($-.822$, .847) & ($-.799$, .797)\\
Prior median of $\sigma_0$, $\sigma_1$ & 4.08 & 1.63 & .82 & 4.08\\
Maximum $\widehat R$ & 1.005 & 1.007 & 1.002 & 1.008\\
Minimum bulk ESS & 805 & 464 & 1{,}049 & 757\\
\bottomrule
\end{tabular}
\end{table}

\subsection{Multiple imputation of leading missing values}\label{supp:app-mice}
We compared the median imputation with multiple imputation by chained equations \citep{vanBuurenGroothuisOudshoorn2011} on the ten folds of the first repetition. Within each training fold, five imputations of each subject's leading-missing entry values were drawn, each after $20$ iterations, from an imputation model that contains the other predictors at entry, the flare indicator and the Nelson--Aalen cumulative hazard at the subject's follow-up time \citep{WhiteRoyston2009}. Held-out subjects were imputed without their outcomes, by a second chained-equation run on the same predictors without the two outcome variables, with the imputation models fitted to the training subjects only. The imputed entry value replaces the subject's leading-missing records, and the not-yet-observed indicators are dropped. TVSurvBART with the piecewise baseline and Cox regression were fitted to each imputed training set, each fit with its own chain seed, and fit $k$ predicts held-out imputation $k$. The held-out predictions are averaged over the five imputations. For the landmark criteria, the survival probabilities at $s$ and $t$ are averaged before their ratio is taken. The median-imputation results are those of the default fits.

The two imputation methods give similar results (Web Table~\ref{tab:app-mice}). For TVSurvBART, multiple imputation changes the path AUC by $+0.004$ (SE $0.011$) at $t=1$, $-0.008$ (SE $0.007$) at $t=2$, and $-0.006$ (SE $0.013$) at $t=3$. For CoxPH, the changes are $-0.001$, $-0.013$, and $-0.018$. The interval-probability score increases by $3.4$ (SE $2.6$) for TVSurvBART and by $6.6$ (SE $3.9$) for CoxPH, so the TVSurvBART advantage over CoxPH in this score is $33.8$ (SE $7.8$) under median imputation and $30.5$ (SE $6.9$) under multiple imputation in this repetition. Multiple imputation changes the landmark AUC of TVSurvBART by $-0.011$ to $+0.001$ and lowers that of CoxPH in all three windows, by $0.007$ to $0.009$ (paired bootstrap standard errors $0.006$ to $0.011$, fitted models held fixed), and changes the landmark Brier score by at most $0.002$. Standard errors of the path criteria and the score are computed over the ten folds with the Nadeau--Bengio correction, and those of the landmark criteria from $2{,}000$ paired bootstrap resamples of the subjects at risk with the fitted models held fixed.

\begin{table}[H]
\caption{Median imputation (the default fits) and multiple imputation by chained equations (MICE, five imputations) of leading missing values, first cross-validation repetition, TVSurvBART with the piecewise baseline and CoxPH. Path AUC is the mean over folds, the landmark AUC and Brier score are computed as for main-text Table~2 (Sections~6.2 and~6.3) on the pooled held-out subjects at risk at $s$ ($438$ at $s=1$, $194$ at $s=2$), with IPCW estimated within the risk set, and the interval-probability score is the cohort total. Differences are MICE minus median imputation, with Nadeau--Bengio standard errors over folds for the path criteria and the score, and paired bootstrap standard errors ($2{,}000$ resamples of the subjects at risk, fitted models held fixed) for the landmark criteria. The median-imputation landmark entries are those of the first of the $100$ repetitions averaged in main-text Table~2, so they differ from the averages there. Differences are computed from unrounded values and can differ by $.001$ from the difference of the displayed entries.}
\label{tab:app-mice}
\centering
\small
\setlength{\tabcolsep}{2.5pt}
\begin{tabular}{@{}lcccccc@{}}
\toprule
 & \multicolumn{3}{c}{TVSurvBART (PW)} & \multicolumn{3}{c}{CoxPH}\\
\cmidrule(lr){2-4}\cmidrule(lr){5-7}
 & Median & MICE & Difference & Median & MICE & Difference\\
\midrule
Path AUC, $t=1$ & .625 & .629 & $+.004$ (.011) & .609 & .608 & $-.001$ (.014)\\
Path AUC, $t=2$ & .608 & .600 & $-.008$ (.007) & .597 & .585 & $-.013$ (.012)\\
Path AUC, $t=3$ & .602 & .597 & $-.006$ (.013) & .625 & .606 & $-.018$ (.014)\\
Landmark AUC, $s=1$, $t=2$ & .575 & .575 & $.000$ (.006) & .566 & .559 & $-.007$ (.009)\\
Landmark AUC, $s=1$, $t=3$ & .575 & .576 & $+.001$ (.007) & .581 & .573 & $-.008$ (.008)\\
Landmark AUC, $s=2$, $t=3$ & .574 & .563 & $-.011$ (.010) & .602 & .593 & $-.009$ (.011)\\
Landmark Brier score, $s=1$, $t=2$ & .219 & .219 & $.000$ (.001) & .223 & .222 & $-.001$ (.002)\\
Landmark Brier score, $s=1$, $t=3$ & .245 & .246 & $+.001$ (.001) & .250 & .252 & $+.002$ (.002)\\
Landmark Brier score, $s=2$, $t=3$ & .213 & .213 & $+.001$ (.001) & .214 & .215 & $+.001$ (.002)\\
Interval-probability score & $-945.7$ & $-942.3$ & $+3.4$ (2.6) & $-979.5$ & $-972.9$ & $+6.6$ (3.9)\\
\bottomrule
\end{tabular}
\end{table}

\bibliographystyle{biom}
\bibliography{coxbart_references}